\documentclass[5p]{elsarticle}

\usepackage{hyperref}

\usepackage{epsf}
\usepackage{graphicx}% Include figure files
\usepackage{psfig}% Include figure files
\usepackage{dcolumn}% Align table columns on decimal point
\usepackage{bm}% bold math
\usepackage{subfigure}
\usepackage{sidecap}
\usepackage{color}
\usepackage{amsthm}
\usepackage{amssymb}
\usepackage{amsmath}
\usepackage{pb-diagram}
\usepackage{algorithm}
\usepackage{algorithmic}
\newcommand{\fn}{{\mathsf{FN}}}
\newcommand{\cn}{{\mathsf{CN}}}
\newcommand{\pd}{{\mathsf{PD}}}
\newcommand{\spd}{{\mathcal{PD}}}

\newtheorem{thm}{Theorem}[section]
\newtheorem{lem}[thm]{Lemma}
\newtheorem{defn}[thm]{Definition}
\newtheorem{cor}[thm]{Corollary}
\newtheorem{prop}[thm]{Proposition}

\newtheorem{rem}[thm]{Remark}

\newcommand{\setof}[1]{\left\{ {#1}\right\}}

\newcommand{\R}{{\mathbb{R}}}

\newcommand{\Z}{{\mathbb{Z}}}

\def\setof#1{\left\{{#1}\right\}}
\def\ang#1{\langle {#1} \rangle}
\def\ip#1#2{\left\langle {#1},{#2} \right\rangle}
\definecolor{gray85}{gray}{0.85} % 15%
\definecolor{gray8}{gray}{0.8} % 20%
\definecolor{gray7}{gray}{0.7} % 30%
\definecolor{gray6}{gray}{0.6} % 40%
\definecolor{gray5}{gray}{0.5} % 50%
\definecolor{gray4}{gray}{0.4} % 60%
\definecolor{gray35}{gray}{0.35} % 65%

\include{correctm}

\begin{document}

\title{
Quantifying force networks in particulate systems
}

\author[ru]{ Miroslav Kram\' ar}
\ead{miroslav@math.rutgers.edu}

\author[ru]{Arnaud Goullet}
\ead{arnaud.goullet@gmail.com}

\author[njit]{Lou Kondic}
\ead{kondic@njit.edu}

\author[ru]{Konstantin Mischaikow}
\ead{mischaik@math.rutgers.edu}

\address[ru]{Department of Mathematics,
Hill Center-Busch Campus,
Rutgers University,
110 Frelinghusen Rd,
Piscataway, NJ  08854-8019, USA}
\address[njit]{Department of Mathematical Sciences,
New Jersey Institute of Technology,
University Heights,
Newark, NJ 07102}

\date{\today}

\begin{abstract}

We present  mathematical models based on persistent homology for analyzing
force distributions in particulate systems. We define three distinct chain complexes
of these distributions:  
{\em digital}, {\em position}, and {\em interaction}, 
motivated by different types of data that may be available from experiments and simulations, e.g.
digital images, location of the particles, and the forces between particles, respectively.
We describe how algebraic topology, in particular, homology allows one to obtain algebraic
representations of the geometry captured by these complexes.
For each  complex we define an associated force network from which persistent 
homology is computed.
Using numerical data obtained from discrete element simulations of a system of particles undergoing
slow compression, we demonstrate how persistent homology can be used to 
compare the force distributions in different systems, and
discuss the differences between the properties of digital, position, and interaction force 
networks.    To conclude, we formulate well-defined measures quantifying
differences between force networks corresponding to different states of a system, and therefore allow 
to analyze in precise terms dynamical properties of force networks.
\end{abstract}

\maketitle

\section{Introduction}

Particulate systems consisting of a large number of particles have attracted significant attention in the last decades. 
Despite significant research on these systems,  their properties are still not well understood and some of them appear to be rather elusive.  
The fact that the forces do not propagate uniformly in  systems made of interacting particles
has been established in a number of different systems, including granular matter, colloids, gels, emulsions and foams, 
see, e.g.,~\cite{brujic03,cates98,liu95,majmudar05a}. It is 
well accepted that the interparticle forces play a key role in determining the mechanical properties of 
static and dynamic systems; see e.g.~\cite{alexander_physrep05} for an extensive review of the role of interaction networks in the
context of amorphous solids.
However there are no universal methods for describing  and quantifying  
relevant aspects of the interparticle  forces.  
For example, even the commonly used notion of `force chain' -- which we take to mean a connected  set of particles interacting  by a larger than average force -- is not generally defined.
One important goal of this paper is to present a method that can be used to describe precisely global 
properties of force networks in both static and dynamic settings.

Forces between interacting particles have been considered extensively from statistical point of view, in particular in the context of
dense granular matter (DGM).   For example, the works by Radjai and collaborators, see, e.g.~\cite{radjai_96b,radjai98b}, 
discussed the differences in the probability density functions  of strong and weak forces (distinguished by the forces being larger
or smaller than the average one) arising in simulations; Behringer and collaborators explored these forces in the 
experimental systems  built from photoelastic particles, see e.g.~\cite{majmudar05a}.   
Possible universality of the force distributions has been considered~\cite{ostojic06},   as well as the connections between
force and contact networks~\cite{tighe_sm10}.  These works have provided
a significant insight into statistical properties of the force distributions but by design do not focus on 
the structural properties of force networks.    

Only recently, attempts have been made to move beyond purely statistical description and consider in 
more detail  the properties of these networks.   Examples of recent studies include works by Tordesillas 
and collaborators, see~\cite{peters05,tordesillas_bob_pre12, tordesillas_pre10} and the references therein.
These studies include extensive discussion of local properties of networks of forces  based on the forces that particles
experience and on their connectivity, including appropriately defined force chains and force cycles with a particular
emphasis on cycles of length 3 and 4.
Furthermore, these studies introduce mesoscopic network properties such as degree, clustering coefficient 
and centrality which describe particle arrangements. Averaging these properties over the entire network allows to discuss
the connection between the changes observed in the macroscopic network properties to the underlying structural 
rearrangements of the material.

Alternative approaches use network-type of analysis to discuss the properties of force networks~\cite{daniels_pre12, herrera_pre11,walker_pre12}.
These works provide a significant new insight and confirm that the properties of force networks are relevant in the 
context of propagation of acoustic signals~\cite{daniels_pre12}, fracture~\cite{herrera_pre11}, and compression and shear~\cite{walker_pre12}.
Topology based approach has been considered as well, with focus on the contact network topology in
isotropically compressed~\cite{arevalo_pre10}  and tapped granular media~\cite{arevalo_pre13}.
A similar approach is considered in our recent work~\cite{epl12}, where we discuss 
connectivity of force networks, including the dependence of the number of connected components and holes/loops (quantified 
by the Betti numbers),   on the  (normal) force between the particles.   While that work uncovered
some intricate properties of force networks and allowed to connect the results of topology based analysis to the ones obtained
using standard percolation-based approach, it was still based essentially on counting components and loops 
at fixed magnitudes of force.  As such, it thus does not provide an understanding of how these geometric structures persist
through different magnitudes of the force.

In \cite{pre13} we introduced the use of persistent homology~\cite{carlsson, edelsbrunner:harer} to DGM. 
More recently,  these ideas have been employed in the context of  tapped systems~\cite{trevijano_14}.
Conceptually persistent homology is preferable to the above mentioned Betti number analysis.  By design persistent homology
measures the same geometric structures as the Betti number analysis, but simultaneously records how these structures
appear, disappear or persist through different magnitudes of the force. Thus, two networks of forces could
produce identical information on the level of Betti numbers, i.e., the number of connected components and loops, but still
have distinct global structures in the sense that as one varies the magnitudes of the forces the relationships between
the connected components and loops are different. 
Therefore, the results presented in~\cite{pre13} provide better quantification  of  the properties of considered force networks 
and shed  new light on the differences between the systems that differ  by their frictional properties and particle size distributions.  

It should be noted however, that persistent homology is an abstract tool. Hence, there is considerable freedom as to how it can be employed. 
In this paper we provide  a firm mathematical background for using persistent homology in the context of DGM.
In addition, we discuss  different concepts for constructing and comparing the persistence diagrams. 
This allows us to compare the features of different force networks both locally and globally and hence
is complementary to the approaches, discussed above, that consider local properties of force networks.
Furthermore, the ability to compare  different force networks is crucial for quantifying the dynamical properties of DGM. 

In the next section we give an overview of persistence homology 
and the structure of the paper.

\section{Overview}
\label{sec:overview}

In this paper we introduce the concept of a force network, which is designed to model force interactions between the particles.
The definition varies depending on available form of the data, but every force network is described by a scalar function 
$f : D \to \R$.  The domain $D$ models the particles and the function $f$ models the forces. 
Persistent homology is used to reduce the function $f$ to a collection of points in the plane.  This collection of points is
called a {\em persistence diagram} and denoted by $\pd(f)$.  Each point in the persistence diagram encodes a well defined geometric 
feature of $f$.

It is useful to view persistent homology as a mapping from scalar functions to persistence diagrams, e.g.\
$f\mapsto \pd(f)$. Stated more formally, persistent homology can be viewed as a function from a space of scalar
functions to a space of persistence diagrams.  A fundamental result is that with appropriate metrics on the space
of functions and on the space of persistence diagrams,  persistent homology is a continuous function \cite{edelsbrunner:harer}.
 At least theoretically this implies that bounded noise or small errors in measurement of the DGM will lead to a small change in 
 the associated persistence diagram.

This theoretical potential combined with the successful applications presented in \cite{pre13} suggests the need for a careful analysis 
of the practical details of applying persistent homology to DGM.
There are at least three specific issues that need to be addressed:
\begin{enumerate}
\item Given a particular form of the experimental or numerical data, how can one perform the persistent homology computations?
\item Having chosen a method by which the persistent homology computations are being performed, how robust
is the resulting persistence diagram as a function of experimental or numerical noise or errors?
\item How can the information provided by the persistence diagrams be used to analyze DGM?
\end{enumerate}
Addressing these issues in the context of DGM is the main focus of this paper.  

The first step in the construction of the force network is to establish the domain $D$ on which the function $f$ representing 
the force interactions is defined.  A contact network seems to be a natural candidate for the domain  $D$.  
Indeed, if  positions and shapes of the particles are known, then one can  construct a contact network. 
If the data is in the form of a digital image, then building a contact network is more complicated.  In 
Section~\ref{sec:particle} we start by introducing {\em digital} and {\em position} networks that are closely related to  
contact networks. We investigate their stability with respect to measurement errors and show that their topology can considerably differ from topology of the physical system they represent.  
Therefore we propose an alternative domain, the 
{\em interaction} network.  This is an abstract mathematical concept and its topology is not related to topology of the physical 
system it represents. However, it provides a fixed domain for describing the force networks in DGM.  
 
 Section~\ref{sec:homology} introduces homology, which can be crudely interpreted as a tool for 
counting connected components, loops and cavities.  The advantages of homology are that it supports efficient algorithms,
can be used in higher dimensions, and allows one to compare components, loops, and cavities over different spaces.   
Section~\ref{sec:force} introduces force networks, clarifying the connection
between the type of available data and formulation of appropriate network.   
Section~\ref{sec:persistence} focuses on persistence homology, our principal 
tool for analyzing the force networks.    The interaction network can be used in the setting of numerical simulations or 
experiments (see, e.g.,~\cite{majmudar05a}), where complete information about the  forces between adjacent particles 
is known.
However, for many experiments only the total force experienced by a particle may be available~\cite{hartley_03}.  This
necessitates the use of a digital or position network.   In Section~\ref{sec:space} we discuss the space of persistence diagrams, in particular the
appropriate metrics on the space, and we provide a theorem that justifies the claim that the interaction force networks
are optimal.

In Section~\ref{sec:InterpPH} we conclude with a review of the developed concepts in the context of DGM
data obtained by discrete element based simulations (DES). We choose to work with
numerical simulations since all the data is available with high precision and therefore we can process them through all 
three force network constructions (digital, position, and interaction). This allows for greater clarity in interpreting
the geometric meaning of the persistence diagrams,  greater ease in comparing  the results of the different networks,
and simplicity in testing for stability  with respect to perturbations.  The reader who is familiar with the language of 
persistent homology may wish to skip directly to this section, before examining the details of the constructions.
 We note that the focus of this section is not on reaching general conclusions about  the force
networks in DGM, but on describing how the tools of persistence homology {\it can} be used to extract detailed information
about these networks.  For this purpose, we provide a few selected examples of the simulation data and discuss application
of persistence homology to these examples.   More interpretation-oriented  discussion that focuses on the
connection between persistence homology and physical properties of DGM is given in~\cite{pre13}.  

There are several  points that we encourage the reader to keep in mind while reviewing Section~\ref{sec:InterpPH}. 
First,  we provide examples of  two dimensional  force networks in order to facilitate the reader's intuition about the described features, however our analysis is based exclusively on the information contained in the persistence diagrams. 
This is particularly important in the context of three dimensional systems, where visual inspection may be impossible. 
Second, in our examples we mostly concentrate on the 
magnitude of the normal force between particles, but in principle any function
that assigns a scalar value to every edge can be used.  We illustrate this point by briefly considering tangential forces.
Finally, although  this paper is focused on DGM, the constructions are
independent of the details of particle-particle interaction and could as well be applied to any other 
system consisting of interacting particles.   
The software  used  to build the various force networks \cite{miro} and to compute persistence diagrams \cite{perseus} is available in 
the public domain.

%%%%%%%%%%%%%%%%%%%%
%
\section{Particle Networks}
\label{sec:particle}
%
%%%%%%%%%%%%%%%%%%%%

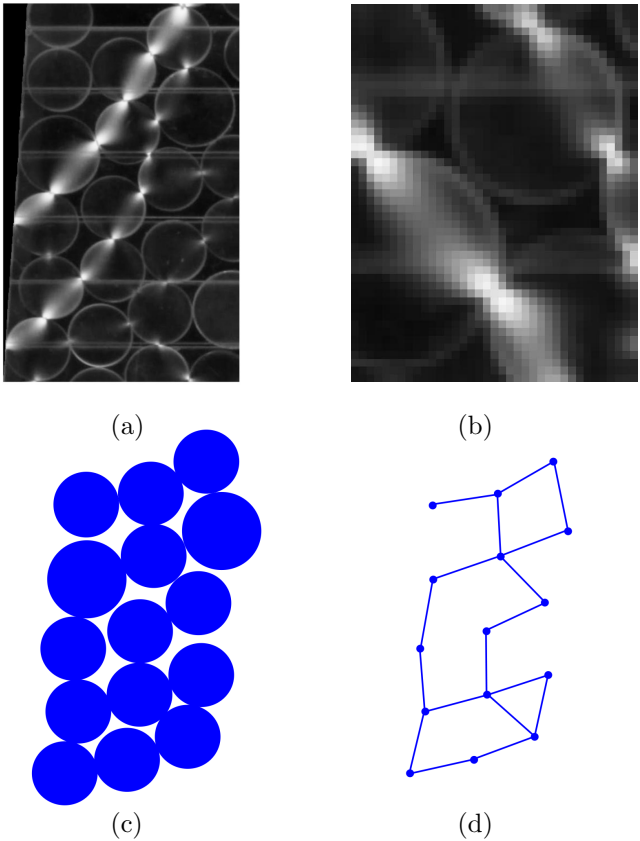
\begin{figure}[t]
\begin{picture}(380,310)
\put(0,155){\includegraphics[width=3.0in]{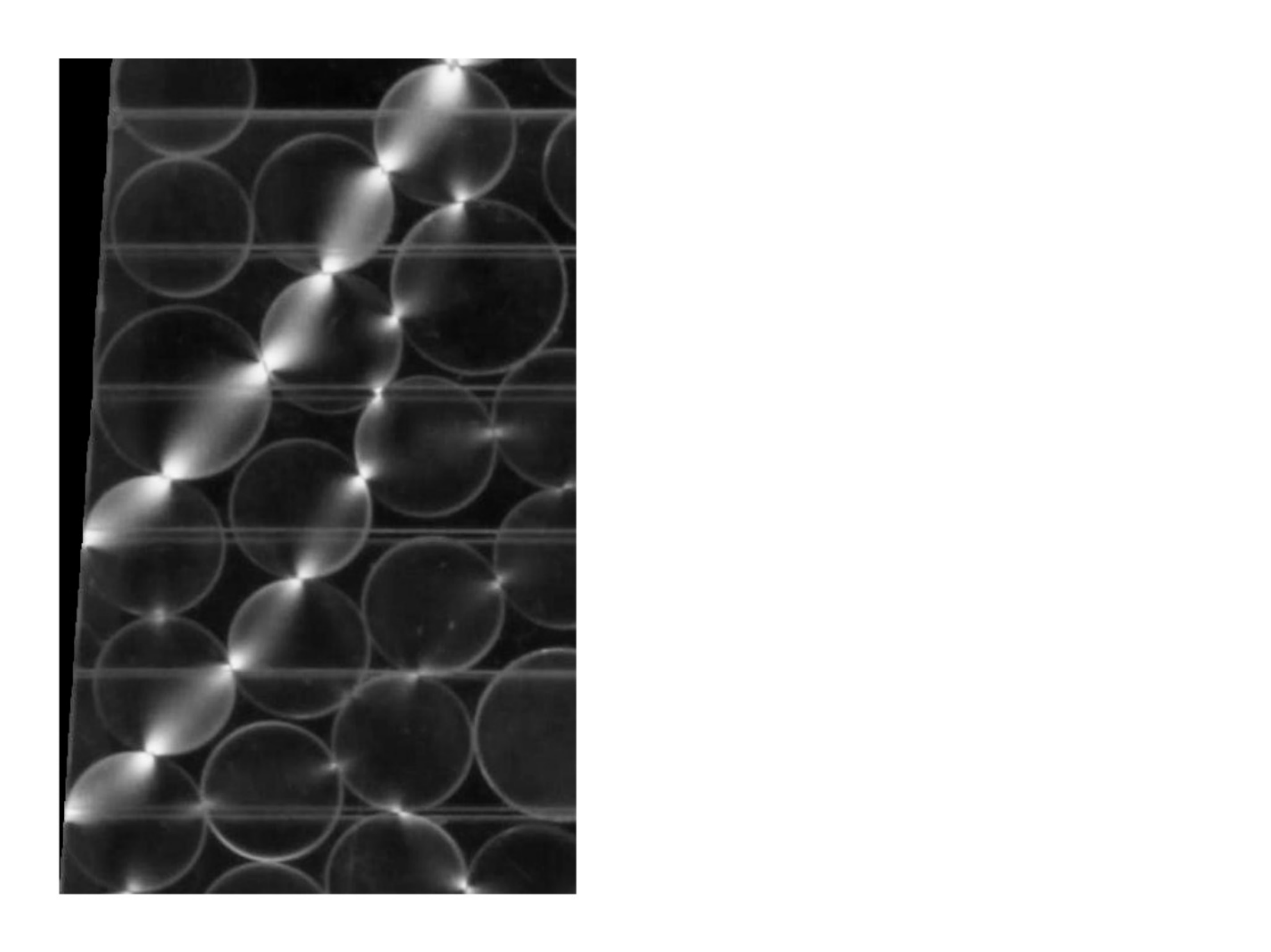}}

\put(140,165){\includegraphics[width=1.97in, angle = 90]{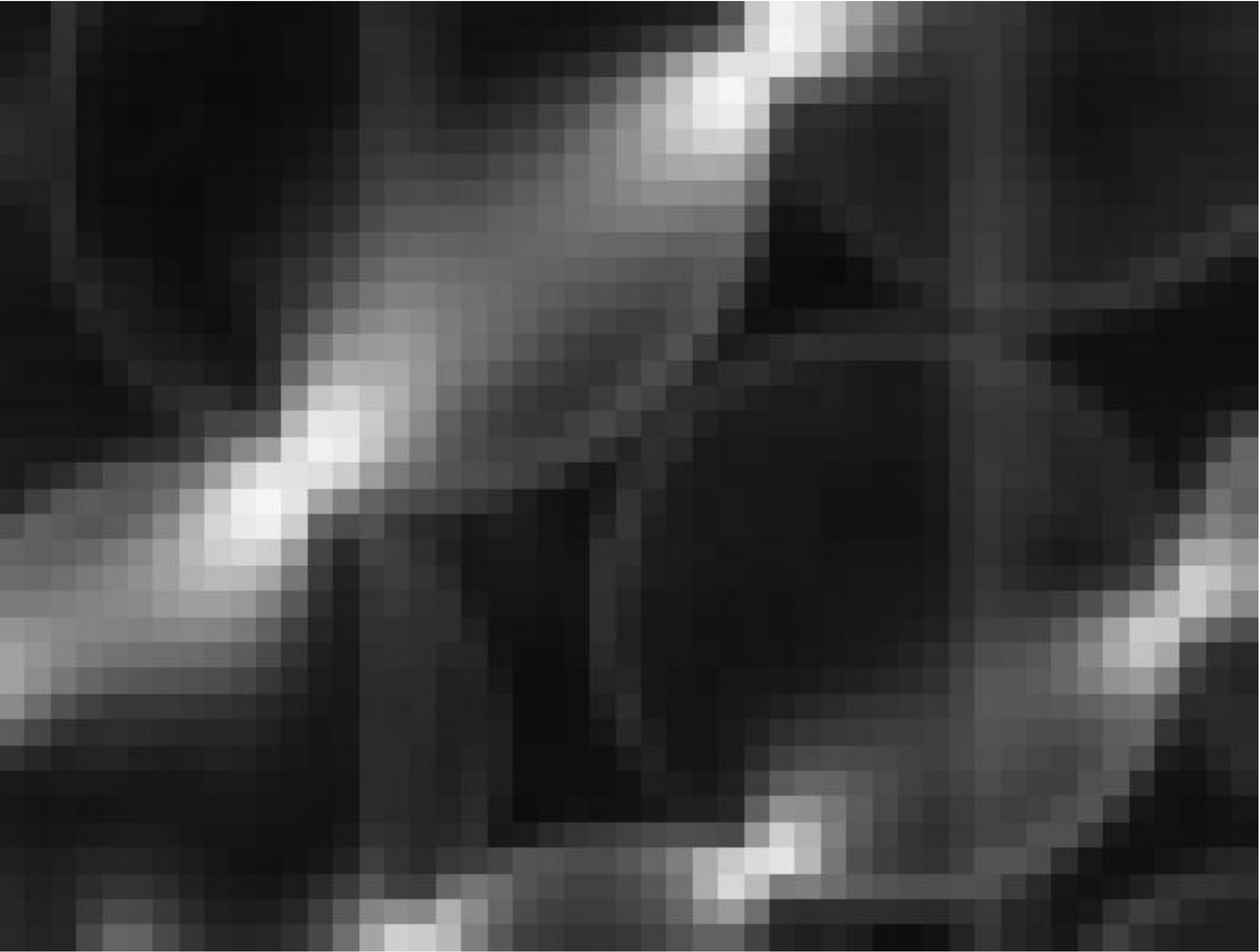}}

\put(10,-5){\includegraphics[width=3.0in]{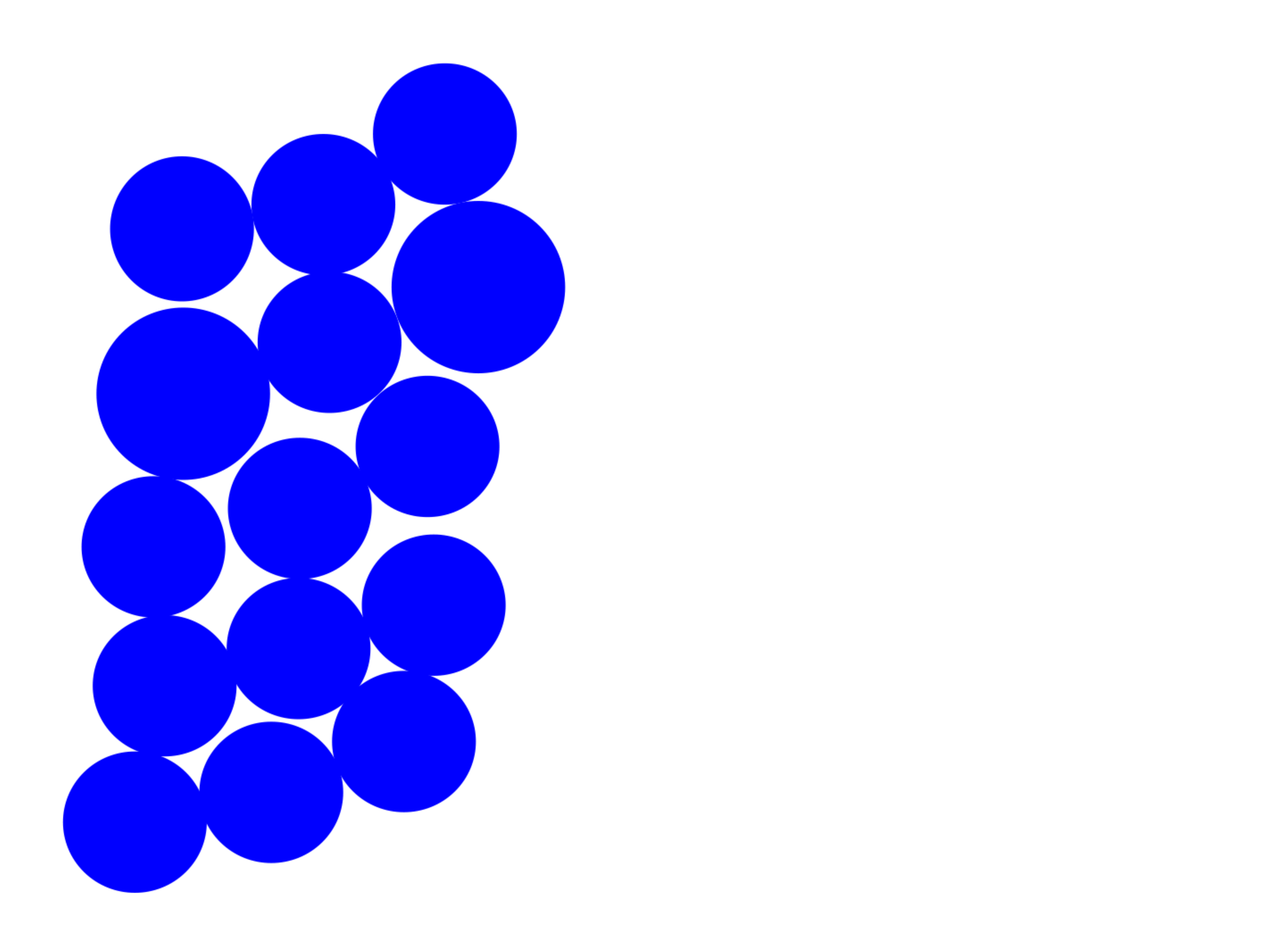}}

\put(140,-5){\includegraphics[width=3.0in]{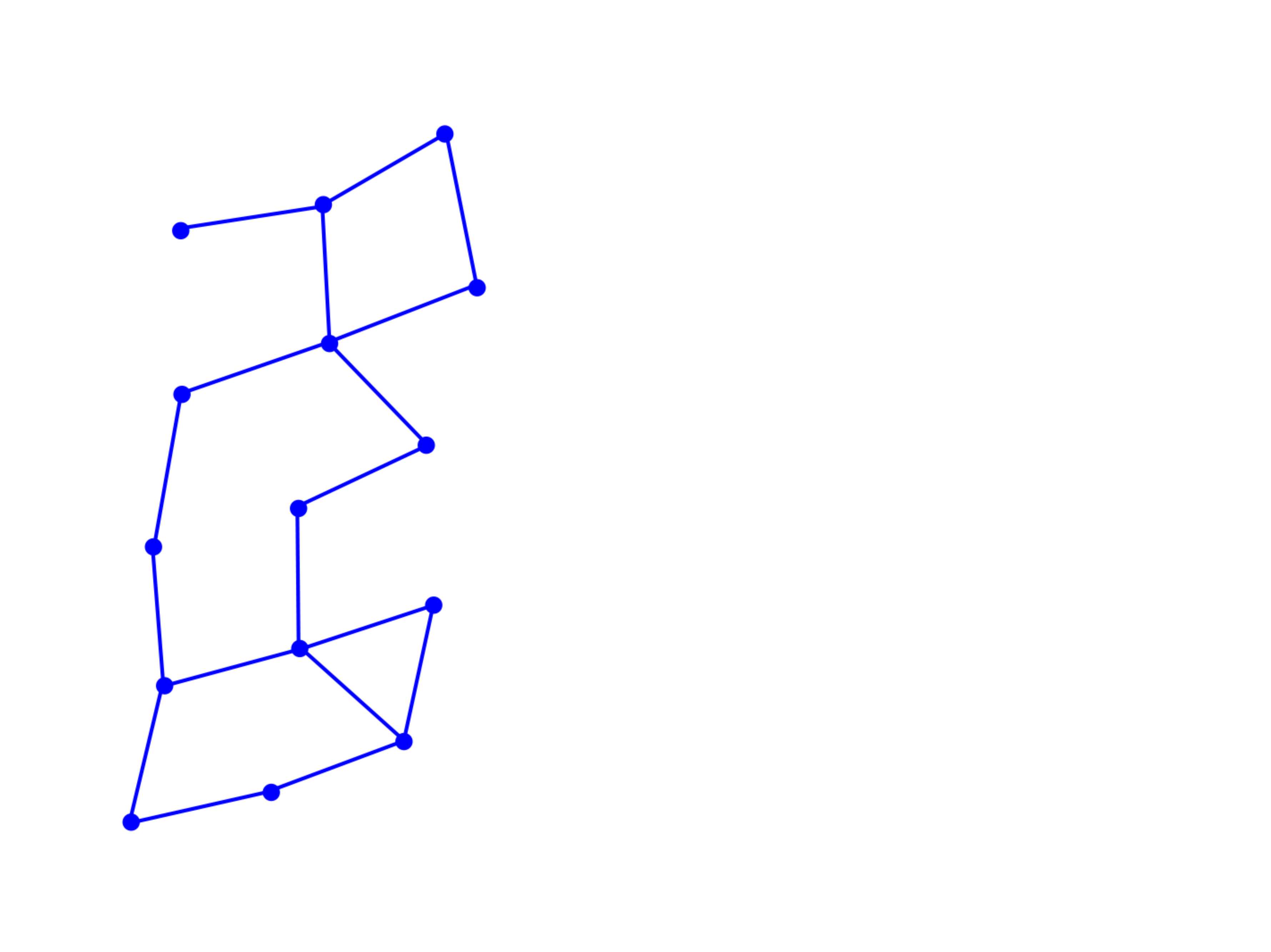}}

\put(50,145){(a)}
\put(180,145){(b)}
\put(50,-5){(c)}
\put(180,-5){(d)}

\end{picture}
\caption{Different representations of the particle networks derived from experimental data. For simplicity we neglect the particles that intersect the edges of the picture.
(a) Small portion of a digital image of an experimental system (courtesy of R.P. Behringer, unpublished results). 
(b) Detail of a digital  image.  (c) Digital complex. The blue pixels represent two dimensional cubes present in the complex.
(d) Position complex.}
\label{fig:complexes}
\end{figure}

In the Introduction the force network is described by a scalar function $f : D \to \mathbb R$. 
The first step towards using algebraic topology to characterize the geometric structures associated with DGM is to
represent the domain $D$ as a finite complex, defined below.  We introduce
three complexes  motivated by the type of data commonly obtained from  experiments or simulations. 
Consider Figure~\ref{fig:complexes}(a) that shows  a
small portion of an image of a collection of photoelastic disks. We interpret this information in three ways:
\begin{description}
\item[Digital]  This figure arises from a digital image and thus the data can be viewed as a collection of a large number of pixels.
\item[Positions] Since this is a controlled experiment  involving circular disks, 
 the configuration of the particles (locations of their center points  and their radii) can be determined.
\item[Interactions] The particles are made of photoelastic material and thus the light intensities within the particles can be used to determine the  normal forces between the particles.  
\end{description}
We note that the information required increases considerably as digital, position, interaction complexes are considered, respectively.
Our approach is to encode each of these data types into  different force networks and one of the goals of this paper is to make 
clear the difference of the geometric information that can be extracted. 
One not particularly surprising conclusion is that  the interaction data provides the best information and the digital data the worst,
but it is worth quantifying these differences.
With this in mind we begin with several formal definitions.  Our focus is on physical systems, thus for the most part we restrict our 
discussion to two and three dimensional complexes (see \cite{edelsbrunner:harer,kaczynski:mischaikow:mrozek} for a more general discussion).

To interpret the pixel data we make use of cubical complexes.  Observe that once
the pixel data is obtained the actual size used to represent each pixel is no longer an issue. Thus, for the sake of 
simplicity of discussion and without loss of generality we  assume
that the pixel data is embedded in $\R^2$ with each pixel being represented by a square 
 defined by the integer lattice.  More precisely, a {\em 2-dimensional cube} ({\em pixel}) is  a
square of the form $[n,n+1]\times [k,k+1]$, a {\em 1-dimensional cube} ({\em edge})  is a unit interval of the form 
$[n,n]\times [k,k+1]$ or   $[n,n+1]\times [k,k]$,  where $n,k\in\Z$, and a {\em 0-dimensional cube} ({\em vertex}) is a point with integer coordinates.  A two dimensional {\em cubical complex} $\cn$ is a collection of  0,1 and 2-dimensional cubes  that satisfy the following property:
if $\sigma\in\cn$ and $\sigma'\subset\sigma$, then $\sigma'\in\cn$.
This property guarantees that for every 2-dimensional cube (edge) in $\cn$ its edges (vertices) belong to $\cn$ as well. 

\begin{defn}
\label{defn:digitalcomplex}
\sloppy
{\em
Given a digital image of particles $\setof{p_i\mid i=0,\ldots, I}$ the {\em digital complex} 
$\cn_D$ is the cubical complex consisting
of squares $\setof{\sigma_j}$ where each square $\sigma_j$  represents  a single pixel associated with some particle.
}
\end{defn}

How pixels are associated with particles is intentionally
left vague in Definition~\ref{defn:digitalcomplex}. The actual association depends on the particular characteristics of the imaging
device, filtering, thresholding, etc used to obtain and process the data. 
Conceptually, the most straightforward approach  is to discretize the domain of the image into squares, identify the squares
with pixels, and declare the pixel to represent a particle if the associated square intersects the particle. A cubical particle network $\cn_D$ is shown in  Figure~\ref{fig:complexes}(c).
We provide more detail about the construction later.

\begin{rem}
{\em
In this paper we consider only two-dimensional examples.  However,  the same ideas can be applied  to particles
in $\R^3$ where the
three dimensional images are represented as voxels.  In this case one builds a cubical complex by representing each voxel as a unit cube  of the integer lattice in $\R^3$ (see
\cite{kaczynski:mischaikow:mrozek} for the general theory). 
}
\end{rem}

In the case of particles with simple shape the
contact network of \cite{walker_pre12} or the unweighted (binary) network of \cite{daniels_pre12} can be used to represent the particles. 
We represent these networks in terms of simplicial complexes which are defined as follows.
W begin with a finite set of vertices $\cn^{(0)}:= \setof{v_i\mid i=0,\ldots, I}$.  An  {\em
$n$-dimensional simplex} in $\cn$ is a subset of $\cn^{(0)}$ consisting of $n+1$ vertices. 
The set of $n$-dimensional simplices in $\cn$ is denoted by $\cn^{(n)}$.  
Given the set of vertices $\cn^{(0)}:= \setof{v_i\mid i=0,\ldots, I}$ it is customary to denote the $0$-dimensional
simplices by $\ang{v_i}$, the $1$-dimensional
simplices by $\ang{v_i,v_j}$, and the $2$-dimensional simplices by $\ang{v_i,v_j,v_k}$.
One and two dimensional simplices are referred to
as  {\em edges} and {\em triangles}. A {\em simplicial
complex}  $\cn$ is a collection of simplicies that satisfies the following property:  if $\sigma\in\cn$ and $\sigma'\subset\sigma$,
then $\sigma'\in\cn$.  

\begin{defn}
\label{defn:positioncomplex}
\sloppy
{\em
Given a collection of  circular disks $\setof{p_i\mid i=0,\ldots, I}$,  location of their centers  $\setof{x_i\in\R^2\mid i=0,\ldots, I}$, and
their radii $\setof{r_i\mid i=0,\ldots, I}$ the associated {\em position complex} $\cn_P$ is the simplicial complex consisting
of vertices $\setof{v_i\mid i=0,\ldots, I}$, where each vertex $v_i$ is identified with particle $p_i$, and edges 
$\ip{v_i}{v_j}$ if and only if $\| x_i - x_j\| \leq r_i + r_j$.
}
\end{defn}

For the sake of clarity Definition~\ref{defn:positioncomplex} is presented in the context of the examples considered in this paper.  
More generally, one can consider spherical particles positioned in $\R^d$, $d=3$ being the most relevant for physical applications.   As presented,
the position complex is an abstract
simplicial complex; that is, there is no specific geometric object associated with it. In the context of this work we can always
geometrize the complex by declaring the vertices to be the points $\setof{x_i\in\R^2\mid i=0,\ldots, I}$, and the edges to be the
line segments connecting the points. From now on we rarely distinguish between the abstract simplicial complex and its 
geometric realization.

Having defined these complexes, a reasonable first question is whether they correctly capture the topology of  the particle configuration. We begin with the following positive result under the assumption that the particles cannot deform under the pressure induced by contacts with other particles.

\begin{prop}
\label{prop:homotopy}
\sloppy
Given a collection of circular hard  disks $\setof{p_i\mid i=0,\ldots, I}$,  location of their centers  $\setof{x_i\in\R^2\mid i=0,\ldots, I}$, and
their radii $\setof{r_i\mid i=0,\ldots, I}$ the associated position complex $\cn_P$ is homotopic to the union   of the regions occupied by the particles, $\bigcup_{i=0}^I p_i$.
\end{prop}

\begin{figure}%[t]
\begin{picture}(400,90)
\centering
\put(20,-15){\includegraphics[width=3.0in, angle = 90]{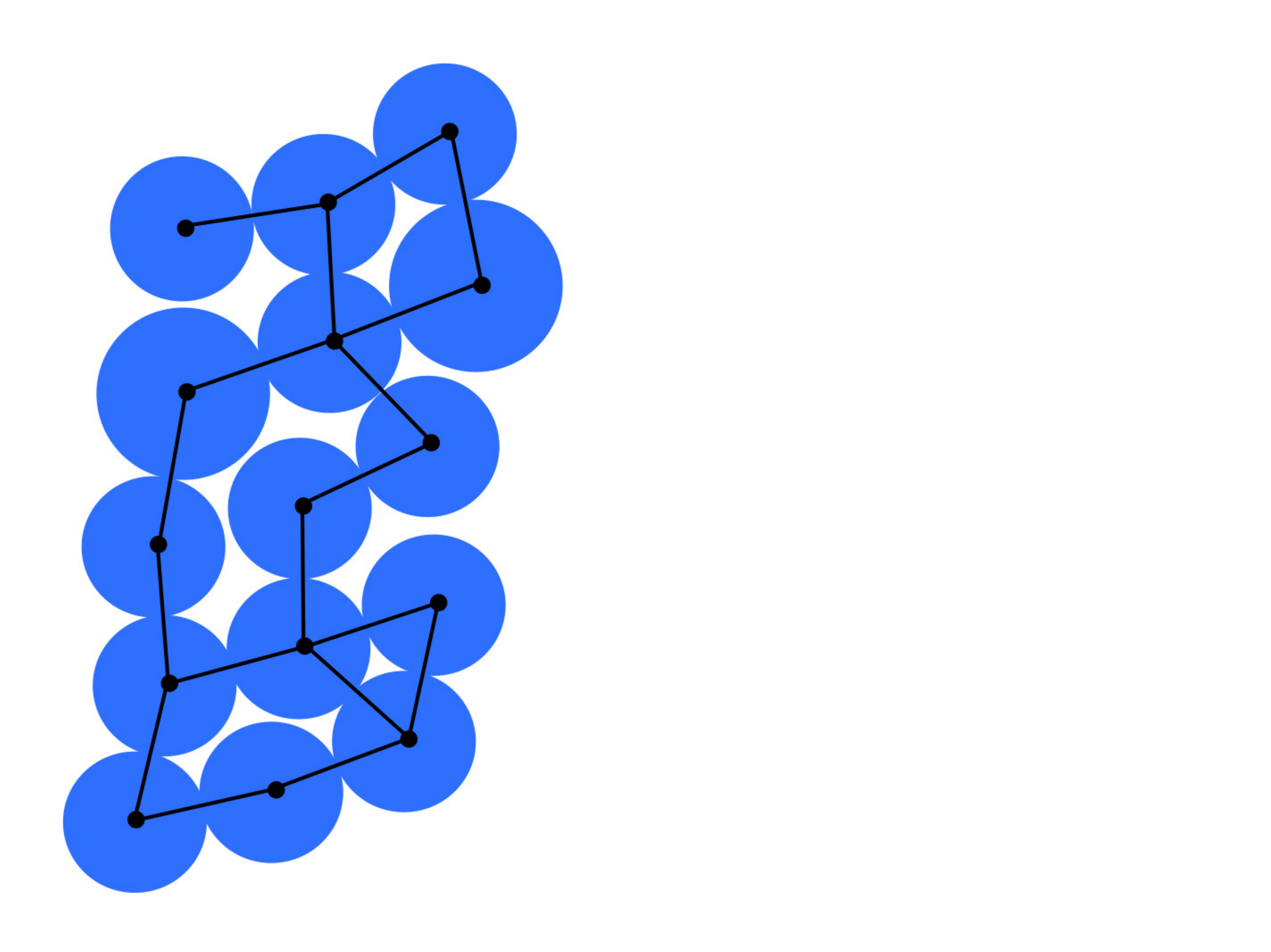}}
\end{picture}
\caption{With complete information the position complex $\cn_P$ has the same homotopy type as the configuration
space of the particles $\cup_{i=0}^I p_i$. The proof involves collapsing the particles  onto graph.}
\label{fig:homotopy}
\end{figure}

The proof follows from retracting the set of particles onto the geometric realization of $\cn_P$ (see for example
Figure~\ref{fig:homotopy}). We do not provide details of the
proof because this result is of  limited importance.  In any experiment or numerical simulation the locations of the
particles can only be given up to some specified precision.  If we assume the particles to be hard, then two particles 
$p_i$ and $p_j$ are
in contact if and only if $\|x_i - x_j\| = r_i + r_j$. Clearly, arbitrarily small errors in $x_i$ and $x_j$ can lead to an inequality
which indicates that the particles are not in contact. The same argument applies to arbitrarily small errors in the measurements
of the radii of the disks. Assuming that the particles are soft makes this result slightly more robust, but this is tempered by the fact that this stability depends on the existence of sufficiently  large normal forces.  We attempt to quantify these comments in 
Section~\ref{sec:stability}.

\begin{figure}[t]
\begin{picture}(400,100)
\put(0,10){\includegraphics[width=1.5in]{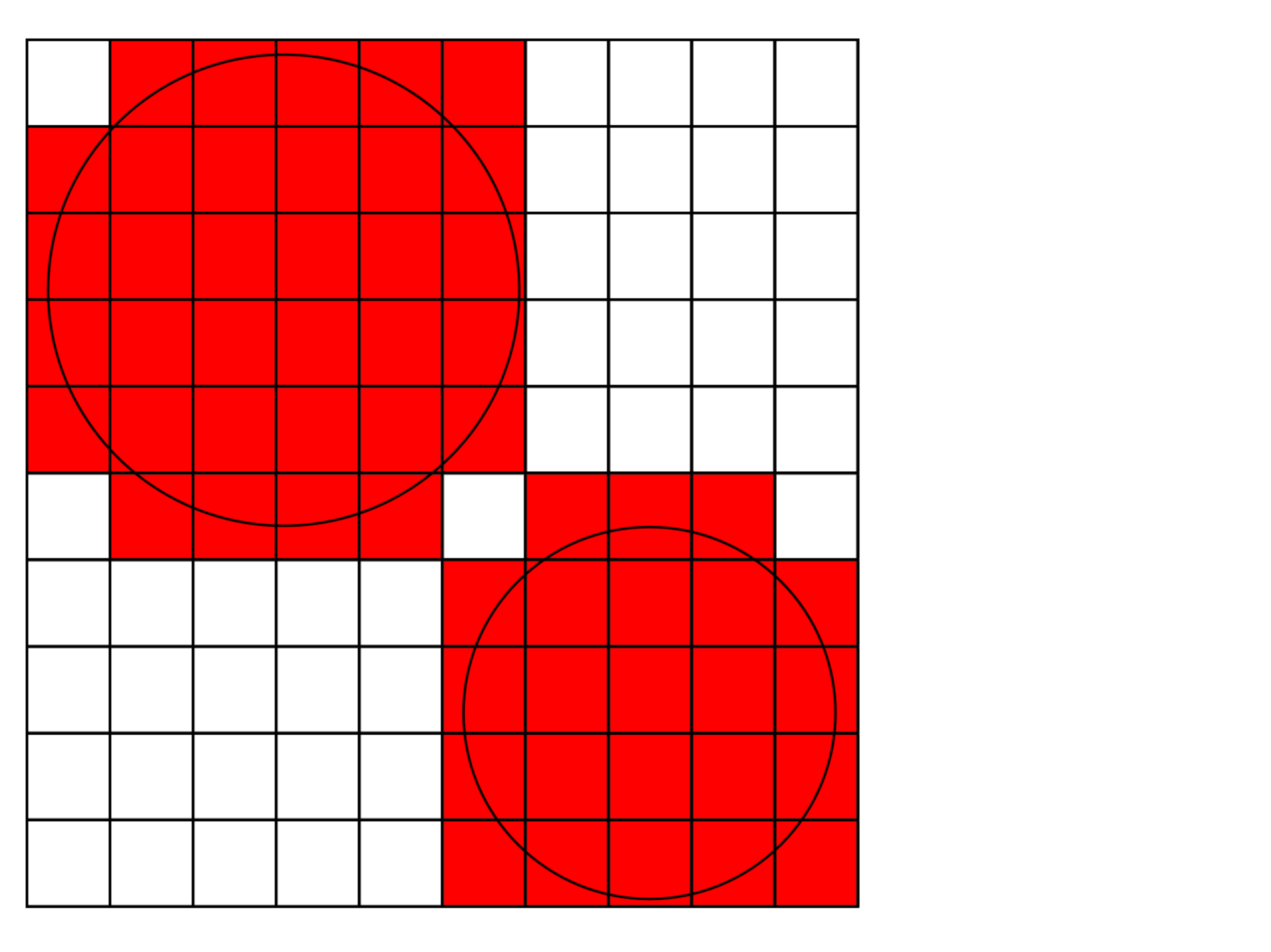}}
\put(85,10){\includegraphics[width=1.5in]{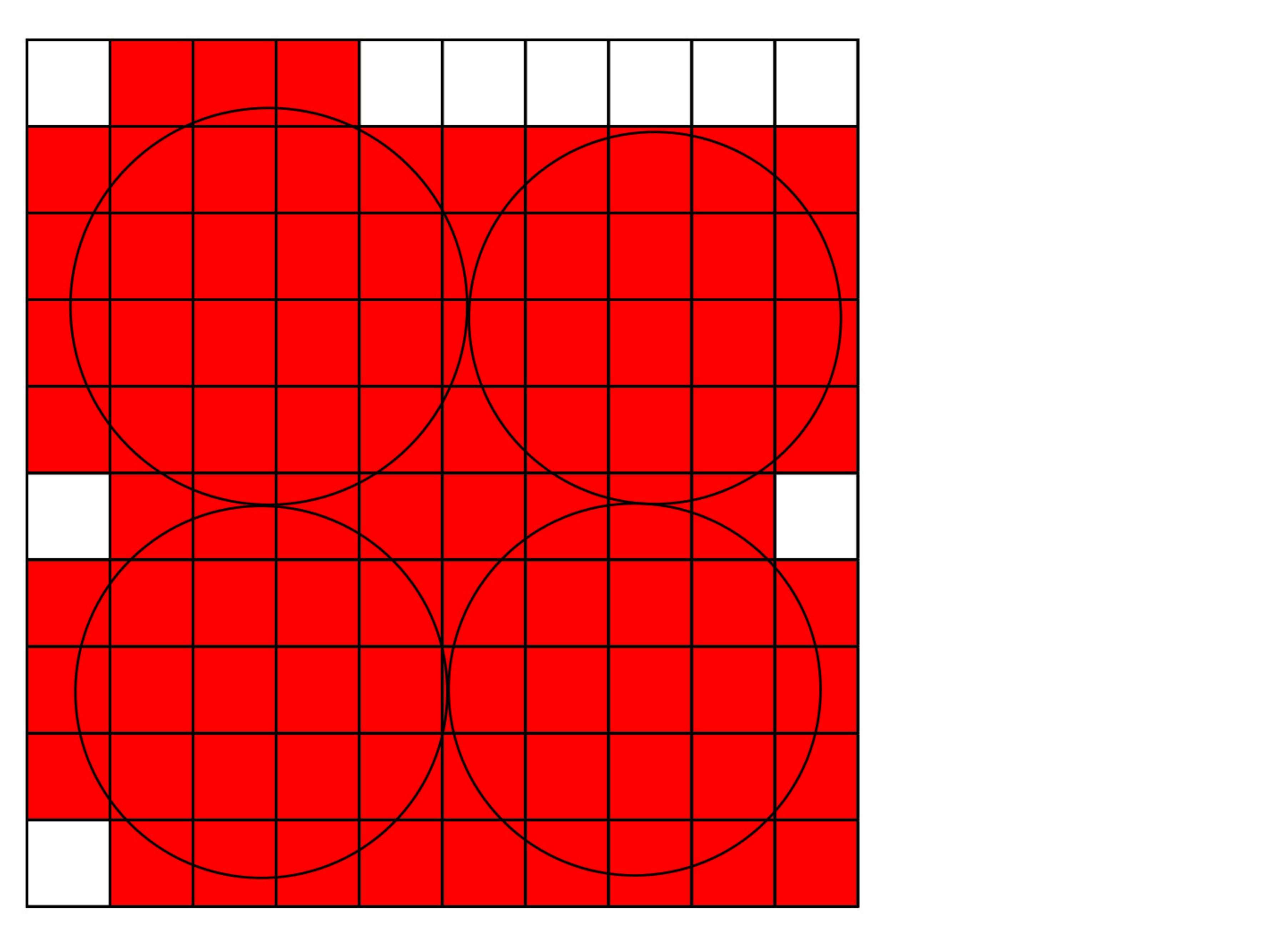}}
\put(170,10){\includegraphics[width=1.5in]{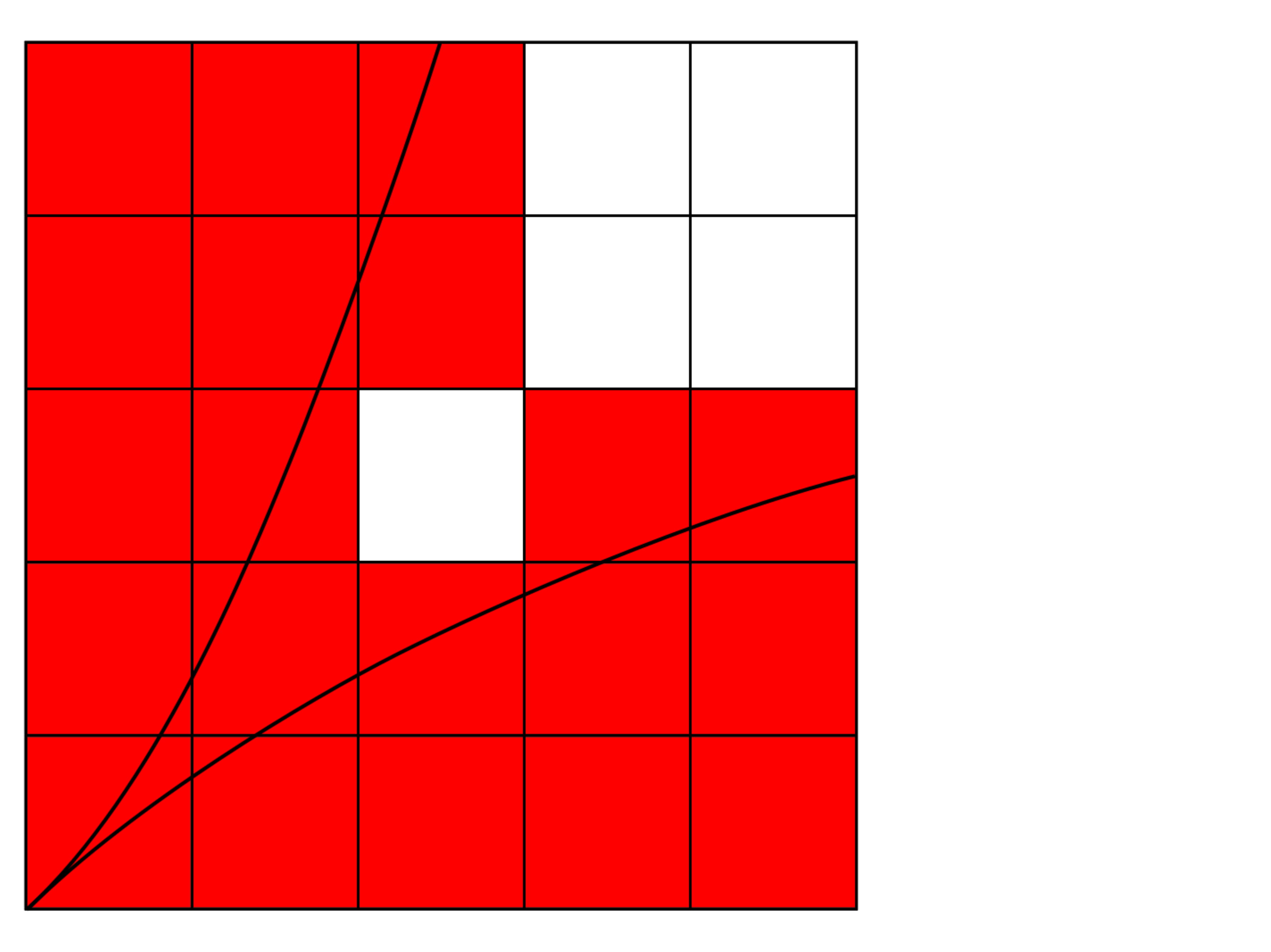}}

\put(30,0){(a)}
\put(115,0){(b)}
\put(200,0){(c)}

\end{picture}
\caption{Failure of the digital complex to correctly capture the topology of the particle configuration. 
(a) The particle configuration consists of two components and contains no loops. The associate digital complex has one component and one loop. (b) The particle configuration  contains one loop while the digital complex contains none.  (c) The particle
configuration contains one component and no loops. The associate digital complex has one component and one loop. Furthermore,
doubling the resolution does not remove the unwanted loop. }
\label{fig:digitalFailure}
\end{figure}
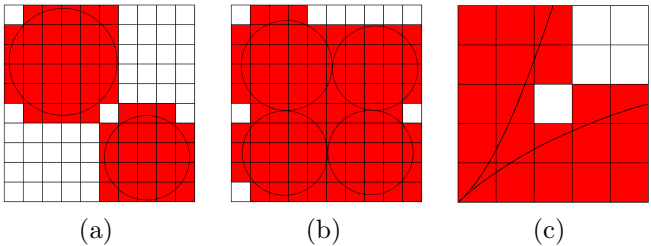

To measure the topological fidelity of the digital complex requires the choice of a rule for determining if a pixel is included
in the complex or not. For the sake of clarity we continue with the conceptually simple rule that the pixel belongs to the complex if  the associated square intersects some particle.

Figure~\ref{fig:digitalFailure} shows the digital complexes associated with different particle configurations. One can see that the failure of the digital complex $\cn_D$ to correctly capture the topology of the particle configuration can be quite dramatic.  Even  the simplest setting of two particles with a high pixel resolutions does not guarantee a correct topological
description. Figures~\ref{fig:digitalFailure}(a) and (c) demonstrates that both the number of  connected components and loops   can be counted incorrectly.
Figures~\ref{fig:digitalFailure}(a) and (b) show that  there is no particular direction to the error in 
the loop count.  The fact that the number of components of the digital complex is never larger than 
that of the configuration of particles
arises from our assumption on how to identify pixels with particles. In particular our approach leads to an artificial expansion of the area covered by each particle. Thus, two separate particles can appear to be  in contact, but two particles that are in contact can
never appear to be separated.

To keep things in perspective we remind the reader that even though it is clear that the digital complex can fail to record the correct
topology in a variety of ways it is the easiest means of collecting data and is applicable in situations in which the forces 
between the grains cannot be directly measured and without a priori assumptions about the geometry and rigidity of the grains.

\begin{figure}[t]
\begin{picture}(400,100)
\put(0,10){\includegraphics[width=1.5in]{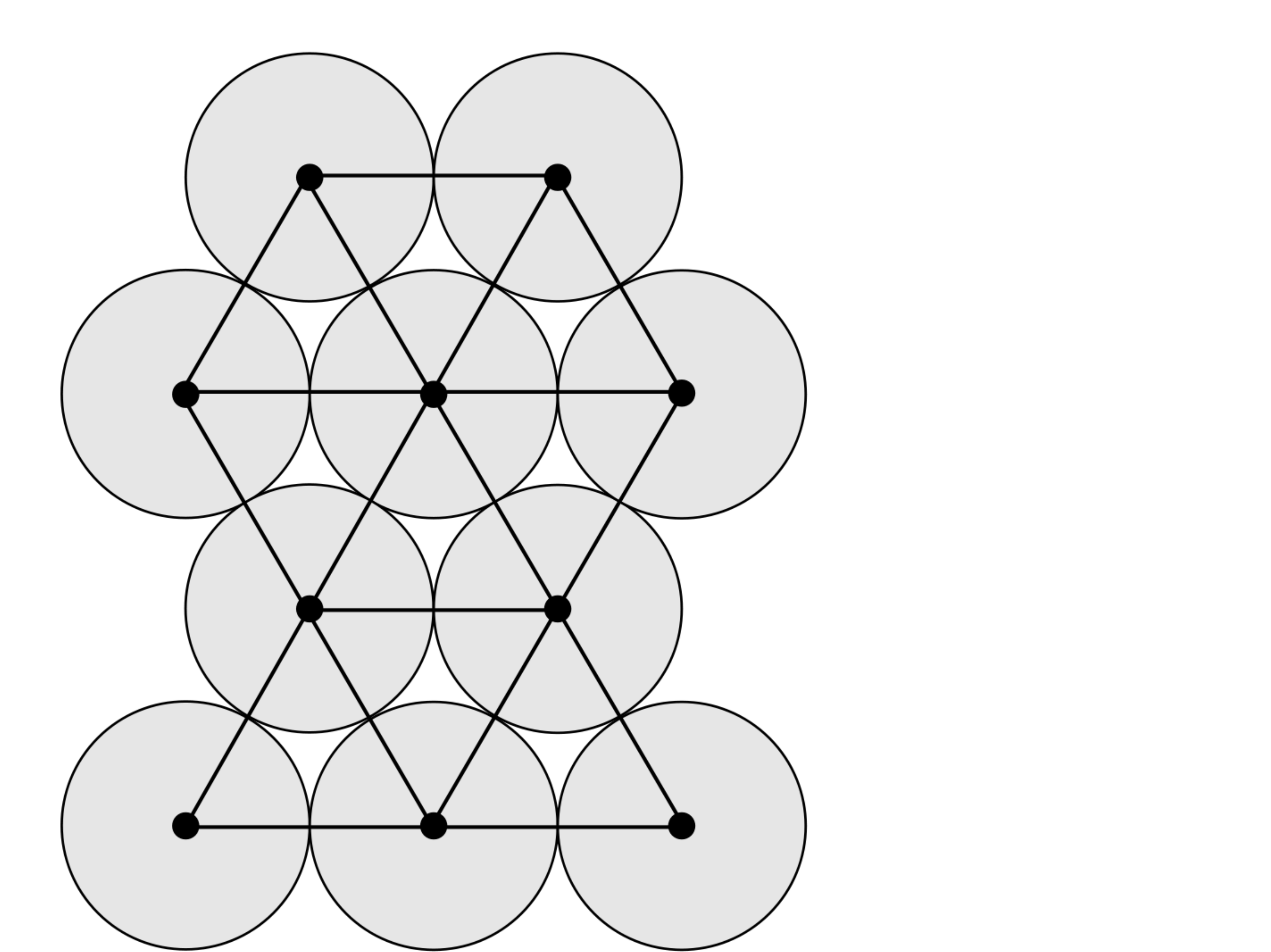}}
\put(80,10){\includegraphics[width=1.5in]{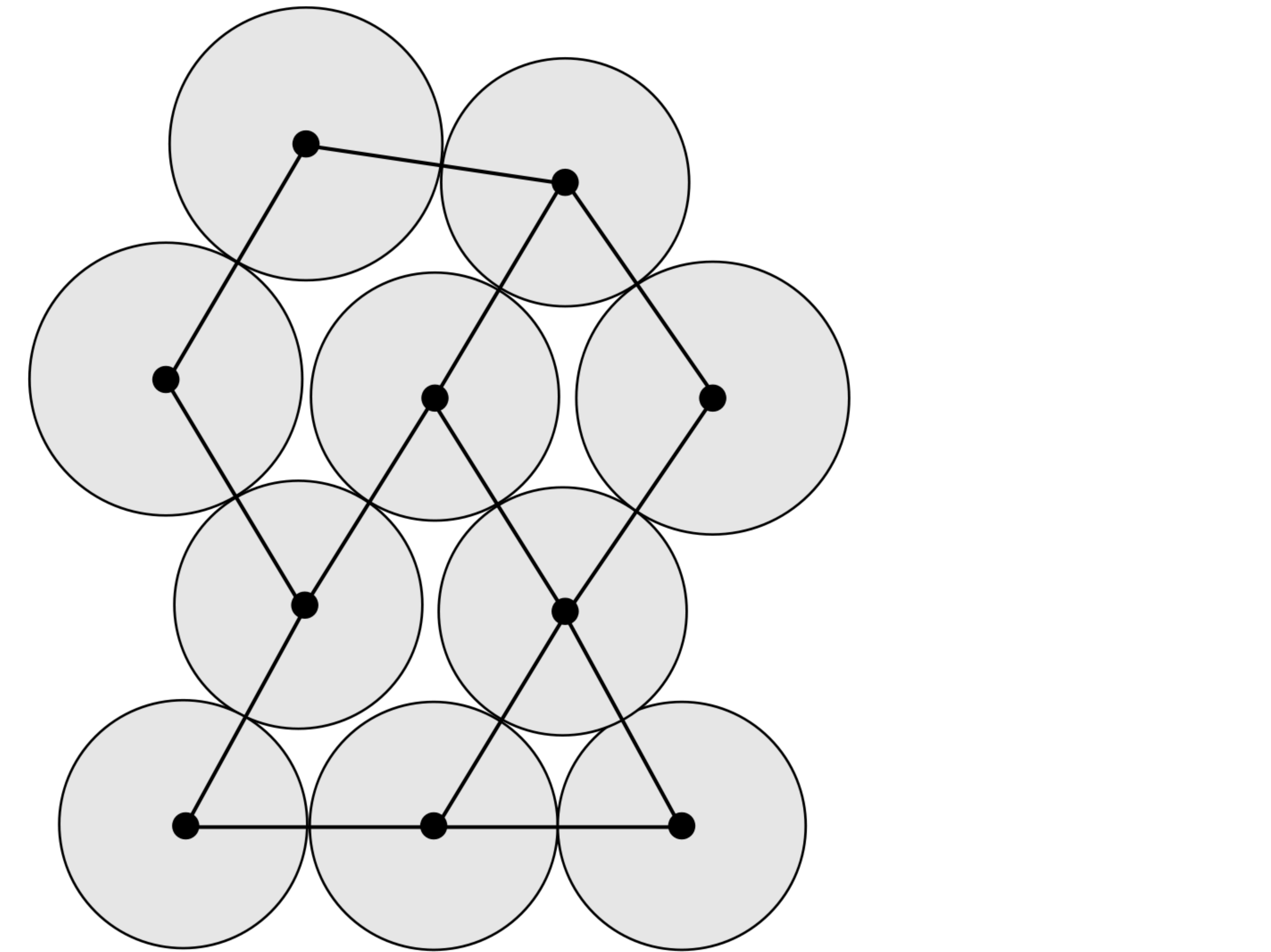}}
\put(165,10){\includegraphics[width=1.5in]{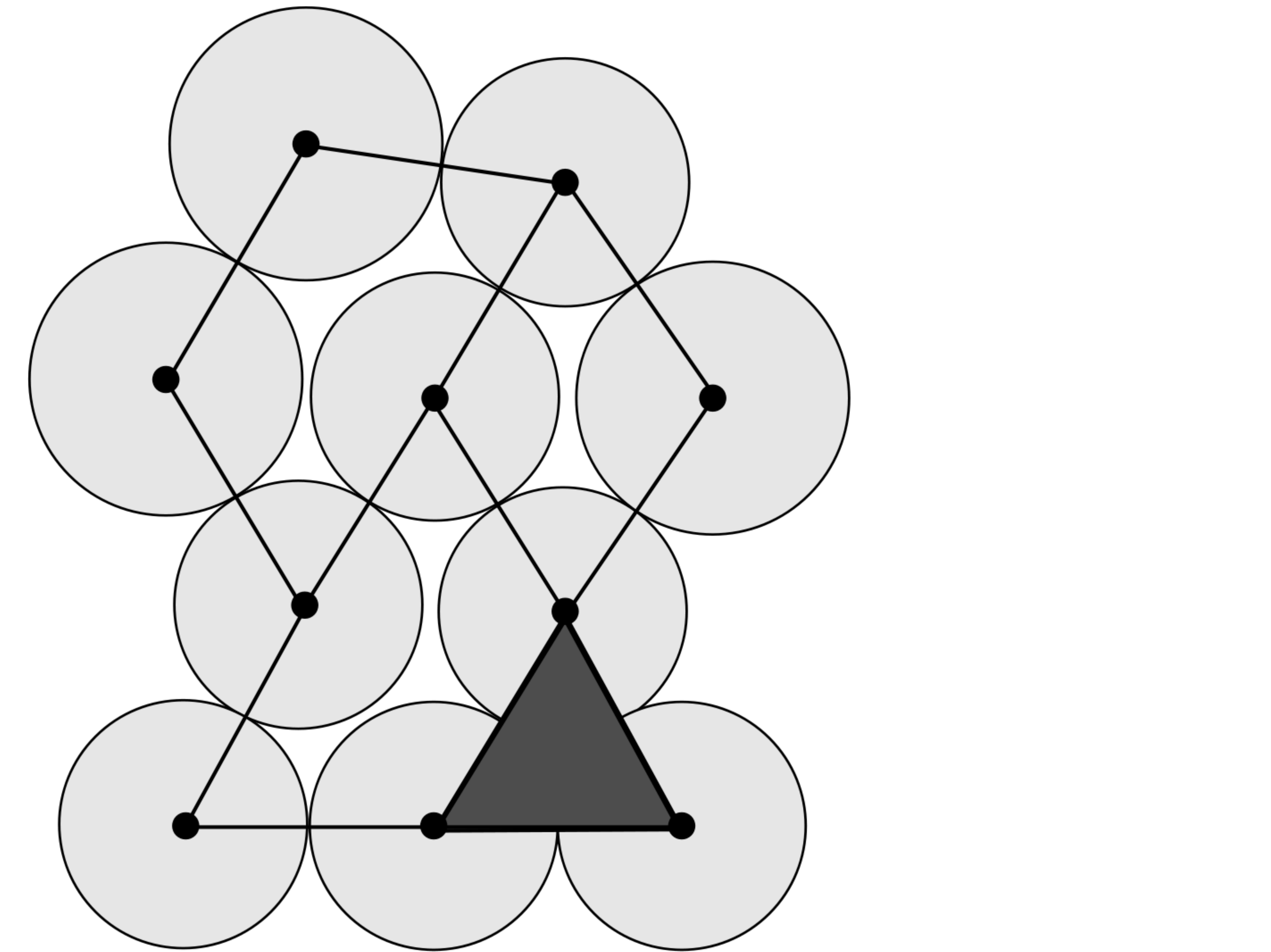}}

\put(31,-2){(a)}
\put(110,-2){(b)}
\put(196,-2){(c)}

\end{picture}
\caption{Two configuration spaces with position complexes $\cn_P$. (a) Crystaline structure with many loops formed by three particles. (b)
Noncrystaline structure with fewer loops than (a). (c) Flag complex $\cn^\blacktriangle_P$ derived from (b). Observe 
that the only loops which remain are associated with defects.}
\label{fig:2cells}
\end{figure}
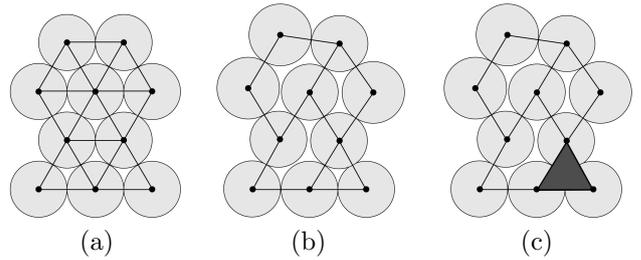

Consider the position network $\cn_P$ for the particle conformation shown in Figure~\ref{fig:2cells}(a), which we refer to as a 
{\em crystalline structure} since the particles are packed as densely as possible.  If we restrict our definition of the position complexes to graphs (one dimensional simplicial complexes), then $\cn_P$ has nine loops all of which involve three particles. This should be contrasted with 
Figure~\ref{fig:2cells}(b) in which there are only 4 loops, but
three of them are associated with particles that are not packed as densely as possible. Since in a perfect densely packed crystalline 
structure  made up of disks of the same size all loops would be made up of exactly $3$ particles we 
refer to a loop involving four or more particles  as a {\em defect}.  We use this definition of a defect 
even for systems built from variable size particles.

In this paper we have chosen to focus on defects, i.e.  we want to avoid counting loops that can be expressed in
terms of three particles. 
This can be done naturally using the flag complex, $\cn^{\blacktriangle}_P$, 
defined as follows. 
Set
\[
\cn_P^{\blacktriangle(0)} := \cn_P^{(0)}\quad\text{and}\quad \cn_P^{\blacktriangle(1)} := \cn_P^{(1)}
\]
and
\[
\ang{v_i,v_j,v_k}\in \cn_P^{\blacktriangle(2)} 
\]
if and only if
\[
\setof{\ang{v_i,v_j}, \ang{v_i,v_k}, \ang{v_j,v_k} }\subset \cn_P^{(1)}.
\]
The newly added triangles $\ang{v_i,v_j,v_k}$ exactly fill in the loops formed by three particles, see Figure~\ref{fig:2cells}(c).

It is worth noting that attempting an analogous construction of a flag complex in the setting of cubical complexes
that arise from digital complexes will not work.A square can be missing
from $\cn_D$ because of the phenomena indicated in Figure~\ref{fig:digitalFailure}(a) or (c), but it can also be missing because
it represents the loop formed by four distinct particles. 

As is discussed at the beginning of this section, the complexes $\cn_D$ and $\cn_P$ are introduced in order to be able to apply 
algebraic topological tools to the characterize the geometric structures of DGM.  Unfortunately, these complexes are not necessarily 
robust with respect to perturbations since arbitrarily  small changes in the locations of the particles can lead to the loss of cubes or 
edges in $\cn_D$ or $\cn_P$, respectively.  
With this in mind we introduce the  following complex.
\begin{defn}
\label{defn:interactioncomplex}
\sloppy
{\em
Given a collection of particles $\setof{p_i\mid i=0,\ldots, I}$ the {\em interaction complex} 
$\cn_I$ is the simplicial complex consisting
of vertices $\setof{v_i\mid i=0,\ldots, I}$ where each vertex $v_i$ is identified with particle $p_i$, all edges
$\ang{v_i,v_j}$, and all triangles $\ang{v_i,v_j,v_k}$.
}
\end{defn}

The interaction complex itself does not have any meaningful geometric interpretation. However, if 
the forces between the grains are available, then we can make use $\cn_I$ as the domain for the force complex (see Section~\ref{sec:persistence}).  
This in turn allows us to prove continuity  for the persistence diagrams (Corollary~\ref{cor:continuity}). 
The implications of  continuity, or lack thereof in the case of $\cn_D$ and $\cn_P$,  is made clear in Section~\ref{sec:stability}.

%%%%%%%%%%%%%%%%%%%%%
%
\section{Homology}
\label{sec:homology}
%
%%%%%%%%%%%%%%%%%%%%%

We pause in our development of the networks to review a few fundamental definitions from the classical theory of homology with
a focus on the  simple setting of the digital,  position and interaction
complexes introduced in Section~\ref{sec:particle}.  For a more general discussion the reader is referred to a standard text
in algebraic topology or to \cite{edelsbrunner:harer,kaczynski:mischaikow:mrozek} for descriptions more closely associated with data analysis.

Recall that position  complexes $\cn_P$ and interaction complexes $\cn_I$ are simplicial complexes. This leads to our use of
simplicial homology. Recall that $\cn^{(n)}$ denotes the set of $n$-dimensional simplices in the simplicial complex $\cn$. The {\em $n$-chains}
of $\cn$ are defined to be the vector space
\begin{equation}
\label{eq:chains}
C_n(\cn) := \setof{\sum_{\sigma\in \cn^{(n)}} m_\sigma \sigma\mid m_\sigma\in\Z_2}.
\end{equation}
Since we are working with planar arrangements of particles it is sufficient  to use $\Z_2$ coefficients,
i.e.\ the set $\setof{0,1}$  with the standard binary addition and multiplication operations.
Observe that $C_n(\cn)$ is the vector space over $\Z_2$ with basis elements consisting of the $n$-dimensional simplices.

The associated {\em boundary maps} are linear maps (these are often represented as matrices using the simplicies as bases)
 $\partial_n\colon C_n(\cn)\to C_{n-1}(\cn)$ ($C_{-1}(\cn) := 0$) defined on the simplices as follows
 \begin{eqnarray*}
\partial_0\ang{v_i} & := &  0 \\
\partial_1\ang{v_i,v_j}  & := & \ang{v_i} + \ang{v_j} \\
\partial_2\langle v_i,v_j,v_k \rangle & := &  \ang{v_i,v_j} + \ang{v_i,v_k} + \ang{v_j,v_k}.
\end{eqnarray*}
A direct calculation making use of the linearity and the use of  $\Z_2$ coefficients show that $\partial_{n-1}\circ \partial_n = 0$, e.g.
\begin{eqnarray*}
&&\partial_1\circ\partial_2   \langle v_i,v_j,v_k \rangle \\
& = & \partial_1( \ang{v_i,v_j} + \ang{v_i,v_k} + \ang{v_j,v_k}) \\
& = & \partial_1\ang{v_i,v_j} + \partial_1\ang{v_i,v_k} + \partial_1 \ang{v_j,v_k} \\
& = & \ang{v_i}+\ang{v_j} + \ang{v_i}+\ang{v_k} + \ang{v_j}+\ang{v_k} \\
&= & 0.
\end{eqnarray*}

The boundary maps can be used to identify components and loops. To do this  we focus on {\em cycles},
these are chains  which are sent to the $0$ vector under $\partial_n$. More formally,
\[
Z_n(\cn) := \ker \partial_n.
\]
Observe that $\ang{v_i}\in Z_0(\cn)$ and $\ang{v_i,v_j} + \ang{v_i,v_k} + \ang{v_j,v_k}\in Z_1(\cn)$.  

\begin{figure}
\centering
	\begin{picture}(300,110)
	\put(0,105){\includegraphics[width=5in, angle = 270]{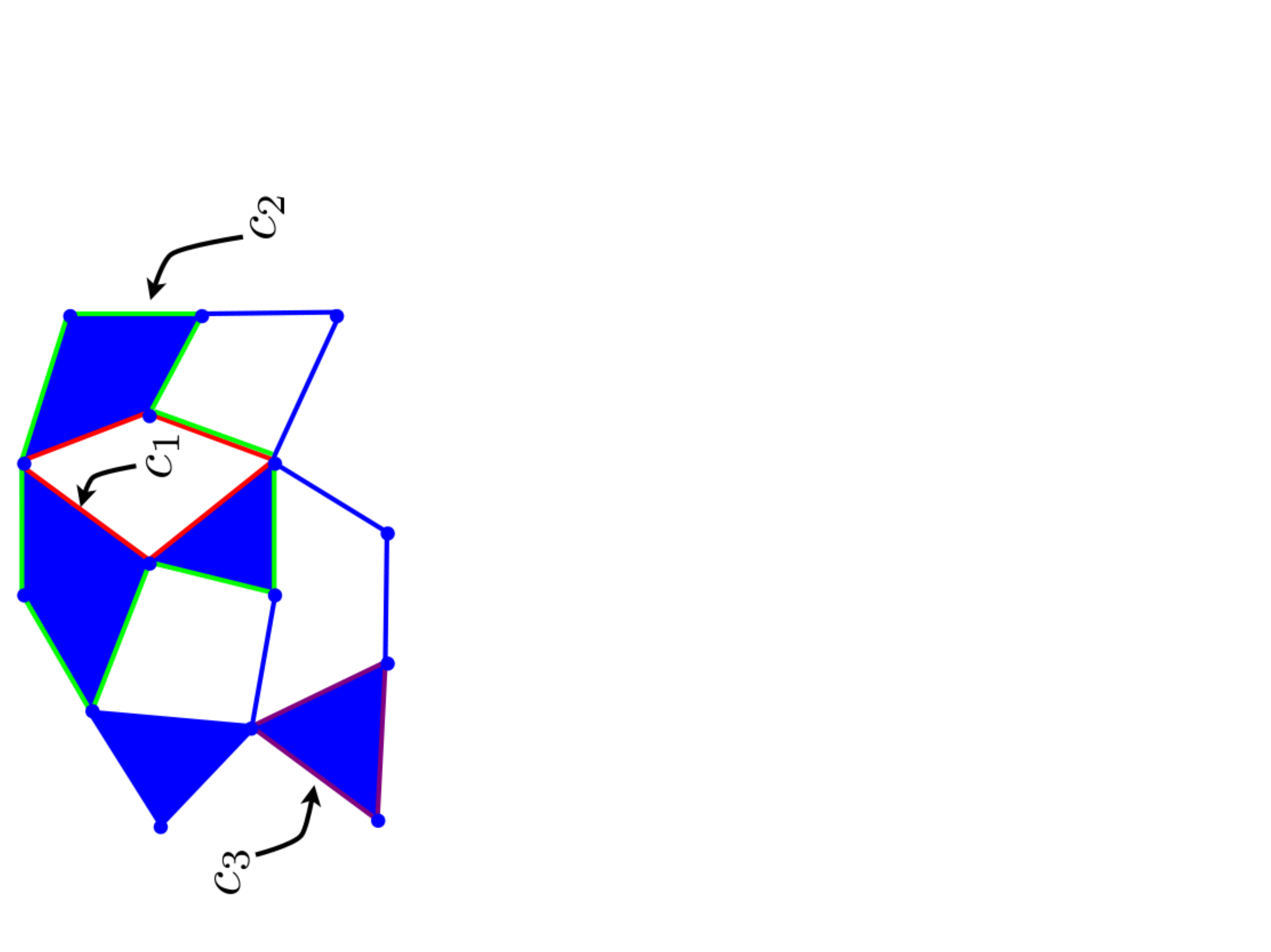}}
	\end{picture}
\caption{
Three different 1-dimensional chains. Each of these chains corresponds to a loop and hence is a cycle. The red chain $c_1$ and the  green chain $c_2$ correspond to the same loop in the particle network. The brown loop $c_3$ does not correspond to any  loop and can be contracted to a point. }
\label{fig:ToyChains}
\end{figure}

The power of homology is that we are able to move from geometric data to an algebraic format from which we can then
extract geometric information. For example, the algebraic statement that  $\ang{v_i}\in Z_0(\cn)$ can be interpreted
as a statement that $\ang{v_i}$ identifies a component of $\cn$. Similarly, $\ang{v_i,v_j} + \ang{v_i,v_k} + \ang{v_j,v_k}\in Z_1(\cn)$
can be identified with the path of edges
$\ang{v_i,v_j},\ang{v_i,v_k}, \ang{v_j,v_k}$ that makes up a loop.  To emphasize the relationship between the algebra
and geometry consider the simplicial complex indicated in Figure~\ref{fig:ToyChains}.  There are three chains that form loops
and hence cycles indicated in red, green and brown.

For obvious reasons it is important not to over count components or loops. 
In particular, if  an edge $\ang{v_i,v_j}$ belongs to $\cn$, then $\ang{v_i}$ and $\ang{v_j}$ belong to the same component and therefore we
wish to identify them. This can be done algebraically  by the relation $\partial_1\ang{v_i,v_j} = \ang{v_i} + \ang{v_j}$. Similarly, if
a 2-dimensional simplex $\ang{v_i,v_j,v_k}\in\cn$, then the loop $\ang{v_i,v_j},\ang{v_i,v_k}, \ang{v_j,v_k}$ does not enclose a 
loop and thus should not be counted. Again, this can be detected algebraically by the relation
$\partial_2\langle v_i,v_j,v_k \rangle =   \ang{v_i,v_j} + \ang{v_i,v_k} + \ang{v_j,v_k}$.
Observe that the relations in these examples are obtained via images of the boundary operator.  This leads to the  definition of
the {\em boundaries} of $\cn$,
\[
B_n(\cn) := \partial_{n+1} \left(C_{n+1}(\cn)\right).
\]
Referring to the complex depicted in Figure~\ref{fig:ToyChains} observe that there exists $\bar{c}$ such  that  $\partial_2 \bar{c} = c_1 + c_2$ which implies that
the cycles $c_1$ and $c_2$ represent the same loop in the complex. This motivates the following definition.
The $n$-th {\em homology group} of the simplicial complex $\cn$ is defined by
\[
H_n(\cn) := \frac{Z_n(\cn)}{B_n(\cn)},
\]
the vector space of equivalence classes of cycles identified by boundaries.  To be more specific given a 
cycle $z\in Z_n(\cn)$ the associated homology class $[z] = [z]_\cn$ is the equivalence class of all cycles of the form
$z+b$ where $b\in B_n(\cn)$.

The dimension of $H_n(\cn)$ is called the $n$-th {\em Betti number} $\beta_n(\cn)$.
$\beta_0(\cn)$ counts the number of components and $\beta_1(\cn)$ counts the number of loops which encircle a void.
If we were working with DGM in three dimensions then $\beta_2(\cn)$ would indicate the number of cavities.

A fundamental property of homology is that if two topological spaces are homotopic, then they have the same homology groups.
A corollary of this is that under the hypothesis of Proposition~\ref{prop:homotopy} the Betti numbers of $\cn_P$ agree with
the Betti numbers of the space defined by $\cup_{i=0}^I p_i$.  Since the hypotheses of this Proposition are rather strong,
e.g.\ exact knowledge of locations and radii, in general, given numerical or experimental data we do not expect that these Betti numbers
agree. 

Recall that the interaction complex does not have a meaningful geometric interpretation. In fact, independent of the 
number and arrangement of the particles, the homology of $\cn_I$ is very simple,
\begin{equation}
\label{eq:beta_incidence}
\beta_n(\cn_I) \cong \begin{cases}
1 & \text{if $n=0$} \\
0 & \text{if $n>0$.}
\end{cases}
\end{equation}
Thus the Betti numbers tell us that $\cn_I$ has a single connected component and does not have any loops that encircle a void.  

We do not present the details of computing homology with cubical complexes. Conceptually the ideas are the same, though
the boundary operators are slightly different. The interested reader is referred to \cite{kaczynski:mischaikow:mrozek} for 
a complete presentation.  Even more generally, simplicial and cubical complexes are examples of chain complexes
and the individual simplices or cubes are examples of  {\em cells}.

The reader may be somewhat underwhelmed by the fact that we have constructed a significant amount of algebra to 
essentially count components and loops, especially since there are extremely efficient graph theoretic algorithms for
performing these operations. 
However, the algebra allows us to relate components and loops in different complexes. 
Recall that components and loops are measured by elements of $H_0(\cn)$ and $H_1(\cn)$.
Thus, given two distinct chain complexes $\cn$ and $\cn'$ we need to be able to relate homology classes of $H_k(\cn)$
with homology classes of $H_k(\cn')$.  This is done via the following algebraic construction.
Linear maps $\phi_n \colon C_n(\cn)\to C_n(\cn')$
are {\em chain maps} if 
\[
\partial'_n \phi_n = \phi_{n-1}\partial_n
\]
for all $n$ where $\partial_n$ and $\partial'_n$ are the boundary maps for $C_n(\cn)$ and $C_n(\cn')$.
A fundamental result is that if $\phi_n$ is a chain map, then $\phi_n$ induces a linear {\em map on homology}
$\phi_n\colon H_n(\cn)\to H_n(\cn')$ defined by
\[
\phi_n([z]_\cn) := \left[ \phi_n(z)\right]_{\cn'}.
\]
As is made clear in Section~\ref{sec:persistence} for the purposes of this paper it is sufficient to note that if $\cn\subset\cn'$, then the inclusion map induces, for each dimension, a chain map  and hence a map on homology.

%%%%%%%%%%%%%%%%%%%%%%
%
\section{Force Networks}
\label{sec:force}
%
%%%%%%%%%%%%%%%%%%%%%%

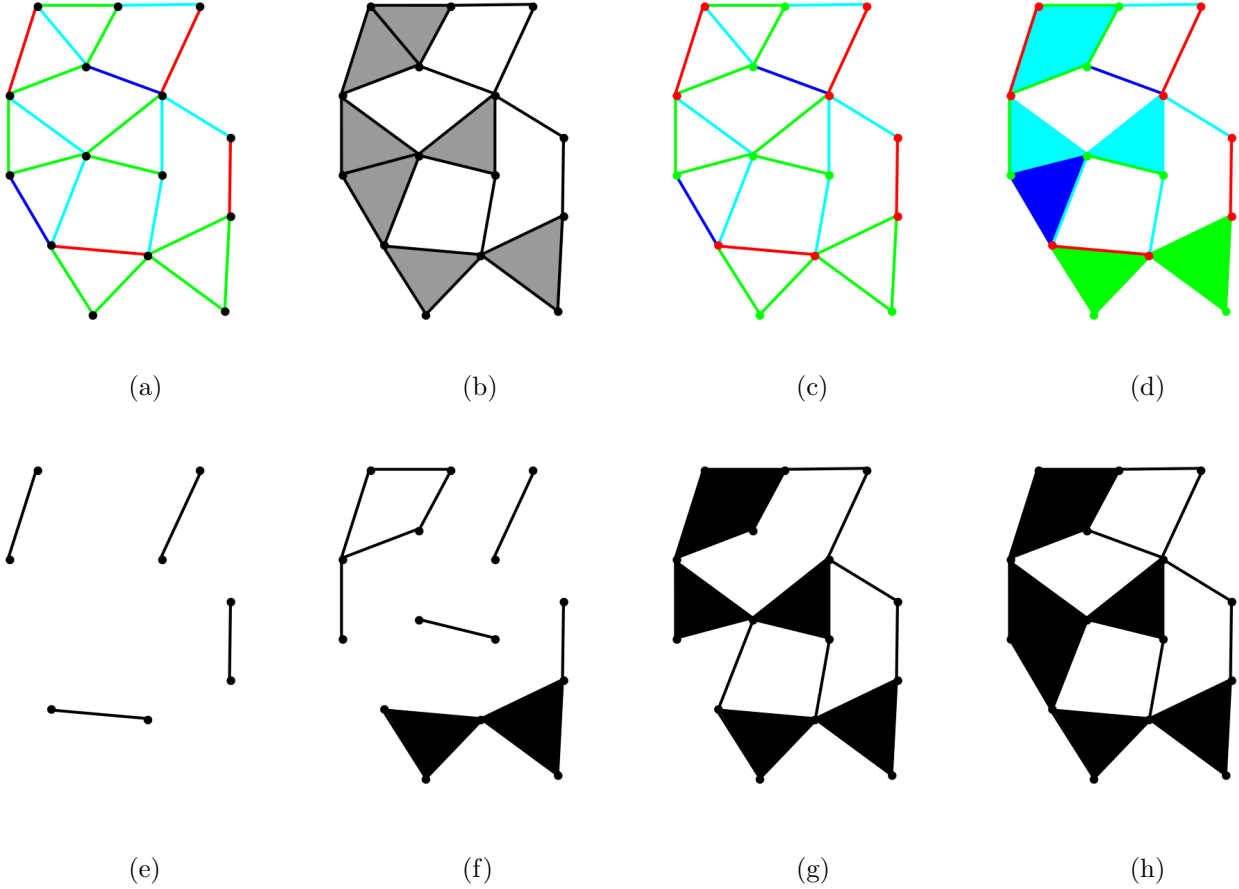
\begin{figure*}[t]
\begin{picture}(400,330)
\put(0,160){\includegraphics[width=4in]{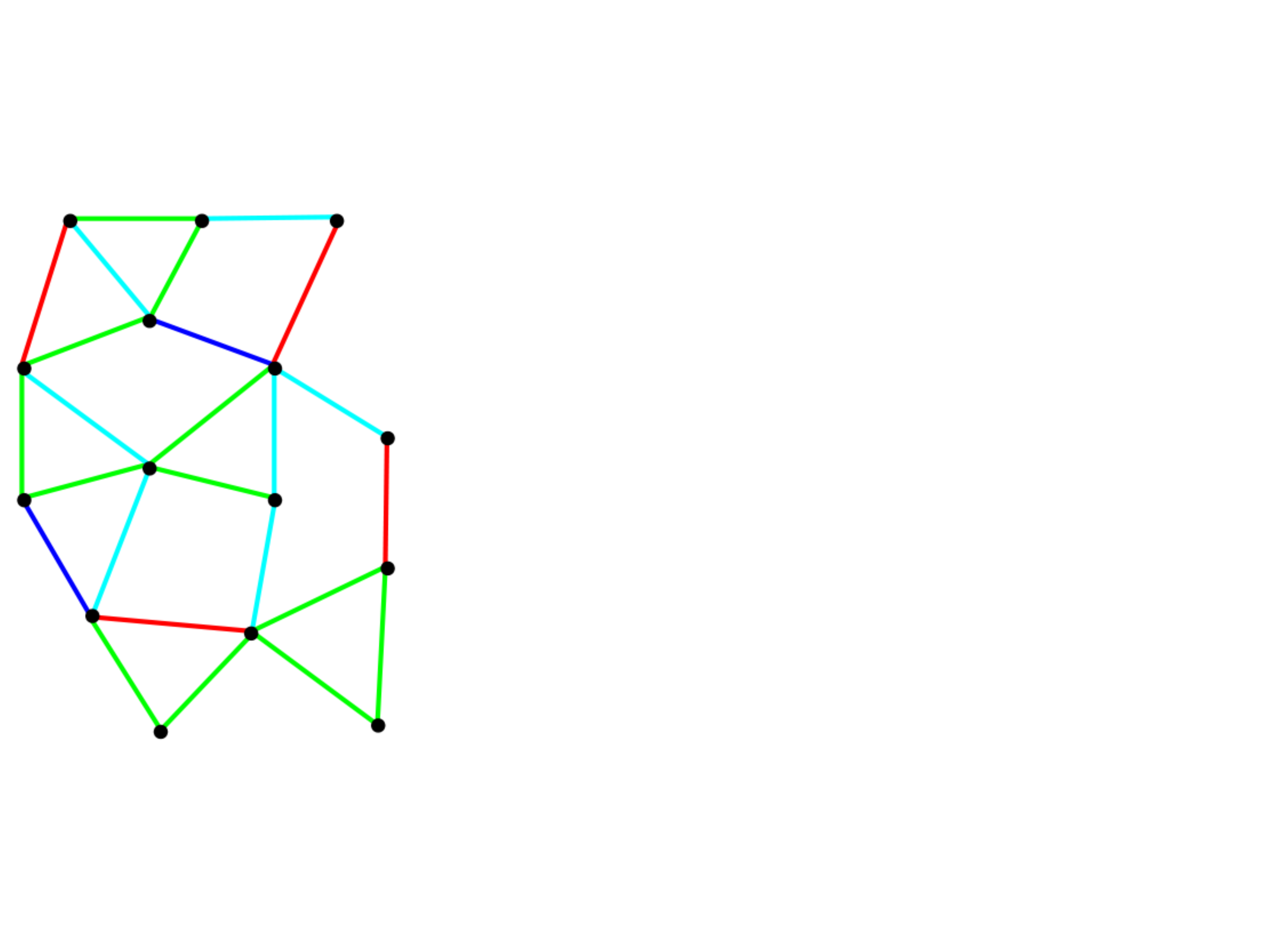}}
\put(125,160){\includegraphics[width=4in]{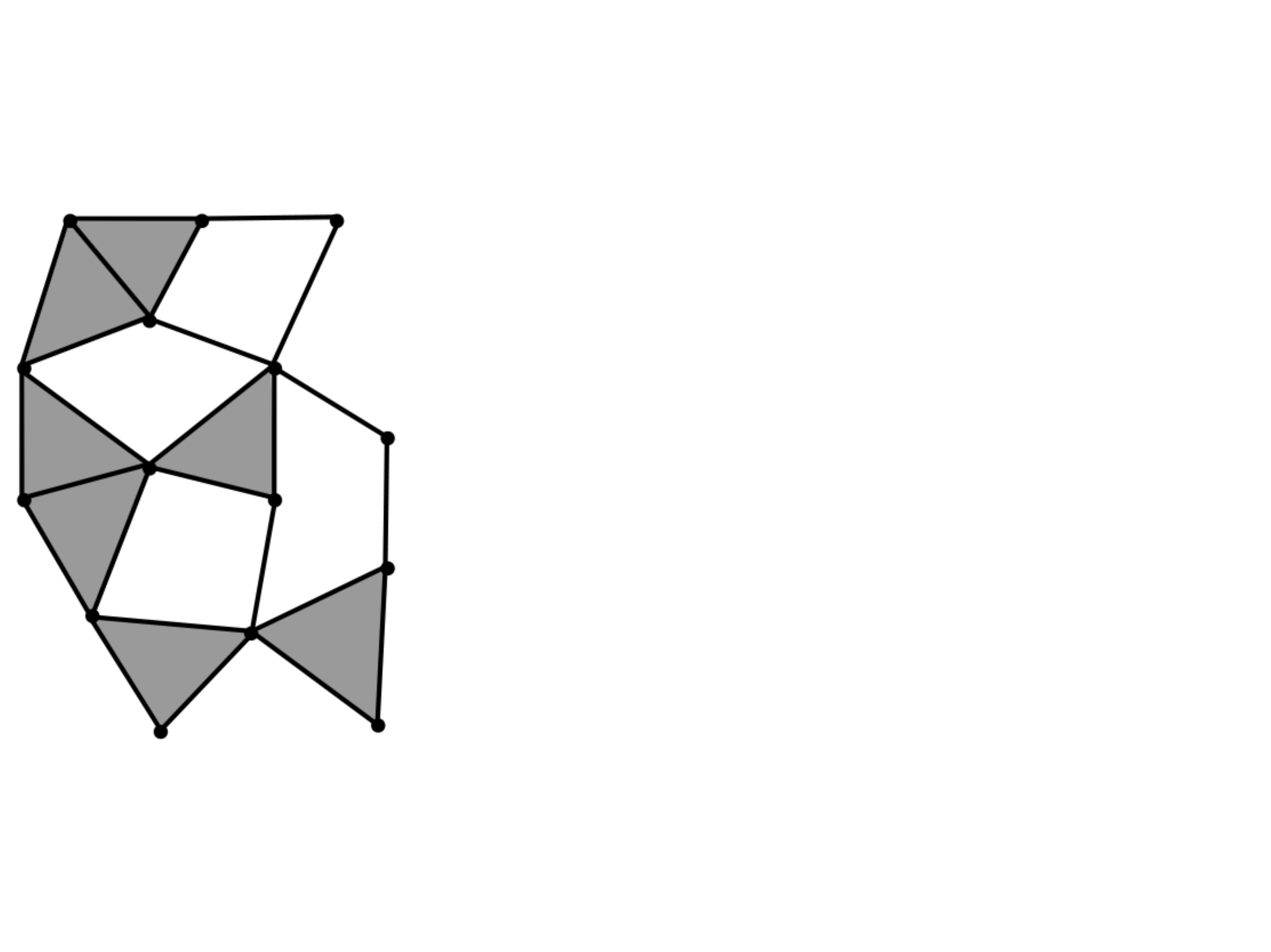}}
\put(250,160){\includegraphics[width=4in]{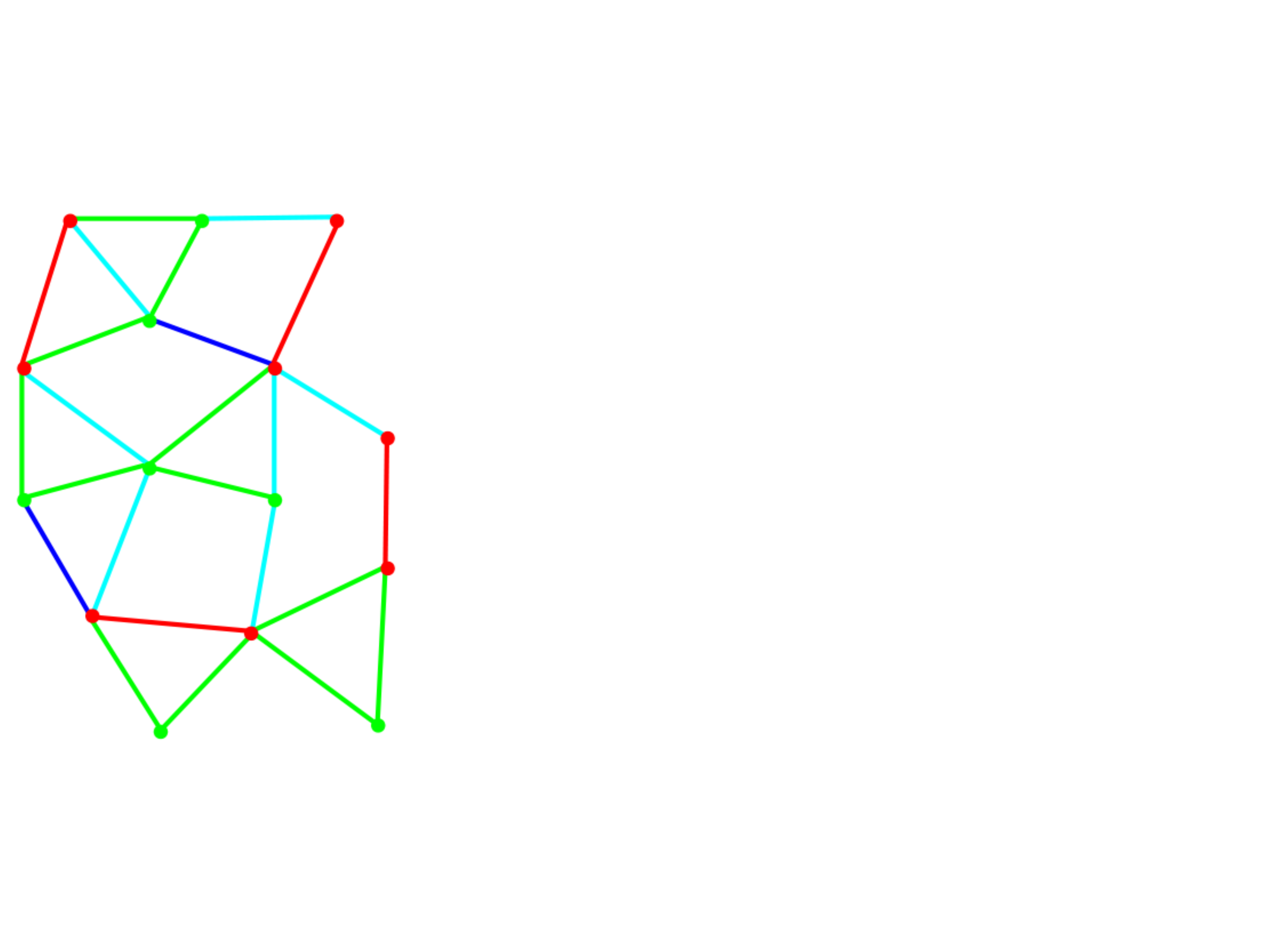}}
\put(375,160){\includegraphics[width=4in]{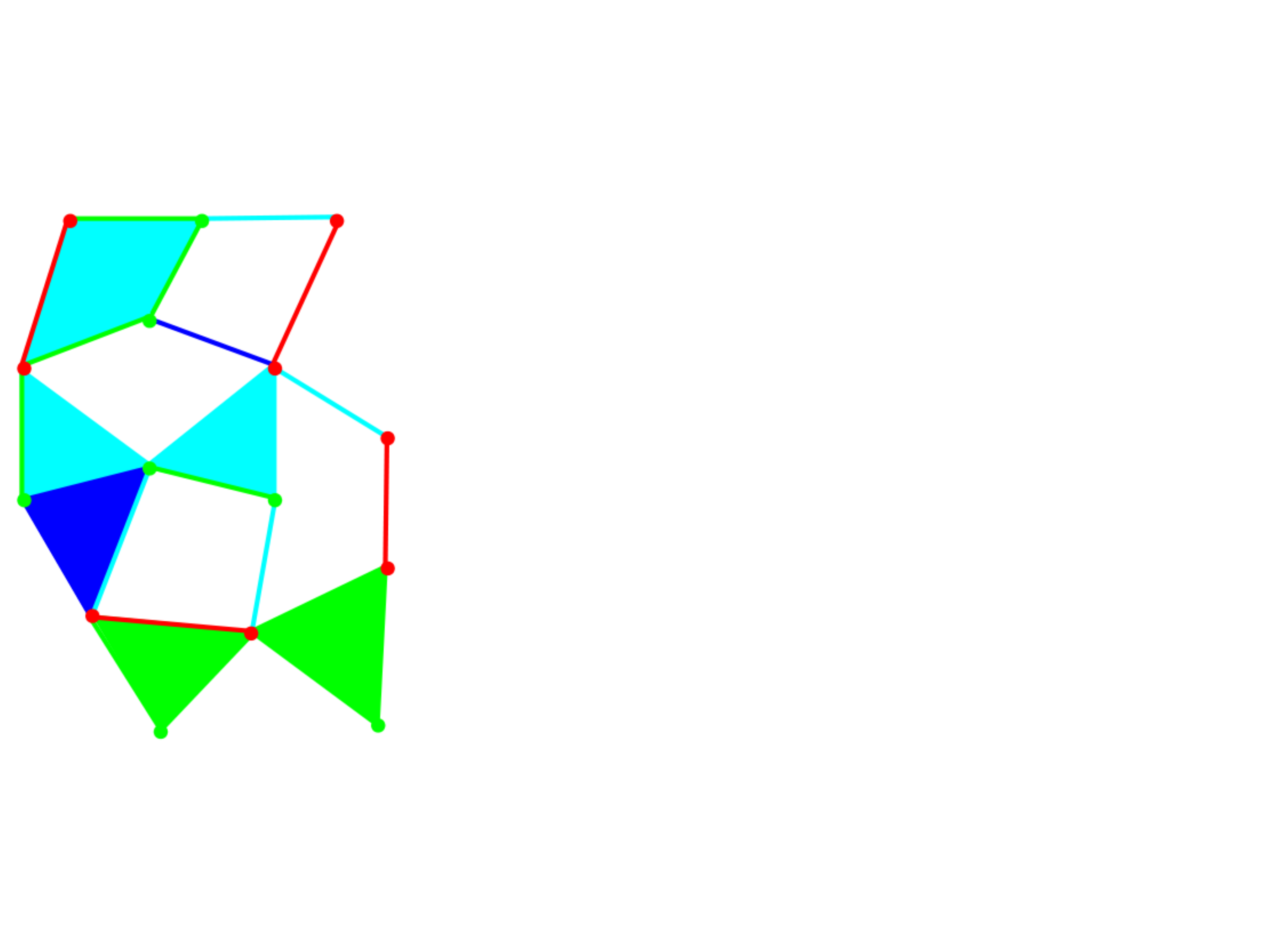}}
\put(0,-15){\includegraphics[width=4in]{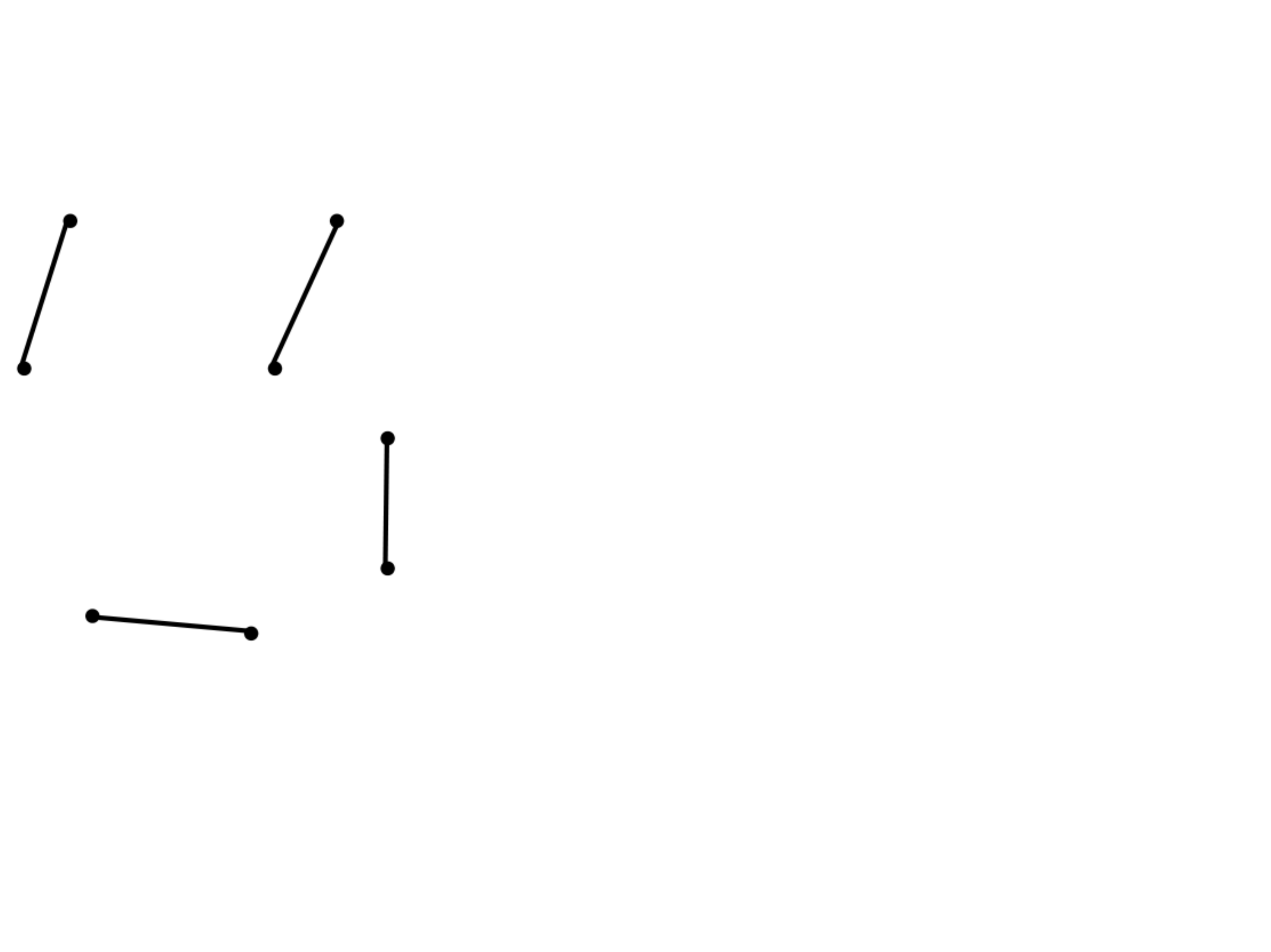}}
\put(125,-15){\includegraphics[width=4in]{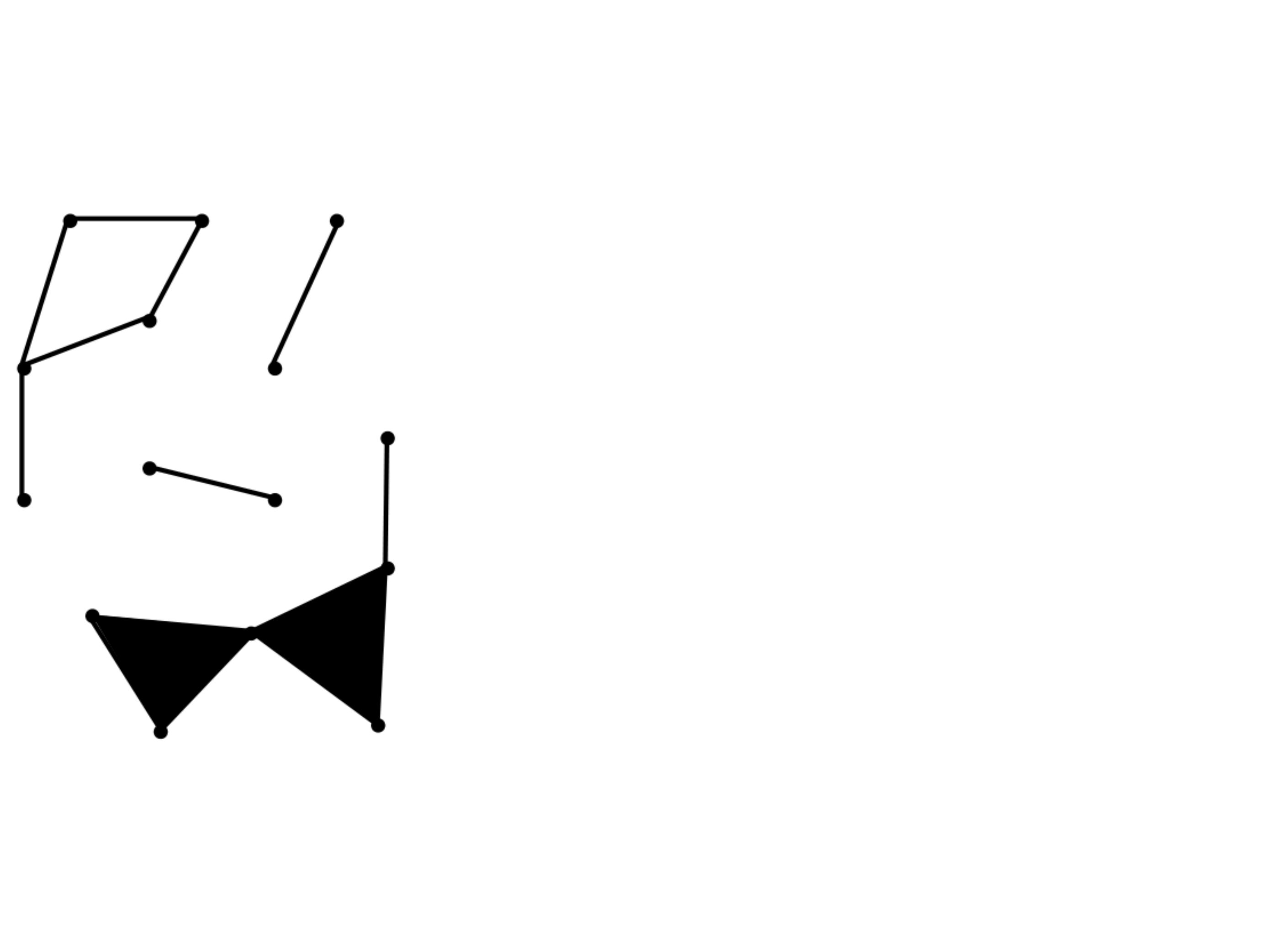}}
\put(250,-15){\includegraphics[width=4in]{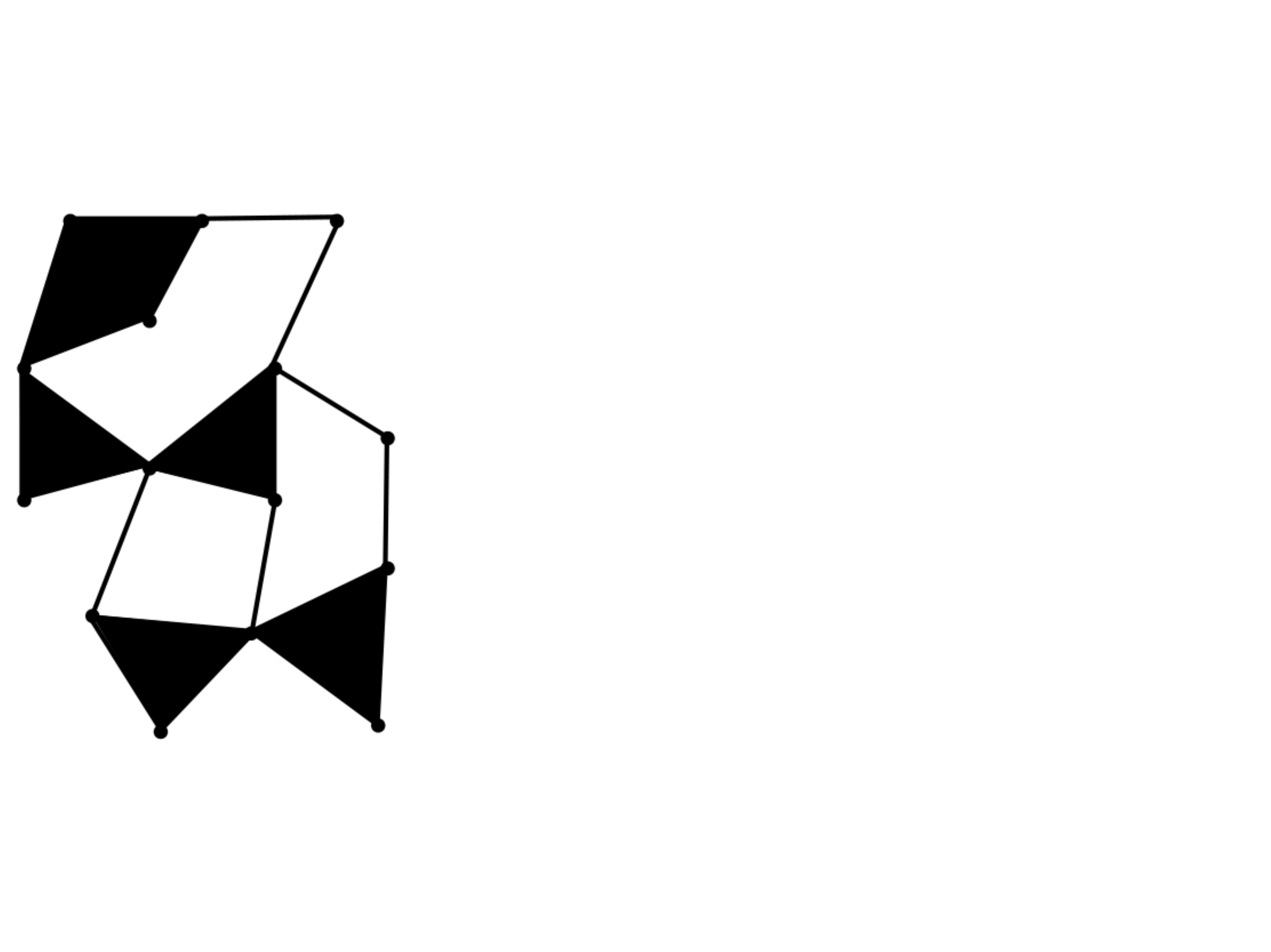}}
\put(375,-15){\includegraphics[width=4in]{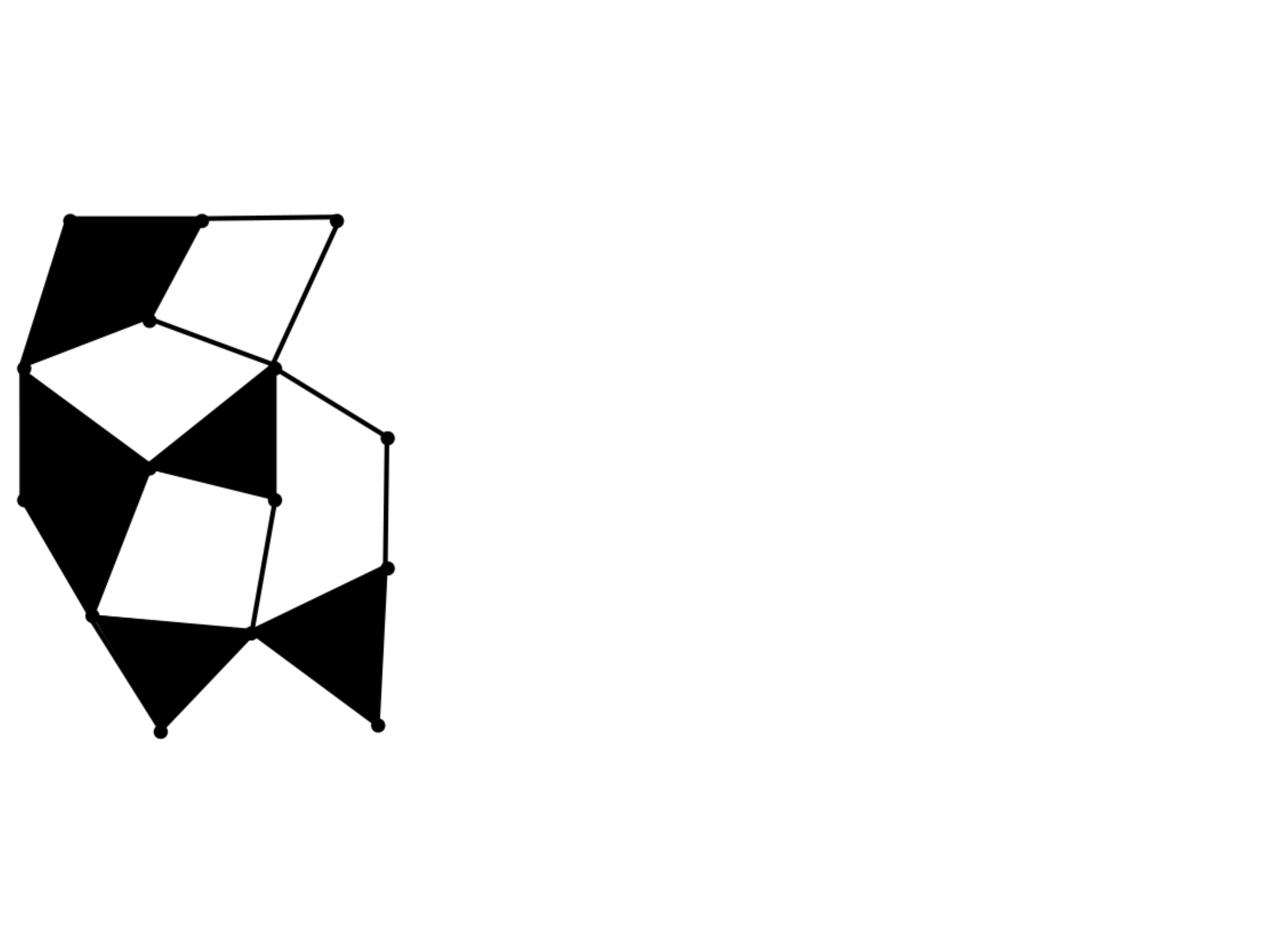}}

\put(50,180){(a)}
\put(175,180){(b)}
\put(300,180){(c)}
\put(425,180){(d)}
\put(50,-2){(e)}
\put(175,-2){(f)}
\put(300,-2){(g)}
\put(425,-2){(h)}
\end{picture}
\caption{A representation of a simple interaction network  $\fn_I$. (a) Vertices represent the particles and the edges correspond to the non zero force between the particles. An increasing value of the force is denoted by  blue, cyan, green, and red. (b) Collection of simplicies on which the function $\cn_I:  f \to \R$ is positive.   Extension of the function $f$ to the vertices  (c) and to the 2-dimensional simplicies (d). (e)-(h) The
complexes  $\fn(f,\theta_i)$ for positive  $\theta$  equal to $\theta_4$ (red), $\theta_3$ (green), $\theta_2$ (cyan) and $\theta_1$ (blue), respectively. 
}
\label{fig:ToyNetwork}
\end{figure*}

As is indicated in the Introduction it is well accepted that the geometry of force chains plays an important role in determining the  macroscopic properties of dense granular material. In this section we expand on the complexes constructed in Section~\ref{sec:particle} to include the forces between the particles into this mathematical framework.   In the present work we mainly  focus on the normal force, that is,  the component of the force projected on the line connecting the centers of interacting particles.   We will view the magnitude of the normal force as a scalar field defined over the complex, i.e., a function $f\colon \cn \to \R$.   There is one  constraint on the definition of $f$ that arises from  the use of persistent homology to capture the geometry of the force chains. 

To understand this constraint assume for the moment that we are given a complex $\cn$ and a scalar field $f\colon \cn \to \R$. 
Since for particulate systems one often considers particles interacting by strong/weak forces, we are 
interested in the geometry of a part of the complex on which the forces exceed a specified level.
Thus we define a {\em force network} to be the super level set
 \begin{equation}
 \label{eq:superlevel}
 \fn(f,\theta) := \{ \sigma \in \cn \mid  f(\sigma) \geq \theta\}
 \end{equation}  
 which corresponds to the part of the particle network experiencing force larger than $\theta$.  We use 
 homology to quantify the geometry of $\fn(f,\theta)$. Hence $\fn(f,\theta)$ has to be a complex for
every value of $\theta$.  This means that  if $\sigma\in\fn(f,\theta)$ and $\sigma'\subset\sigma$,
then $\sigma'\in\fn(f,\theta)$. 
Thus, in our construction of $f$ we need to insure that this condition is satisfied. This leads to the 
following definition.

\begin{defn}
\label{defn:monotone}
{\em
Given a  complex $\cn$,  a function $f\colon \cn \to \R$ is {\em algebraically  monotone} if $f(\sigma')\geq f(\sigma)$ for every $\sigma' ,\sigma\in\cn$ such that $\sigma'\subset\sigma$.
}
\end{defn}

\noindent It is left to the reader to check that if $f$ is algebraically monotone, then  $\fn(f,\theta)$ is a complex for every value of $\theta$. 

\begin{defn}
\label{defn:forceFiltration}
{\em
Given a complex $\cn$ and an algebraically monotone function $f\colon \cn \to \R$, the
associated {\em force network filtration} is the collection of all force network complexes
\[
\setof{ \fn(f,\theta) \mid \theta\in\R}.
\]
}
\end{defn}

The construction of  $f\colon \cn \to \R$ depends on  the available information. The weaker assumption, which we associate with digital or position complexes, is that for each particle $p_i$ we can estimate the   magnitude of the force $\psi_i$ on $p_i$. The stronger assumption, which leads to the use of an interaction complex, is that we can estimate the   magnitude of the force $\psi_{i,j}$ between any two particles $p_i$ and $p_j$.  The function $f$ is defined in two steps. First we define $f$ for cells of a certain dimension depending on the type of complex. Then we uniquely extend the definition to all the cells. The construction of the extension guarantees that $f$ is an algebraically monotone function.

\paragraph{Digital Force Networks} 
Recall that a 2-dimensional cube  $\sigma\in\cn_D^{(2)}$ if it intersects at least one particle $p_i$. 
We define
\[
f(\sigma) = \max\setof{\psi_i\mid \sigma\cap p_i \neq \emptyset}.
\]
for $\sigma\in\cn_D^{(2)}$ .

\paragraph{Position Force Networks} For each
$\ang{v_i}\in\cn_P^{(0)}$ corresponding to the particle $p_i$ we define
\[
f(\ang{v_i}) := \psi_i.
\]

\paragraph{Interaction Force Networks}  For the interaction network $\cn_I$ the natural starting point for the definition of $f$ is on the edges $\ang{v_i, v_j} \in \cn_I^{(1)}$,
\[
f(\ang{v_i, v_j}) := \psi_{i,j}.
\]

For a complex  $\cn_\bullet$, $\bullet \in\setof{P,I}$ we extend the definition of the function $f$ from $\cn_\bullet^{(i)}$ to the cells $\sigma \in \cn_\bullet^{(j)}$ for $j<i$ by 
\[
f(\sigma) = \max\setof{f(\sigma')\mid \sigma\subset \sigma',\ \sigma'\in \cn_\bullet^{(i)}}.
\]
Extension to cells  $\sigma \in \cn_\bullet^{(j)}$ for $j>i$ and $\bullet \in\setof{P,I}$ is defined by 
\[
f(\sigma) = \min\setof{f(\sigma')\mid \sigma'\subset \sigma,\ \sigma'\in \cn_\bullet^{(i)}}.
\]

We use the following proposition  to summarize the above discussion and constructions.

\begin{prop}
Given a complex $\cn_\bullet$, $\bullet \in\setof{D,P,I}$ and  $f$ defined as above, the associated super level set $\fn_\bullet(f,\theta)$ is a complex for all
values of $\theta\in \R$.
\end{prop}

Since we are assuming that there is only a finite number of particles in our system, any force network filtration $\setof{ \fn_\bullet(f,\theta) \mid \theta\in\R}$ contains only finitely many distinct complexes.  We can use homology, in  particular the Betti numbers, to characterize the geometry of each of the distinct force networks in the force network filtration.

To gain intuition into the force networks consider  the interaction force network   indicated in  Figure~\ref{fig:ToyNetwork}. Figure~\ref{fig:ToyNetwork}(a) represents a collection of particles. The particles are represented by the vertices and the shown edges correspond to the non-zero forces between the particles. Figure~\ref{fig:ToyNetwork}(b) shows only the simplices of $\cn_I$ for which the value of the function $f : \cn_I \to \R$ is positive. The value of $f$ on the edges is determined by the forces between the particles. In Figure~\ref{fig:ToyNetwork}(a) the non-zero forces are color coded. In order of  increasing value, the force is denoted by   $\theta_1$ (blue), $\theta_2$ (cyan), $\theta_3$ (green)  and $\theta_4$ (red).
The value of the function $f$ is extended to the vertices in Figure~\ref{fig:ToyNetwork}(c) and to the $2$-dimensional simplicies in Figure~\ref{fig:ToyNetwork}(d).  The figures (e)-(h) indicate the  associated force network for non negative values of $\theta$. For $\theta \leq 0$ the complex $\fn_I(f,\theta) = \cn_I$, and  as explained in Section~\ref{sec:homology}, it consists of a single connected component that does not encircle any loops.

For example, referring to the force network filtration of Figure~\ref{fig:ToyNetwork} we can extract the following data:
\begin{eqnarray*}
(\beta_0(\fn_I(f, \theta_4)),\beta_1(\fn_I(f,\theta_4))) & = & (4,0) \\ 
(\beta_0(\fn_I(f,\theta_3)),\beta_1(\fn_I(f,\theta_3))) & = & (4,1) \\
( \beta_0(\fn_I(f,\theta_2)),\beta_1(\fn_I(f,\theta_2))) & = & (1,3)\\ 
(\beta_0(\fn_I(f,\theta_1)),\beta_1(\fn_I(f,\theta_1))) & = & (1,4).
\end{eqnarray*}
It is worth noting that the $H_0$ homology information for $\fn_I(f,\theta_4)$ and $\fn_I(f,\theta_3)$ agree and yet the
structure of the components has changed dramatically. Two distinct connected components become one and a new connected component is formed. To capture this information we make use of the fact that these
complexes are nested by inclusion.  This leads to the concept of persistent homology.

%%%%%%%%%%%%%%%%%%
%
\section{Persistent Homology}
\label{sec:persistence}
%
%%%%%%%%%%%%%%%%%%%

\begin{figure*}[t]
\begin{picture}(0,120)
{\includegraphics[width=7.2in]{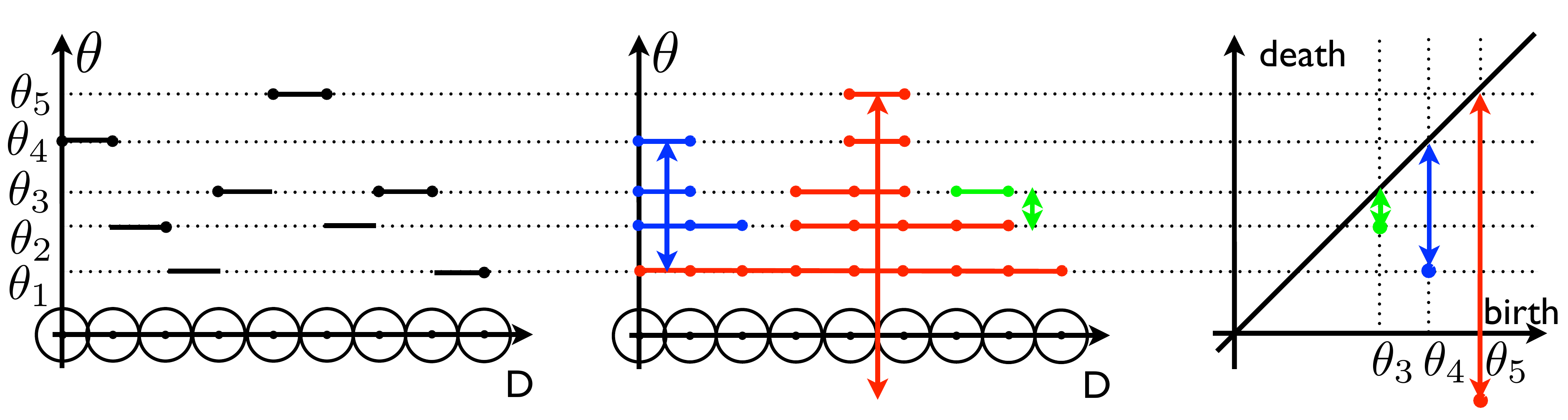}}
\put(-430,-8){(a)}
\put(-240,-8){(b)}
\put(-70,-8){(c)}
\end{picture}
\caption{(a)
A linear chain of particles is shown on the horizontal axis. Magnitude of the normal force between the adjacent particles is given by the value of the step function above the edge connecting the particles' centers.  (b)  Structure of   $\fn_I(f,\theta)$ for different values of $\theta$.  For example,
 $\fn_I(f,\theta_5)$ consist of a single connected component formed by one edge and two vertices. The set $\fn_I(f,\theta_3)$ contains three distinct connected components corresponding to different geometric features indicated by the double arrows.  (c) $\beta_0$ persistence diagram for the interaction force network of the system shown in (a).  The color of the points matches the color of the features they represent.}
\label{fig:ToyDiagram1D}
\end{figure*}

Given a force network filtration 
\[
\fn(f,\theta) := \setof{\sigma\in\cn\mid f(\sigma)\geq \theta}
\]
generated by a finite number of particles, there is a finite number of values 
\[
0 = \theta_0 < \theta_1 <\cdots < \theta_K = \max_{\sigma\in\cn}f(\sigma) 
\]
such that $\theta_k = f(\sigma)$ for some $\sigma\in \cn$.
Though the Betti numbers  characterize the topology of a given force network $\fn(f,\theta_k)$,
 the vector space structure of homology plays an essential role in that it allows us to compare
the topology of $\fn(f,\theta_k)$ with any other  force network $\fn(f,\theta_j)$.  
Given $\theta_i <\theta_j$,
$\fn(f,\theta_j)\subset \fn(f,\theta_i)$ and hence there is an inclusion map
\[
\iota_{\theta_i,\theta_j}\colon \fn(f,\theta_j)\to \fn(f,\theta_i).
\]
As is indicated at the end of Section~\ref{sec:homology}, this defines maps
\[
\iota_{\theta_i,\theta_j\, *}\colon H_*(\fn(f,\theta_j))\to H_*(\fn(f,\theta_i))
\]
on each homology group $H_*$. It is important to note that $\iota_{\theta_1,\theta_2\, *}$ need not be an inclusion
map on the level of the homology groups.

Persistent homology makes use of the above mentioned  maps  to compare topological features within different force networks. 
The first observation, while trivial, is essential for our discussion and follows directly from the fact that $\fn(f,\theta)=\emptyset$
for all $\theta > \theta_K$.

\begin{lem}
\label{lem:0}
If $\theta > \theta_K$, then  $H_*(\fn(f,\theta))= 0$. 
\end{lem}

Now consider a value $\theta_k$ such that $v\in H_n(\fn(f,\theta_k))$ and let $v\neq 0$.
If $n=0$ or $1$, then $v$ provides information about the existence of components or loops, respectively, in $\fn(f,\theta_k)$.
In light of Lemma~\ref{lem:0},  there exists
a unique largest threshold $\theta_b(v) \geq \theta_k$ with the property that there exists $v_b\in H_n(\fn(f,\theta_b))$ such that
$\iota_{\theta_k,\theta_b \, *}(v_b) = v$. The geometric feature associated with $v$ is said to have been {\em born} at level
 $\theta_b(v)$.
 
It is also possible that for some $\theta < \theta_k$, $\iota_{\theta,\theta_k \, *}(v) = 0$.
In this case we define 
\[
\theta_d(v) := \max\setof{\theta_j\mid \iota_{\theta_j,\theta_k\, *}(v)=0}
\]
and
we say that the geometric feature associated with $v$ {\em dies} at level $\theta_d(v)$.  Given our construction, not every
geometric feature needs to die. In particular, for $n=0,1$,
\[
H_n(\fn(f,0))\cong H_n(\cn)
\]
which, as the examples in this paper indicate, need not be trivial for digital and position complexes.
We make use of the following convention
\[
\text{if}\ \iota_{0,\theta_k\, *}(v) \neq 0,\ \text{then}\ \theta_d(v)=-1.
\]
A remarkable fact \cite{ carlsson, edelsbrunner:harer} is that given a finite filtration it is possible to choose a consistent set of bases for
$H_n(\fn(f,\theta_k))$, $k=-1,\ldots, K$ such that each basis element has a well defined birth and death level $(\theta_b,\theta_d)$.  
By equation \eqref{eq:beta_incidence}, if $v\in H_1(\fn_I(f,\theta_k))$, then $\theta_d(v) \geq 0$ and there exists a
unique element $\bar{v}\in H_0(\fn_I(f,\theta_k))$ such that $(\theta_b(v),\theta_d(\bar{v})) = (\theta_K,-1)$.

The collection of all pairs  $(\theta_b,\theta_d)$ associated with the $n$-th homology group for the force networks  are used to construct the $\beta_n$ persistence diagram for the scalar field $f\colon\cn\to [0,\infty)$.  
To provide some intuition concerning the process of going from an interaction network $\cn_I$ to a persistence diagram,
consider a single chain of particles shown along the horizontal axis in Figure~\ref{fig:ToyDiagram1D}(a).  If two particles are
not in contact, then the force acting between them is zero.  Stated more formally,  for any edge $\ang{v_i,v_j}\in \cn_I$
if the particles corresponding to the vertices $v_i$ and $v_j$ are not in contact, then $f(\ang{v_i,v_j}) = 0$.
Otherwise $f(\ang{v_i,v_j}) $ is defined by a step function such as that shown in Figure~\ref{fig:ToyDiagram1D}(a).
The extension of $f$ to the vertex  $\ang{v_i}$ is obtained via the definition in Section~\ref{sec:force} and given by the black dot above the center of the particle corresponding to $\ang{v_i}$. Note that $f(\ang{v_i,v_j,v_k})$ is always zero. 
Figure~\ref{fig:ToyDiagram1D}(b) shows the sets $\fn_I(f,\theta)$ for different values of $\theta$. The set $\fn_I(f,\theta) = \emptyset$ for $\theta > \theta_5$ and $\fn_I(f,\theta_5)$ consist of a single connected component formed by one edge and two vertices. Finally for $\theta \leq 0$ the set $\fn_I(f,\theta) = \cn_I$ has a single connected component. The $\beta_0$ persistence diagram for the interaction force network is shown in Figure~\ref{fig:ToyDiagram1D}(c).

We now explain what can be inferred about  the interaction force network from the $\beta_0$ persistence diagram shown in Figure~\ref{fig:ToyDiagram1D}(c).  The fact that there are no points with birth coordinate larger than $\theta_5$
indicates the absence of components experiencing force larger than $\theta_5$.
The set $\fn_I(f,\theta_5)$ consists of a single connected component. Another connected component appears at $\theta_4$. These two components merge together at $\theta_1$.  In the language of persistent  homology the connected component is born at $\theta_4$ and dies at $\theta_1$  as indicated by the point $(\theta_4,\theta_1)$ in the persistence diagram. The connected component that appears at $\theta_5$ persists  for all values $\theta \leq \theta_5$. In the persistence diagram it is represented by the point $(\theta_5,-1)$.  There is one more geometric feature of the function $f$ described by the point $(\theta_3, \theta_2)$.  This  geometric feature corresponds to a pair consisting of a local maxima with value $\theta_3$ and a local minima with value $\theta_2$. This pair is visualized by the shortest double arrow in Figure~\ref{fig:ToyDiagram1D}.
Also the point  $(\theta_4, \theta_1)$ corresponds to a pair consisting of a local maxima and minima. The special point $(\theta_5,-1)$ encodes the value of the global maxima. 

\begin{figure}
\centering
	\begin{picture}(400,130)
	\put(-18,0){\includegraphics[width=2.6in]{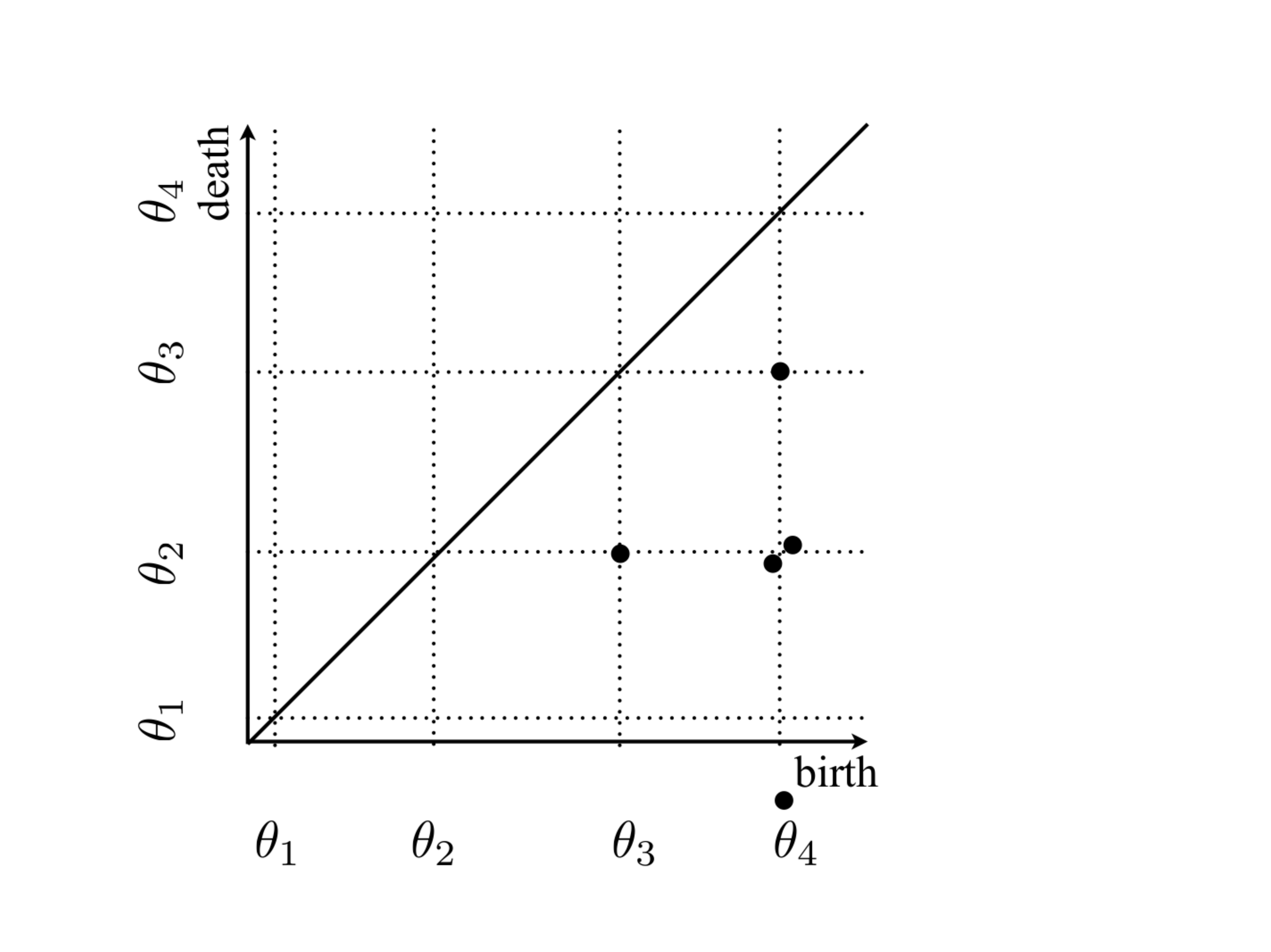}}
	\put(65,-5){(a)}
      	\put(115,0){\includegraphics[width=2.6in]{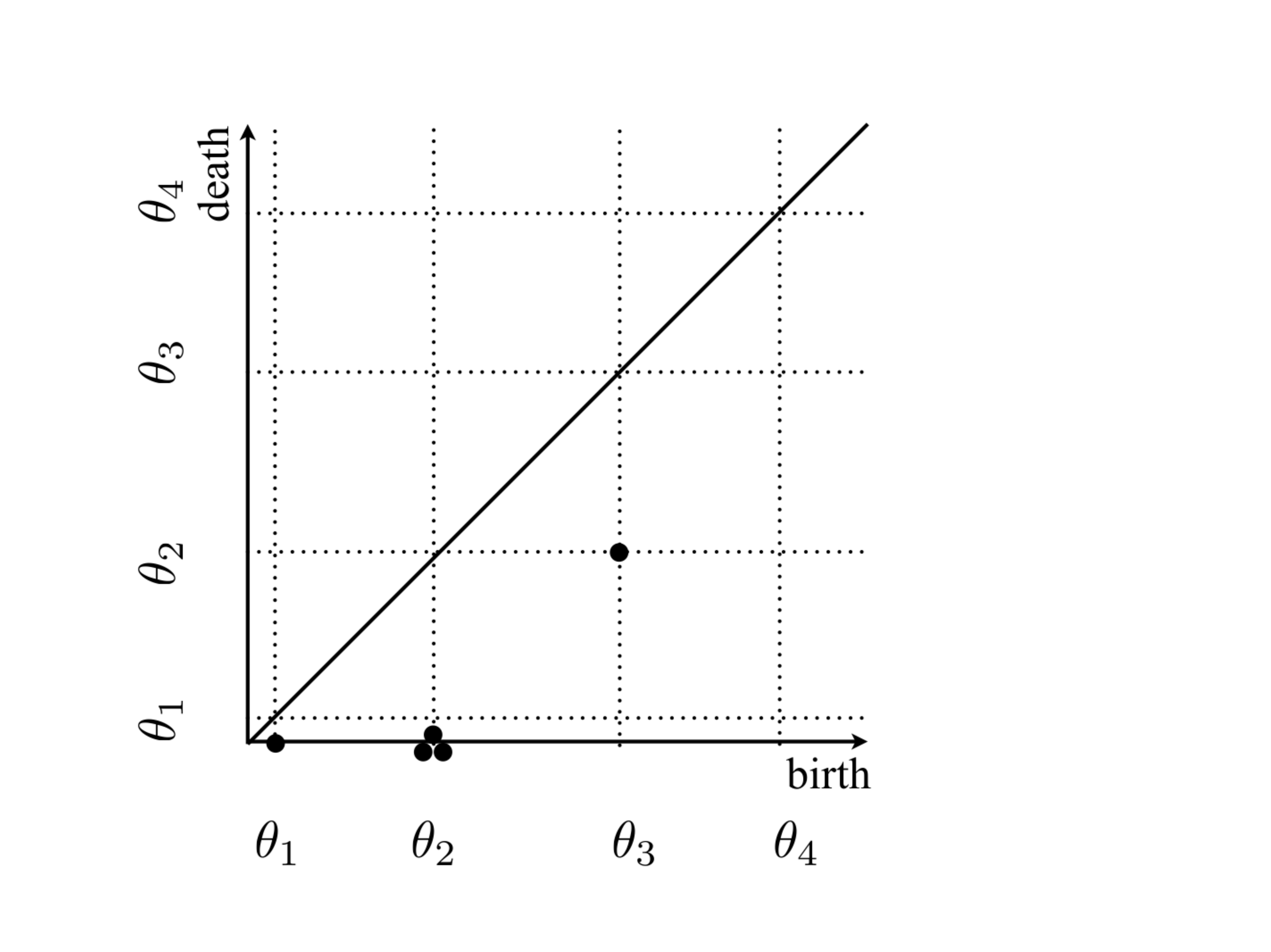}}
	\put(190,-5){(b)}
	\end{picture}
\caption{(a) $\beta_0$ and (b) $\beta_1$ persistence diagrams for the  force network shown in Figure~\ref{fig:ToyNetwork}(c) }
\label{fig:PeristenceDiag}
\end{figure}

We now return to the example of the interaction force network associated with Figure~\ref{fig:ToyNetwork} of 
Section~\ref{sec:force}.
This network is more complex than the one analyzed above. Not surprisingly the persistence diagrams for this network,  shown in Figure~\ref{fig:PeristenceDiag}, contain more points.
For the same reason as before, the $\beta_0$ persistence diagram, see Figure~\ref{fig:PeristenceDiag}(a), does not contain any points with the birth coordinate larger than $\theta_4$.   Four points with the birth coordinate $\theta_4$ correspond to the four connected components that appear  in $\fn_I(f,\theta_4)$ (Figure~\ref{fig:ToyNetwork}(e)). The death coordinates of the points differ. This indicates that the components merge for different values of $\theta$. The first merging appears for $\theta_3$ (Figure~\ref{fig:ToyNetwork}(f)) and is represented by the dot $(\theta_4,\theta_3)$. Moreover a new component appeared at the level $\theta_3$ and consequently merged with a preexisting component at $\theta_2$ (Figure~\ref{fig:ToyNetwork}(g)) as indicated by the dot $(\theta_3,\theta_2)$.  Also another two components  that appear at $\theta_4$ disappear at $\theta_3$ hence there are two copies of the point $(\theta_4,\theta_2)$. Finally there is only one connected component for  all $\theta < \theta_2$. This component appears for $\theta = \theta_4$ 
and does not disappear.  In the persistence diagram it is represented by $(\theta_4,-1)$.

The fact that the interaction force network contains loops can be inferred from  the $\beta_1$ persistence diagram. The first loop appears at $\theta_3$ (Figure~\ref{fig:ToyNetwork}(f)) and is filled by triangular cells at $\theta_2$ (Figure~\ref{fig:ToyNetwork}(g)) as shown by the point $(\theta_3,\theta_2)$ in Figure~\ref{fig:PeristenceDiag}(b). Another three loops appear at $\theta_2$ and persist for all positive thresholds.  Due to the definition of $\fn_I(f,\theta)$ all the loops are filled in for $\theta = 0$. So  these three  loops are represented by  three copies of the point  $(\theta_2, 0)$. The last loop appears at $\theta_1$ (Figure~\ref{fig:ToyNetwork}(h)) and also persists for all positive thresholds hence the  point $(\theta_1,0)$ belongs to the $\beta_1$ persistence diagram.

In general for the $\beta_0$ diagram a birth level $\theta_b(v)$ corresponds to the value of  a local maximum that is associated with the birth of a connected component  measured by an element $v\in H_0(\fn_I(f, \theta_b))$.
As $\theta$ decreases  this component grows until it meets, at a point associated with a local minimum or saddle,  another component. Assume this other  component is measured by the homology class $v'$ and that the value of the local minimum (or saddle ) is $\underline{\theta}$.   If $\theta_b(v)<\theta_b(v')$, then $\theta_d(v)= \underline{\theta}$.  In this case,  $\theta_b(v)-\theta_d(v)$ measures the difference in height between the local maximum and local minimum and hence this difference can be used as  a measure of how robust a feature is.

For the $\beta_1$ diagram a birth level $\theta_b(v)$  of the loop corresponding to an element $v \in _0(\fn_I(f, \theta_b))$  is the smallest value of $f$ along this loop.  If the loop is filled in by particles forming a crystalline zone, then $\theta_d(v)$ is the smallest value of $f$ inside the region encompassed by the loop. If the interior of the loop is not completely filled in by a crystalline structure, then there must be a defect encircled by this loop. So the loop cannot be filled in with the triangles.
Therefore,  if we use a digital or position force network, then the loop  never dies and $\theta_d(v) = -1$.  For the  interaction network  all the loops are filled in at $\theta = 0$ and $\theta_d(v) = 0$.

We close this section with a formal definition of the persistence diagrams.
\begin{defn}
\label{defn:PD}
{\em
Let $\Theta = \setof{\theta_k \mid  k = -1,\ldots K}$ and
\[
\setof{\fn(f,\theta_k) \mid \theta_k\in\Theta }
\] 
be a force network filtration over a complex $\cn$. The associated {\em $n$-th persistence  diagram} $\pd_n(f,\cn,\Theta)$ is the multiset
consisting of the following points:
\begin{enumerate}
\item one point for each $n$-th persistence point $(\theta_b,\theta_d)$; 
\item infinitely many copies of points $(\theta,\theta)$ on the diagonal.
\end{enumerate}
}
\end{defn}

Condition (1) of Definition~\ref{defn:PD} arises because distinct geometric features can appear and  disappear at the same thresholds and thus there may be multiple copies of the same persistence pair.  The necessity of condition (2)  is made clear in Section~\ref{sec:space}.

We conclude this section with an observation. Let $\spd_n$ denote the set of all $n$-th persistence diagrams and 
$\spd$ the set of all persistence diagrams.  Given a chain complex $\cn$, let $M(\cn,[0,\infty))$ denote the set of monotone maps on $\cn$.
We can view persistence diagrams as a function 
\begin{equation}
\label{eq:pd}
\pd\colon M(\cn,\R) \to \spd
\end{equation} 
or equivalently a collection of functions
$\pd_n\colon M(\cn,\R) \to \spd_n$ defined by
\[
\pd_n(f) = \pd_n(f,\cn,\Theta)
\]
where $\Theta = \setof{\theta_k \mid  k = -1,\ldots K}$ consists of the finite set of values obtained by $f$ along with the convention
that $\theta_{-1}=-1$ and $\theta_0=0$.

\section{The space of Persistence Diagrams}
\label{sec:space}

The results concerning
topological  fidelity
of the complexes $\cn_{\bullet}$ have, up to this point, been mostly negative. The introduction of persistence 
allows us to present positive results. In and of itself this suggests that force network filtrations and their associated 
persistent homology provide more appropriate metrics for understanding force networks than measurements performed at single thresholds.  
To obtain continuity results that guarantee that small changes in measurement or forces in the DGM lead to small 
changes in the persistence diagrams requires us to be able to measure the distance between persistence diagrams.

\begin{defn}
\label{defn:bottleneck}
{\em
Let $\pd = \{\pd_i\}_{i=0}^n$ and $\pd' = \{\pd'_i\}_{i=0}^n$ be two collections of persistence diagrams. 
The {\em bottleneck distance} between $\pd$ and $\pd'$ is defined to be 
\[
d_B(\pd,\pd') = \max_{0 \leq i \leq n}\inf_{\gamma\colon \pd_i \to \pd'_i} \sup_{p\in\pd_i} \| p -\gamma(p)\|_\infty , 
\]
where $\|(a_0,b_0)-(a_1,b_1)\|_\infty := \max\setof{|a_0-a_1| , |b_0-b_1|}$ and $\gamma$  ranges over all bijections.
Similarly, the
{\em degree-$q$ Wasserstein distance} is defined as
\[
d_{W^q}(\pd,\pd') = \left[ \sum_{i = 0}^n \inf_{\gamma\colon \pd_n \to \pd'_n} \sum_{p\in\pd_n} \| p -\gamma(p)\|^q_\infty \right]^{1/q}. 
\]
}
\end{defn}

As is indicated in \cite{edelsbrunner:harer}, equipped with
either the bottleneck or degree-$q$ Wasserstein distance  $\spd$
is a metric space.  From now on we always assume $\spd$ is one of these metric spaces.

\begin{figure}[t]
\centering
	\begin{picture}(400,115)
	\put(2,0){\includegraphics[width=1.7in]{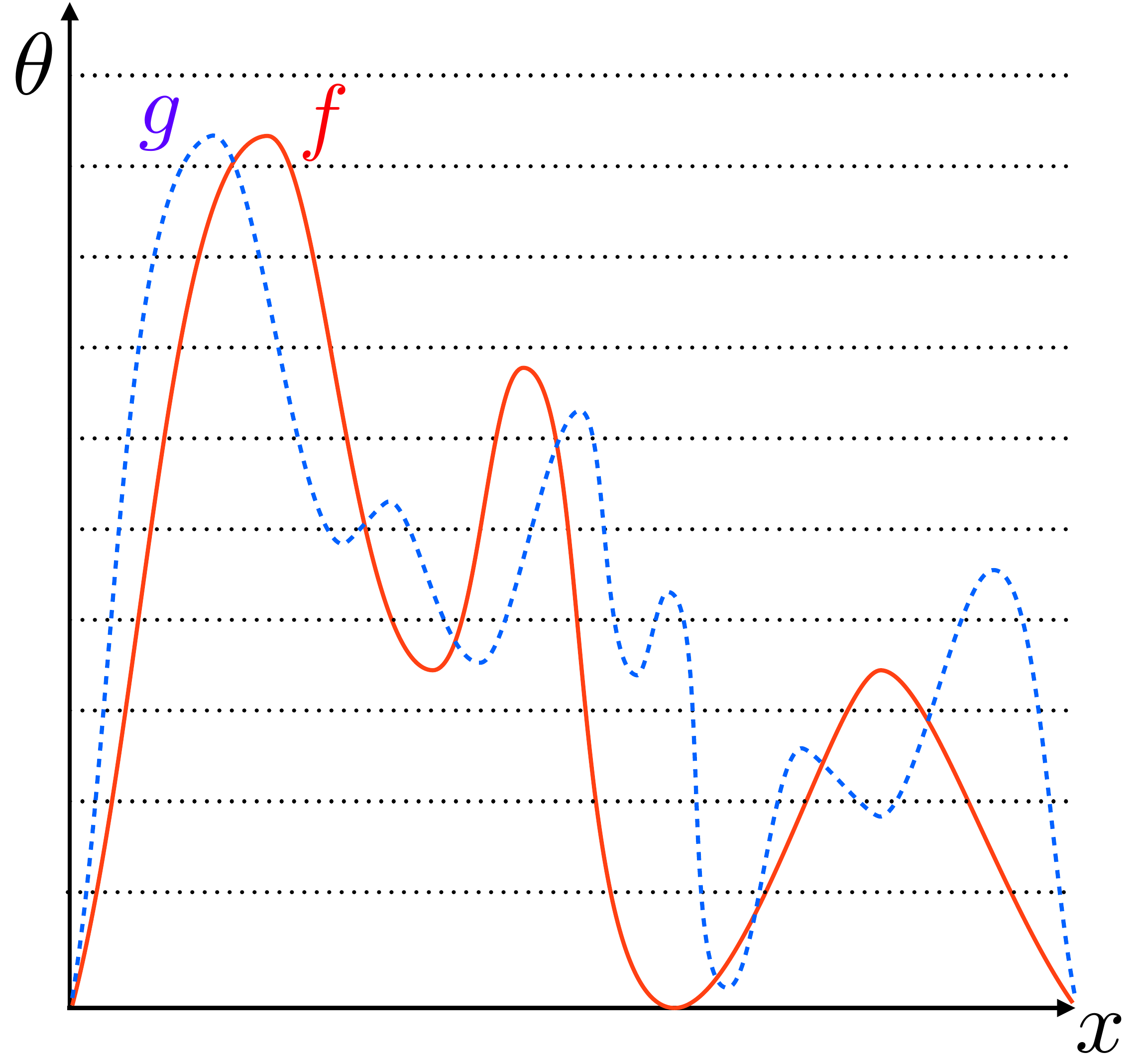}}
	\put(57,-10){(a)}
      	\put(125,-4){\includegraphics[width=1.8in]{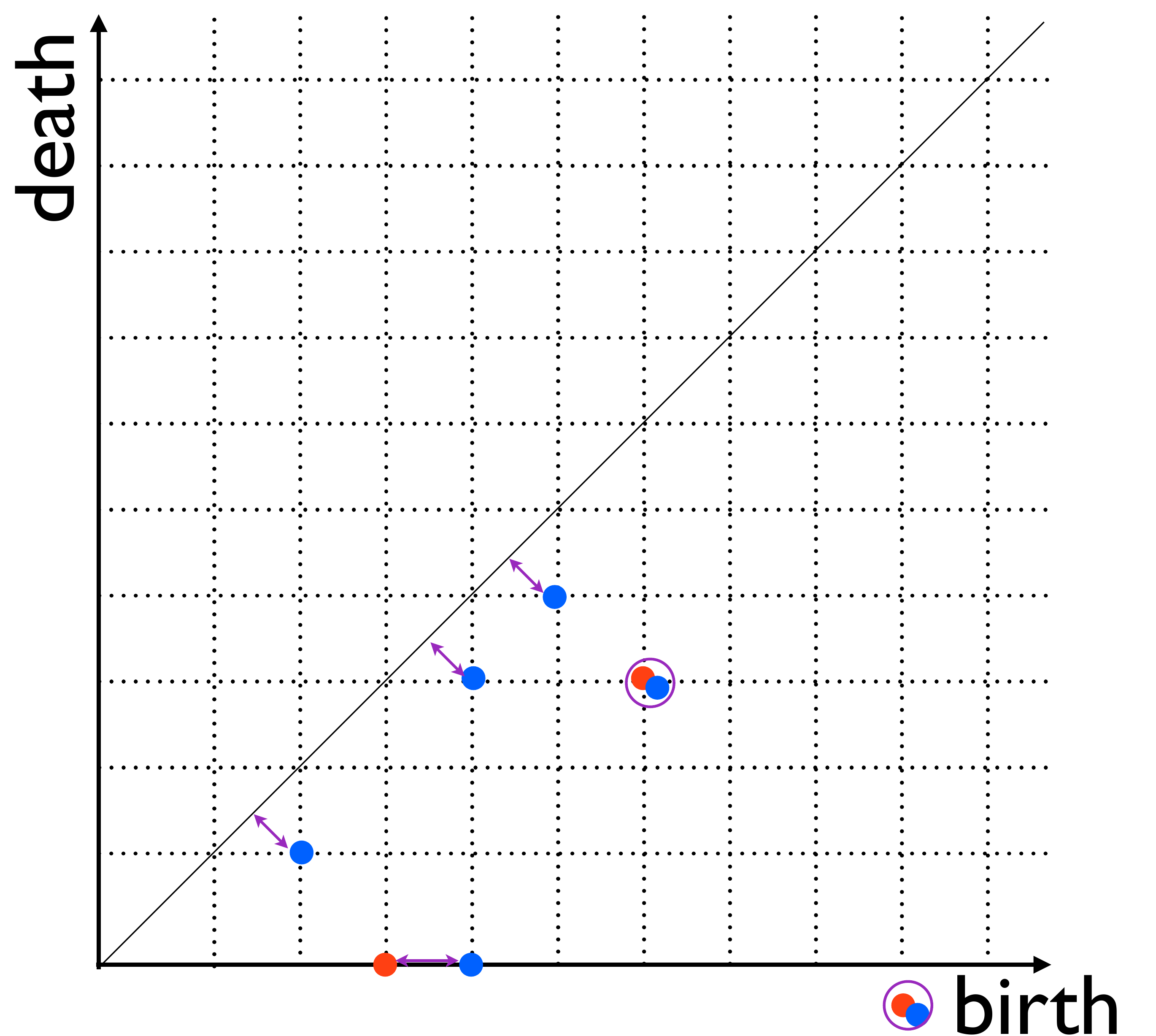}}
	\put(185,-10){(b)}
	\end{picture}
\caption{(a) Two functions. Dashed (blue) represents a noisy perturbation of solid (red). (b) Associated persistence diagrams along
with matching of persistence points satisfying the definition of bottleneck distance. 
}
\label{fig:bottleneck}
\end{figure}

Figure~\ref{fig:bottleneck} shows two functions and their persistence diagrams. The function $g$ is a small perturbation of $f$. 
The ability to match points in persistence diagrams with points on the diagonal, as shown in Figure~\ref{fig:bottleneck}(b),
suggests that small perturbations lead to small  distances between persistence diagrams. In fact,  it is proven in 
\cite{edelsbrunner:harer}  that given a complex $\cn$ and two monotone functions $f,g \colon \cn \to \R$ the bottleneck distance satisfies 
\begin{equation}
\label{eqn::Stability}
d_B(\pd(f),\pd(g)) \leq \sup_{x\in X} | f(x)- g(x)|.
\end{equation}
A similar result holds for the degree-$q$ Wasserstein distance \cite[Section VIII.3]{edelsbrunner:harer}.  A more
formal statement is as follows.

\begin{thm}
\label{thm:continuity}
Given a complex $\cn$ let $M(\cn,\R)$ denote the set of monotone functions on $\cn$ equipped with the sup norm $\|\cdot \|_\infty$.
Then 
\[
\pd\colon M(\cn,[0,\infty)) \to \spd
\] 
defined by (\ref{eq:pd}) is a Lipschitz continuous map.
\end{thm}

\begin{cor}\label{cor:continuity}
The map $\pd\colon M(\cn_I,[0,\infty)) \to \spd$ is Lipschitz continuous.
\end{cor}

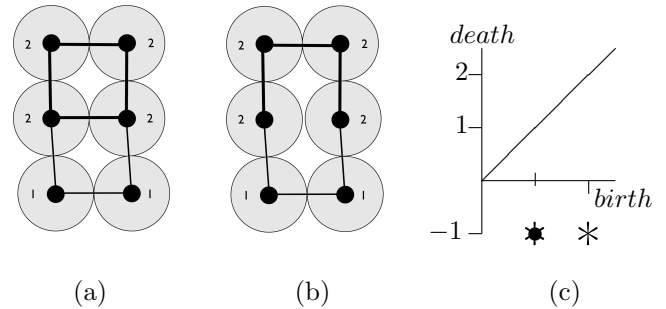
\begin{figure}[t]
\centering
	\begin{picture}(400,110)
	\put(0,20){\includegraphics[width=1.6in]{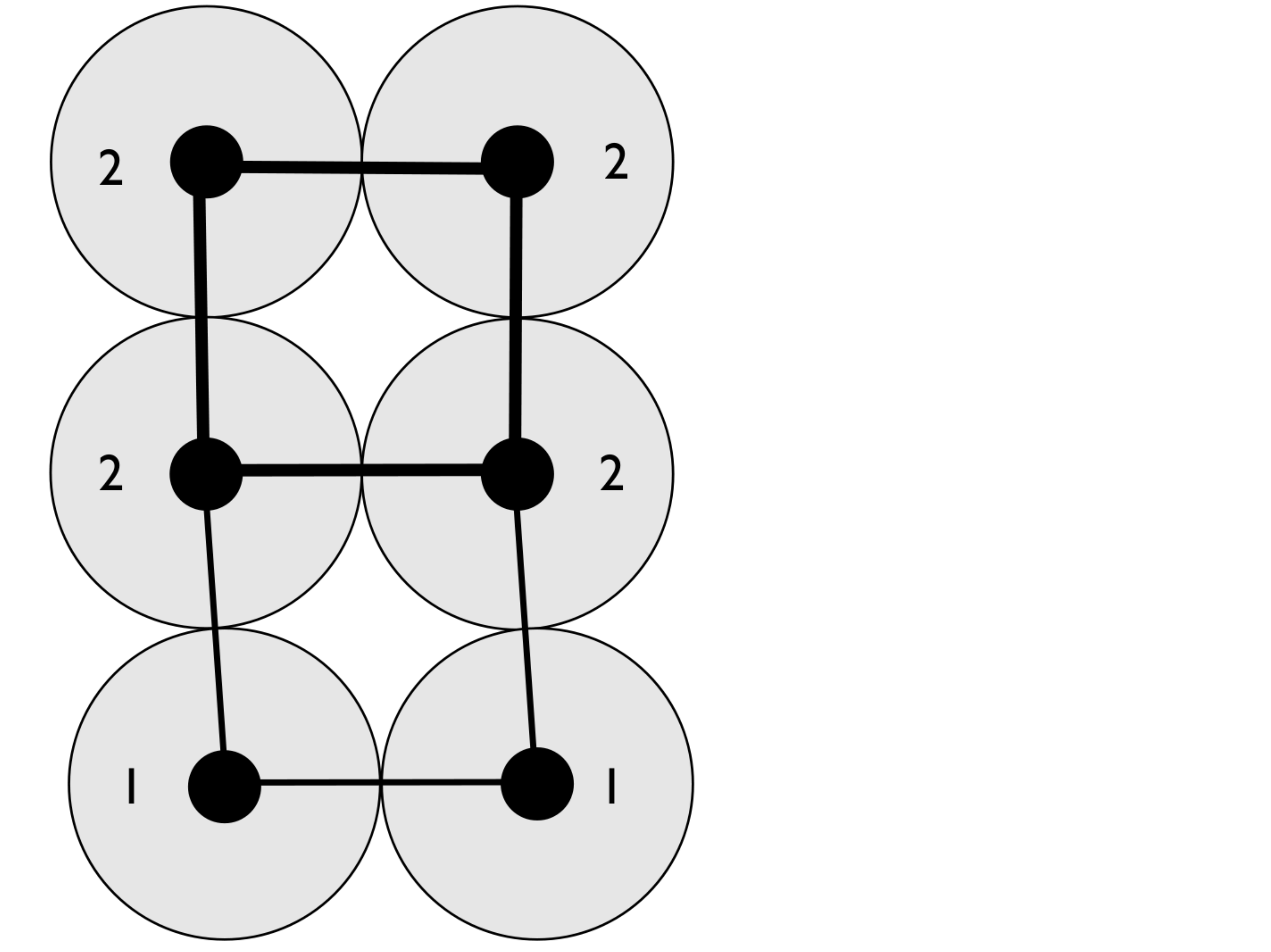}}
	\put(27,-5){(a)}
      	\put(80,20){\includegraphics[width=1.6in]{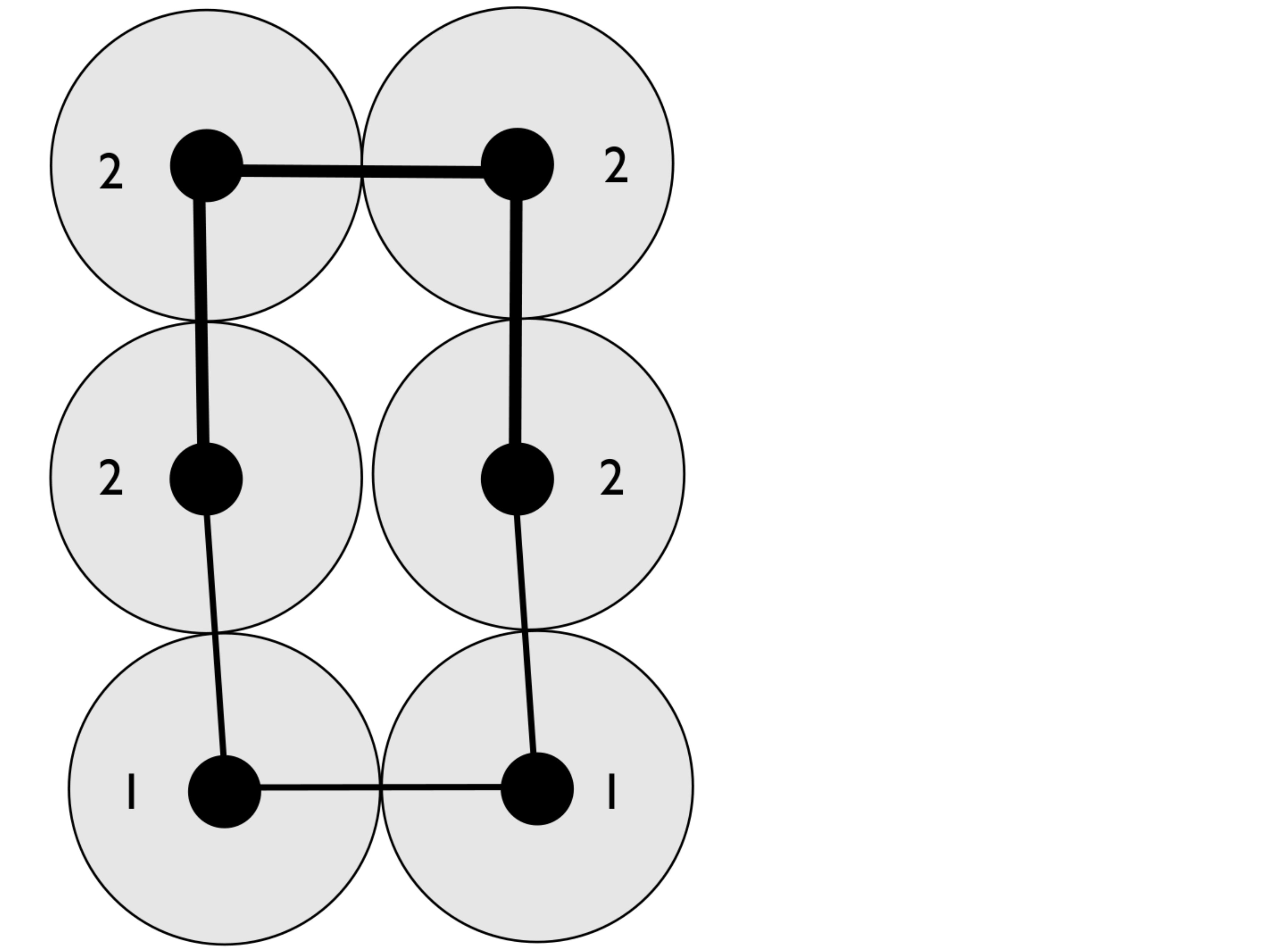}}
	\put(110,-5){(b)}
	\put(205,-5){(c)}
	\put(180,40){\line(1,0){50}}
	\put(180,20){\line(0,1){70}}
	\put(180,40){\line(1,1){50}}
	\put(175,60){\line(1,0){5}}
	\put(170,58){$1$}
	\put(175,80){\line(1,0){5}}
	\put(170,78){$2$}
	\put(175,20){\line(1,0){5}}
	\put(200,20){\circle*{5}}
	\put(195,9){\huge *}
	\put(215,9){\huge *}
	\put(160,18){$-1$}
	\put(222,30){$birth$}
	\put(168,92){$death$}
	\put(200,38){\line(0,1){5}}
	\put(220,35){\line(0,1){5}}

	\end{picture}
\caption{
An arbitrarily small change in positions can lead to an order one change in the bottleneck distance.  Six particles with
the magnitude of the force field indicated. Thick edges have force value $2$, thin edges have force value $1$. (a) Because
of configuration of particles we see two loops. One loop appears at $\theta =2$, the second loop appears at $\theta = 1$. (b) A perturbation of the configuration in (a) but the forces  on the particles do not change.   Only one loop appears at $\theta =1$.
(c) $\beta_1$ persistence diagrams.  The stars at $(1,-1)$ and $(2,-1)$ correspond to the persistence points for (a) and the dot 
at $(1,-1)$ is the single persistence point for (b).}
\label{fig:positionfail}
\end{figure}

Corollary~\ref{cor:continuity} implies that a small change in the forces, either through perturbation of the system or
experimental error, results in a small change in the associated persistence diagrams. This is the long promised stability
result. The failure of $\cn_P^\blacktriangle$ and $\cn_D$ to be stable with respect to perturbations follows from the fact
that small changes of particle positions can result in changes of the underlying complex and thus Theorem~\ref{thm:continuity} is not applicable. 
Figure~\ref{fig:positionfail} demonstrates that it is possible, using  $\cn_P^\blacktriangle$,  for an arbitrarily small change in the position of the particles to lead to an order one change in the bottleneck distance.

%%%%%%%%%%%%%%%%%%%%%%
%
\section{Application of Persistent Homology to the Results of Discrete Element Simulations}
\label{sec:InterpPH}
%
%%%%%%%%%%%%%%%%%%%%%%%

The discussions of the previous sections provide a mathematical framework for studying force networks  associated with
DGM.  In this section we apply these concepts to simulated data. 
We begin with a brief review of the numerical simulations and computational tools employed in Sec.~\ref{sec:simulations}. We then
analyze persistent homology  on a variety of levels. First, in Sec.~\ref{sec:stability}, we consider the stability  of the persistence diagrams
obtained from the digital, position,  or interaction networks with respect to  numerical error.   Then, in Sec.~\ref{sec:function_complex} 
we discuss how the results depend on the choice of complex used, and therefore on the quality of input data.
Sec.~\ref{sec:comparison_diagrams} presents few examples that outline how 
the information contained in  individual persistence diagrams can be related to physically observable properties of the DGM network 
considered.     We note that while the focus of our discussion is on the networks defined using normal force between the particles, 
we also briefly discuss persistence diagrams obtained using tangential forces  (in frictional systems). 
Finally, in Sec.~\ref{sec:distances} we briefly illustrate application of the concept of distance between persistence diagrams
to DGM networks.   

\subsection{Simulations Used}
\label{sec:simulations}

We perform a series of discrete element simulations (DES) similar to our previous works \cite{epl12,pre13}. For the present 
paper, we consider 
a set of about $2,000$ circular particles contained in a square domain with rough walls composed of monodisperse particles. 
The system is slowly compressed allowing for a change of packing fraction, $\rho$, between $0.6$ and $0.9$.  Initially the particles are given
random velocities and are placed on a square lattice.  The equations of motion are integrated using a fourth order predictor-corrector 
scheme. We implement the Cundall-Strack model for static friction which includes normal and tangential forces at the 
contact \cite{cundall79}. For frictionless system, the contact force reduces to a normal force with a spring and viscous damping term. 
In general, we use polydisperse particles where the particle sizes are chosen from a uniform distribution with width 
$r_p=(r_{max}-r_{min})/r_{ave}$, where $r_{ave}$ is the mean particle radius.   The coefficient of restitution measuring energy loss 
is given the value of $e_n = 0.5$, and the
coefficient of static friction is either $\mu = 0.5$ for the frictional case, or $\mu = 0$ for the frictionless one.  
See~\cite{epl12,pre13} for more details.

We focus on a system of particles with $\rho \approx 0.86$, except if specified differently.  This $\rho$ is beyond $\rho_c$, at which jamming transition
occurs. (Note that the $r_p = 0$, $\mu  = 0$ system has the highest $\rho_c$, which is under the implemented protocol, and
for the considered realization at $\rho \approx 0.85$.  More extensive discussion of transition through jamming, as well as averaged (over
initial conditions) values for $\rho_c$, are given in~\cite{pre13}.)   Therefore, the
particles are packed close enough so that most of the particles belong to the same 
connected component of the position network $\cn_P$. 

For the $\rho$'s  of interest, we extract the magnitude $\psi_{i,j}$ of the normal force between any two particles $p_i$ and $p_j$. 
The values $\psi_{i,j}$ completely determine the interaction network $\fn_I$. To construct $\fn_P$ and $\fn_D$ the positions of the particles 
need to be extracted as well.  The value $\psi_i$ assigned to the particle $p_i$ is the total force experienced by this particle, i.e.,
\[
\psi_i := \sum_{\setof{j\mid \ang{i,j}\in\cn^{(1)}_P}} \psi_{i,j}  = \sum_j \psi_{i,j}.
\]
For consistency with our previous works~\cite{epl12,pre13}, 
we normalize  the function $f\colon \cn\to \R$, defined in 
Section~\ref{sec:force}, by dividing it by the average force $\hat{f}$ defined as follows: for the  interaction force network 
\begin{equation}
\label{eq:avefint}
\hat{f}_I =\frac{1}{M}{\sum_{i,j = 1}^N \psi_{i,j}},
\end{equation}
and for the
position and digital force networks
\begin{equation}
\label{eq:avefpos}
\hat{f}_P = \frac{1}{N}\sum_{i = 1}^N \psi_{i} =  \frac{1}{N}\sum_{i,j = 1}^N \psi_{i,j},
\end{equation}
where   $M$ is the number of non zero force interactions, $\psi_{i,j}$,
 and $N$ is the number of particles. Note that the average number of contacts is $Z ={M/N}$ and hence
\begin{equation}
\label{eq:Z}
\hat{f}_P = Z\hat{f}_I.
\end{equation}

We have produced open source software \cite{miro} that is used to encode this procedure and produce a 
force networks filtration $\left\{ \fn(f,\theta_k)\mid \theta_k\in\Theta \right\}$.
The persistent homology of each filtration is computed using the open source software Perseus \cite{mischaikow:nanda:11,perseus}.
We note that the size of the digital complex, $\cn_D$, is 
considerably larger than the size of the position or interaction complexes, 
implying that the computational cost of analyzing $\cn_D$ is much larger as well.
To give a sense of the time needed to perform the types of computations we remark that using a  $2.53$ GHz processor to  
compute the persistence diagrams for the position, digital, and interaction  force network required $25$, $97$, and $43$ seconds, respectively.
The worse case complexity of computing the bottleneck distance between the persistence diagrams 
$\pd_1$ and $\pd_2$ is $O((n_1+n_2)^2)$ where $n_i$  is the number of  generators in $\pd_i$.  
For the Wasserstein distance the complexity is even higher  $O((n_1+n_2)^3)$. 
Therefore the time required for computing the distances strongly depends on the number of generators 
in the persistence diagrams. In practice the number of generators is rather small for $\cn_I$ and $\cn_P$. 
In our case there are typically  a few hundred  persistence generators. Thus both distances can be 
computed in a couple of seconds.  However the digital networks contain a large amount of artificial 
loops shown in Figure~\ref{fig:digitalFailure}(a) and Figure~\ref{fig:digitalFailure}(c) (this number typically 
increases with the resolution) and the runtimes are much longer. For the resolution $1000\times1000$ runtime 
required to compute the bottleneck distance is $10$ minutes and for the resolution $2000\times2000$ it is $30$ minutes. 
Finally we needed three weeks to compute the Wasserstein distance between two persistence diagrams for digital 
networks with the resolution $1000\times 1000$. We stopped computation of this distance for the digital network with 
higher resolution after it had not  terminate within three  weeks.

\subsection{Stability of Persistence Diagrams}
\label{sec:stability}

\begin{figure*}[t]
\centering
	\begin{picture}(400,165)
      	\put(-50,5){\includegraphics[width=2.15in]{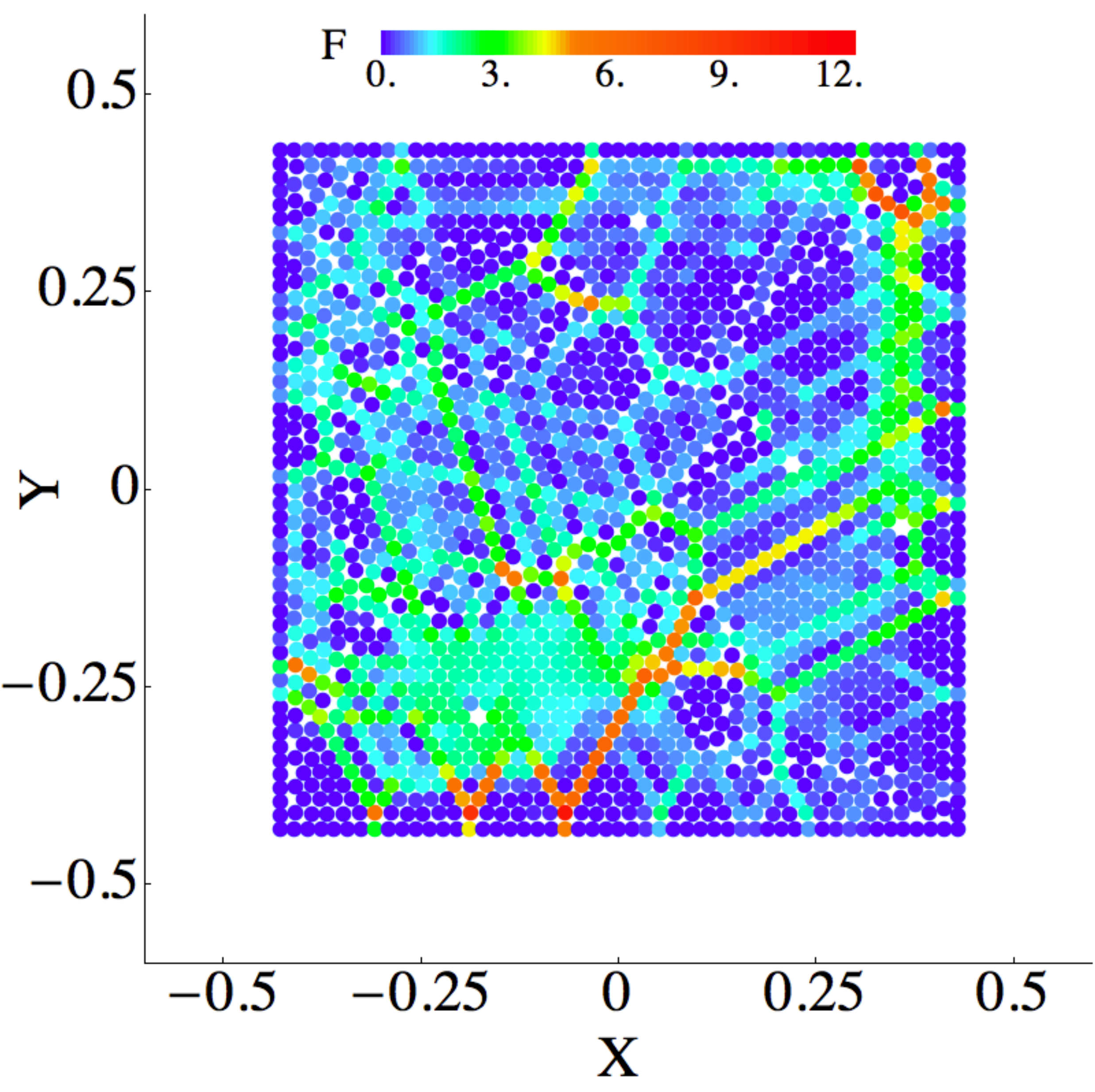}}
	\put(120,5){\includegraphics[width=2.15in]{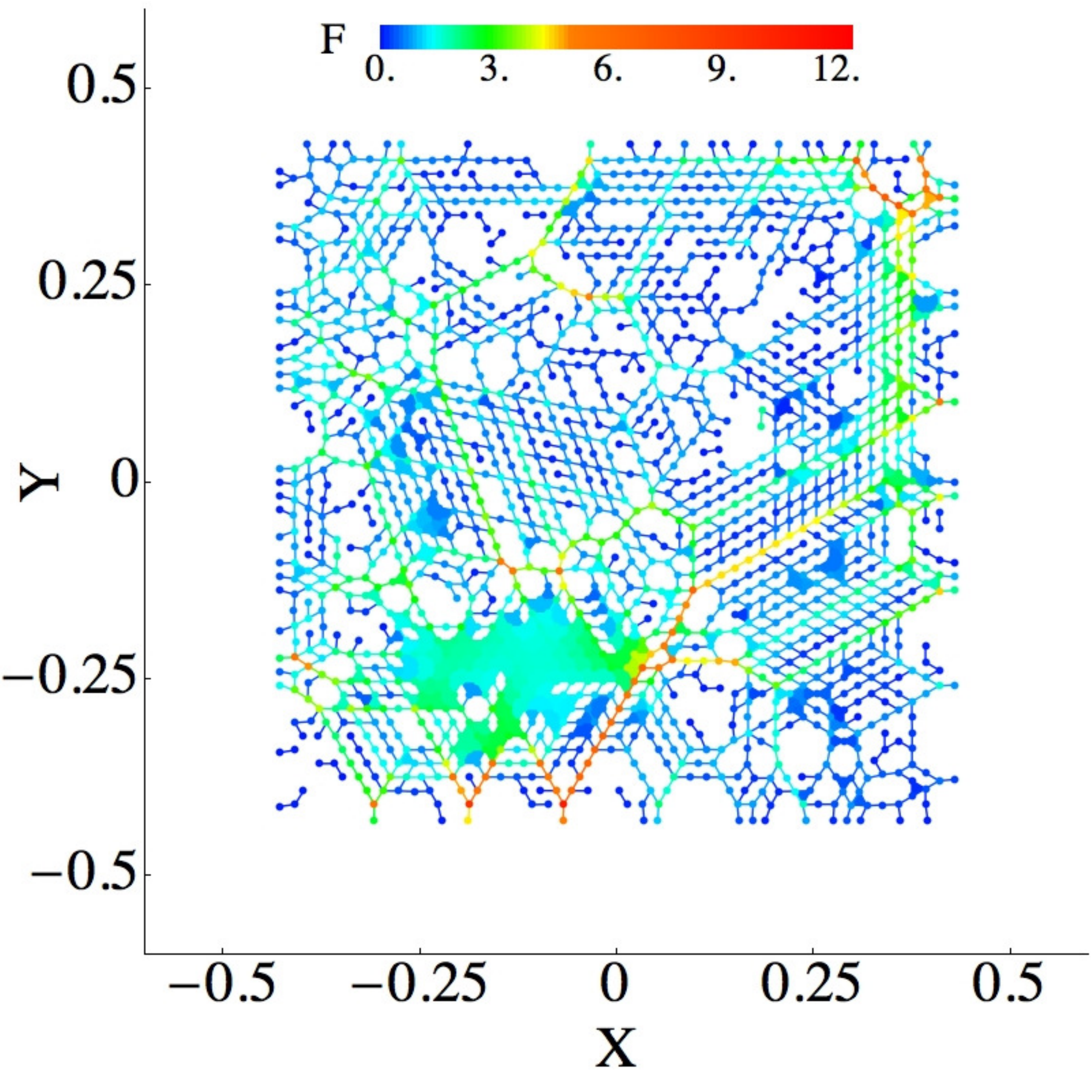}}
	\put(300,5){\includegraphics[width=2.15in]{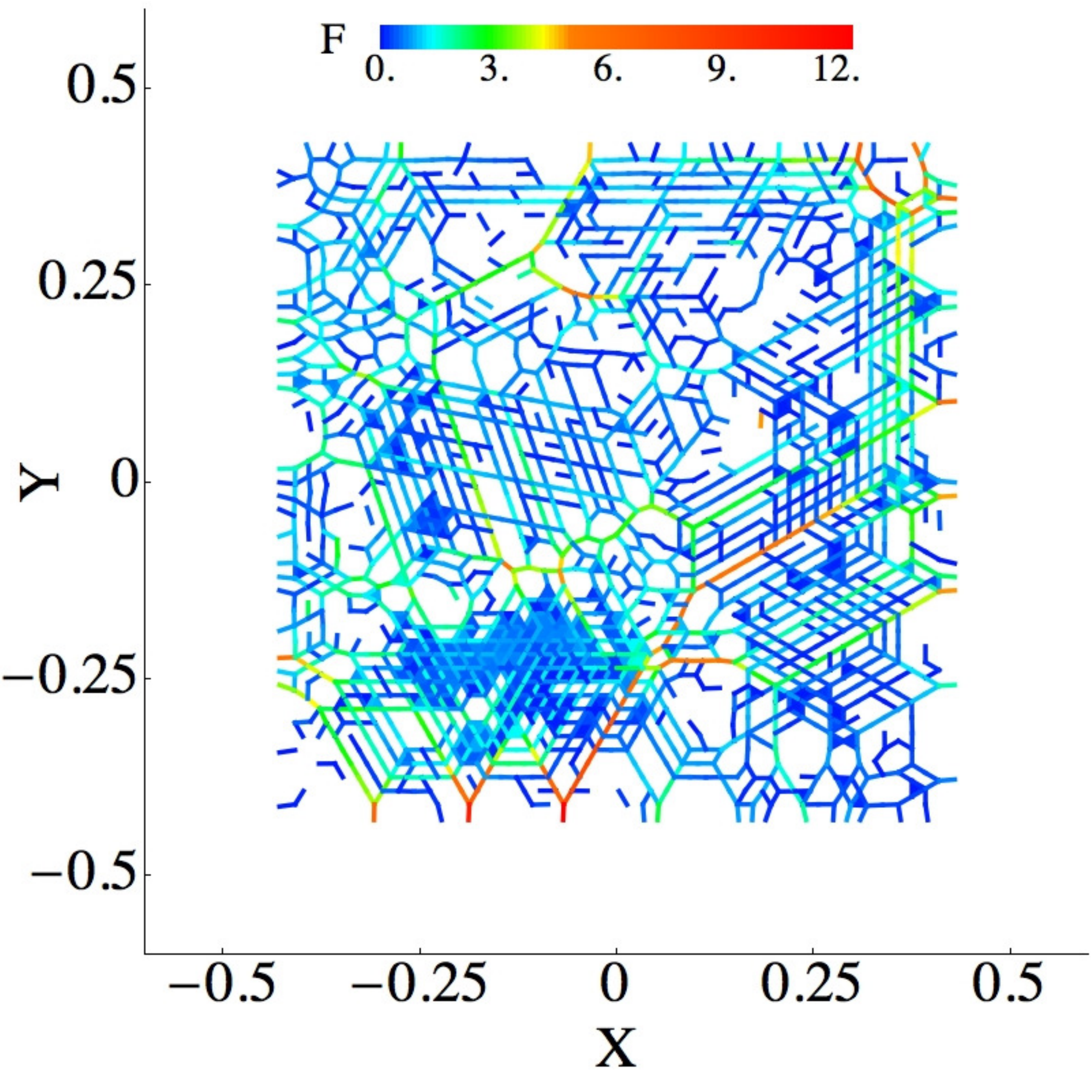}}
	
	\put(-50,0){(a)}
	\put(110,0){(b)}
	\put(300,0){(c)}
	\end{picture}
\caption{
Different force networks  for mono disperse, $r_p = 0$, frictionless, $\mu =0$, system at the 
packing fraction, $\rho = 0.86$.  
(a) digital force network (b) position force network (c) non zero simplices of the interaction force network.}
\label{fig:r00mu00_networks}
\end{figure*}
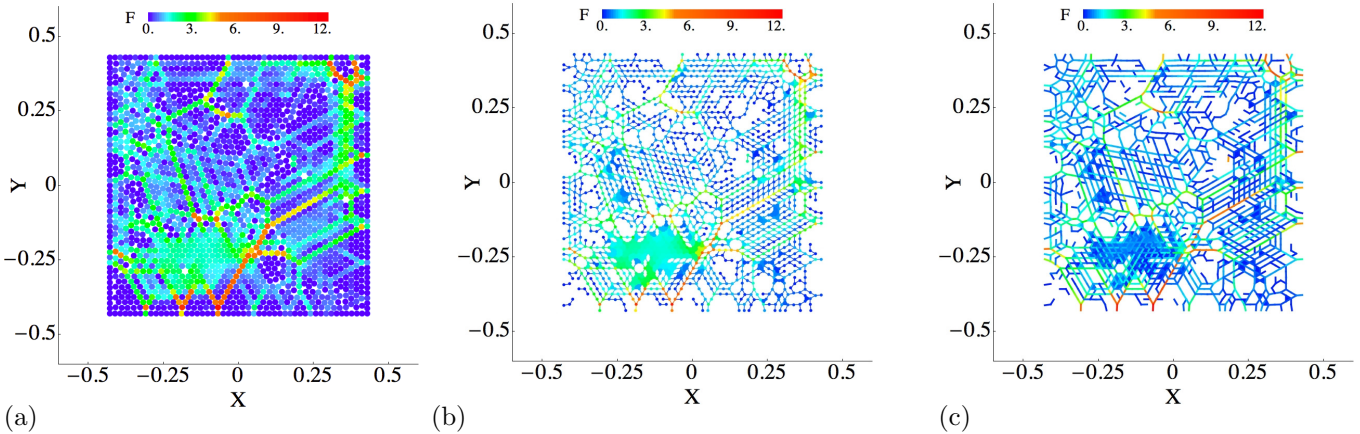

\begin{table*}
\centering 
\begin{tabular}{c c | c c c c }
& & \multicolumn{3}{c} {Network Type}\\
\hline
metric & truncation  & interaction  & position  & digital (1000) & digital (2000)  \\
\hline
$d_B$ & 3  & 0.0004 & 2.6151 & 1.2946 & 2.3339  \\ 
	& 2  & 0.0035 & 2.6387 &  3.4672 &  3.4672  \\ 
\hline
$d_{W^1}$ & 3  & 0.1806 & 439.01 & 659.4 & - \\
	& 2   & 1.8447 & 689.79 &  2226.1 & - \\
\end{tabular}
\caption{The distance between the persistence diagram of original networks shown in Figure~\ref{fig:r00mu00_PD},  and the persistence diagram after truncation of numerical
data to 2 or 3 significant digits. The bottleneck distance $d_B$ measures the single largest difference between the persistence diagrams while the Wasserstein $d_{W^1}$ distance is a sum of all differences between the diagrams (see Definition~\ref{defn:bottleneck}). Computation of $d_{W^1}$ for the digital (2000) network did not terminate within three weeks using $2.53$ GHz processor.}
\label{table::Perturbations}
\end{table*}

Figure~\ref{fig:r00mu00_networks}  shows the three networks for $r_p = 0$, $\mu = 0$ system.
The associated persistence diagrams are depicted in Figure~\ref{fig:r00mu00_PD}. We will
discuss some features of these networks and diagrams in what follows; to start with, we ask
the following question:
how stable is the information contained in the persistence diagrams  with respect to an error in input data? 

The numerical simulations and the extraction of particle positions and normal  forces are done using double 
precision floating point numerics.  We then compute $d_B$ and $d_{W^1}$ distances between the original and perturbed persistence diagrams.  
The results are given in Table~\ref{table::Perturbations}. 
The relatively small values associated with the interaction network are predicted by Corollary~\ref{cor:continuity}. 
The fact that the values for $d_{W^1}$ are significantly larger than  $d_B$ for each type of network is not surprising 
since $d_B$ distance measures the single largest change in the network while $d_{W^1}$ is 
sensitive to all local perturbations that may be occurring.

Behavior of the position network with respect to small perturbations lacks the stability of the interaction network. 
A measurement error at the third decimal place radically changes the network, with the $d_B$ distance three orders of
magnitude larger than the error introduced.  This large difference is caused by 
 the phenomena shown in Figure~\ref{fig:positionfail}. To show this, note that 
$d_{W^1}$ is several orders of magnitude larger than $d_B$, implying that there must be
 many locations in the network at which the local maxima and minima of the forces change due to introduced
 error.

We have no theoretical results that explain the relative differences in perturbations of distances between the persistence diagrams
associated with the position and digital force networks.  The digital force network was constructed using  resolutions of $1000\times 1000$
and $2000\times 2000$ pixels. Given the size of the domain, each pixel in $1000\times 1000$ case  represents a measurement to 
approximately three significant figures. We hypothesize that this explains the relatively small  (as compared with the position network) 
$d_B$ distance.  The $d_{W^1}$ distance for truncated data is larger for the digital (1000) force network than for the position network. This seems to be connected to a significant drop in the number of $\beta_1$ generators corresponding to artificial loops (see Figure~\ref{fig:digitalFailure}) in the perturbed digital (1000) force network. The original network contains $6344$ $\beta_1$ generators while the network obtained by truncation to three significant digits  only $4487$. In contrast,
the  difference between the  number of generators for the position networks is $340$. Thus the difference is much smaller and so is the $d_{W^1}$ distance.

The sensitivity of digital complexes to small perturbations demonstrated in Figure~\ref{fig:digitalFailure} suggests that the larger distance value for the 
$2000\times 2000$ digital complex with truncation at three significant digits should not come as a surprise. 
The different distance values for different digital complexes raises another issue; how sensitive is  the persistence diagram to the 
resolution of the digital network? We consider this issue using the system shown in   Figure~\ref{fig:r00mu00_networks}. 
Computing with the original numerical data at a resolution of  $2000\times2000$ pixels, we find that  the $\beta_0$ persistence 
diagrams for the digital networks are almost identical.   Comparison of the $\beta_1$ persistence  diagrams reveals  that the number of loops is 
around $30\%$ larger for the higher resolution. 
We have verified that this increase is caused by formation of extra loops at the places where the particles are close to 
each other; essentially the phenomenon indicated in Figure~\ref{fig:digitalFailure}(a).

\subsection{Force Networks as a Function of  Complex Type}
\label{sec:function_complex}

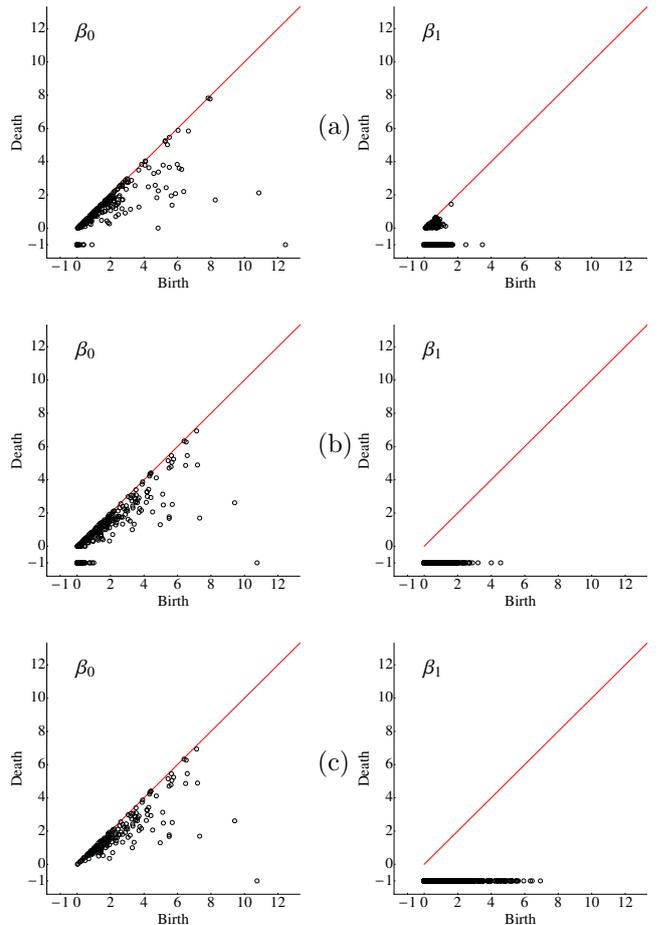
\begin{figure}[t]
\centering
	\begin{picture}(220,355)
	\put(-15,0){\includegraphics[width=1.5in]{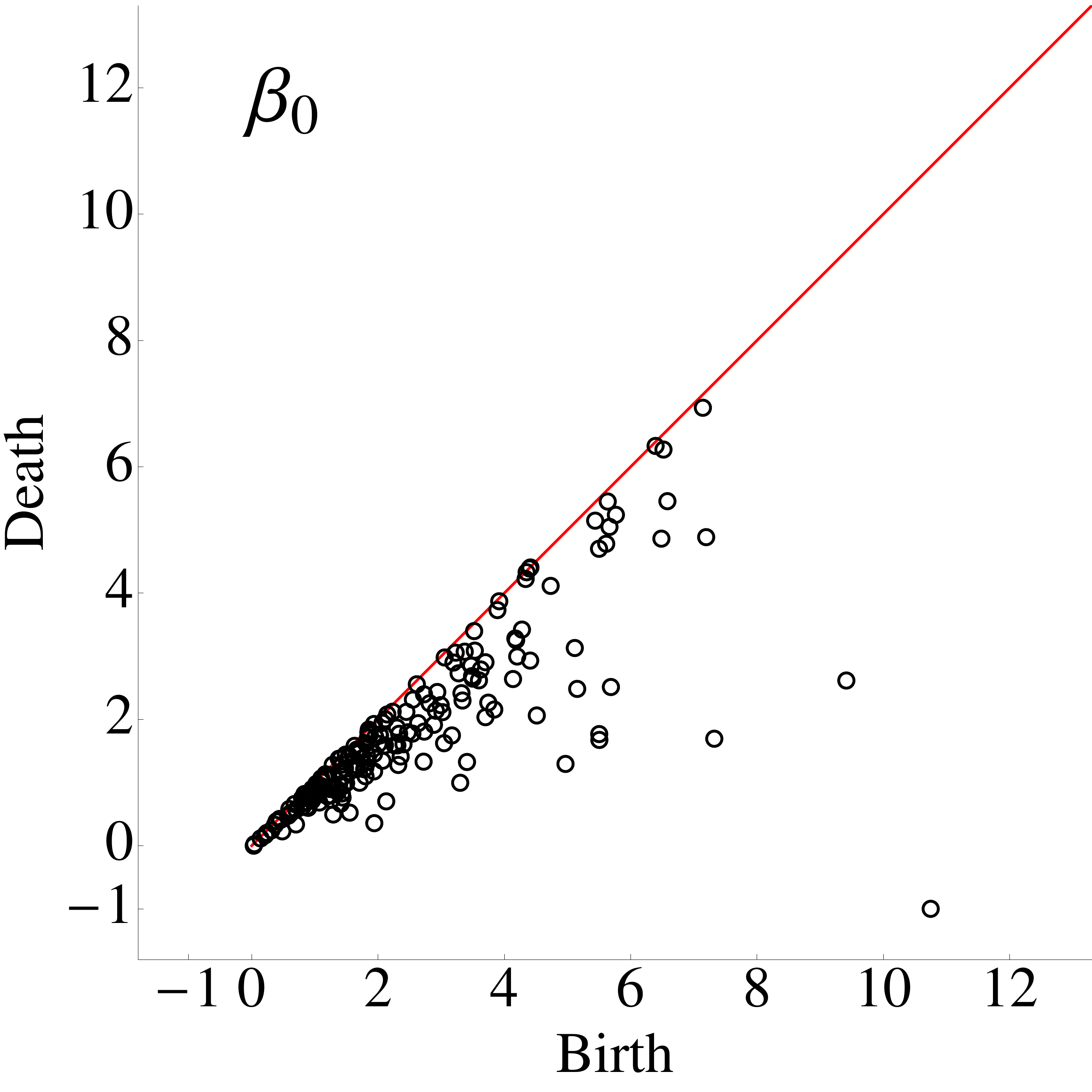}}
	\put(115,0){\includegraphics[width=1.5in]{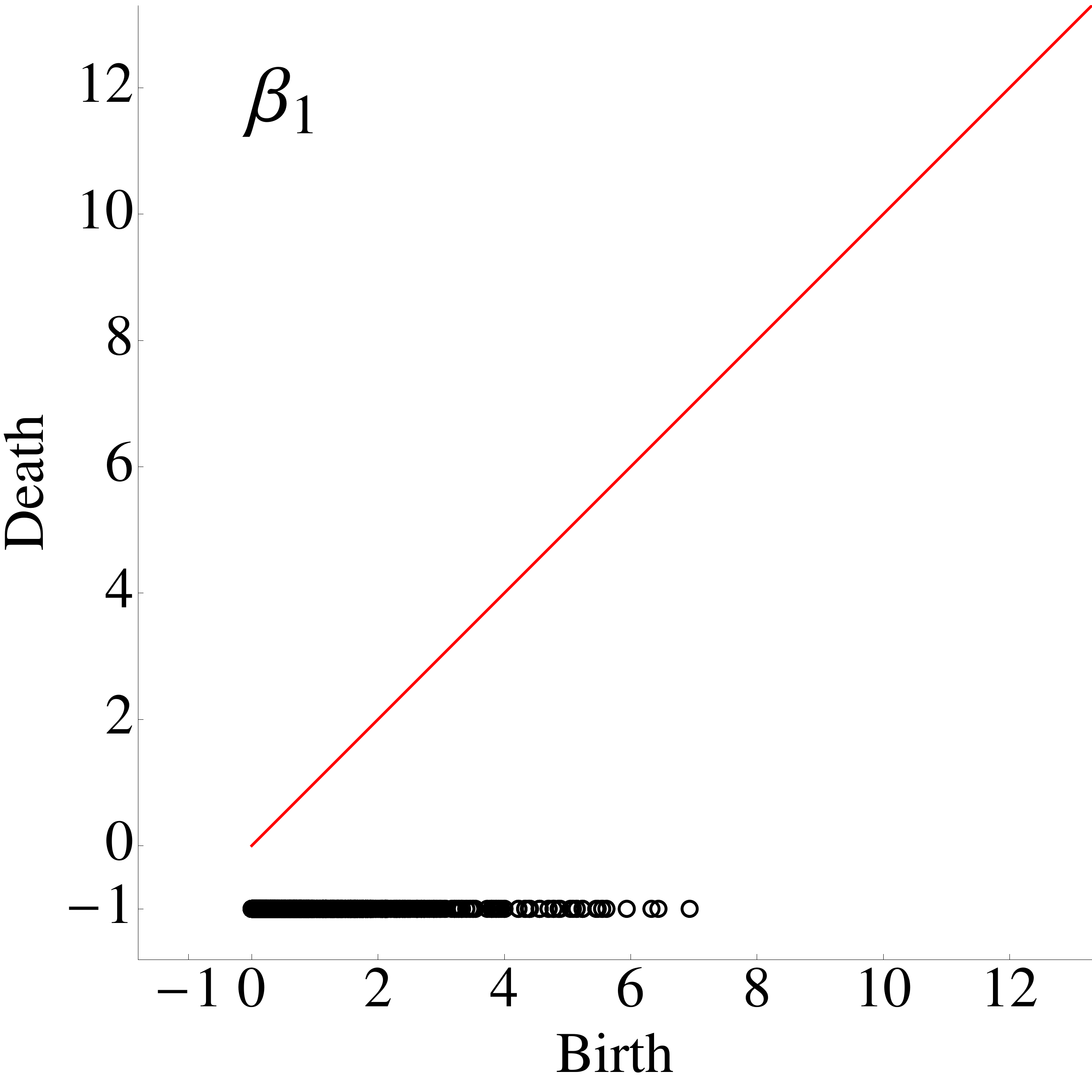}}
      
      	\put(-15,120){\includegraphics[width=1.5in]{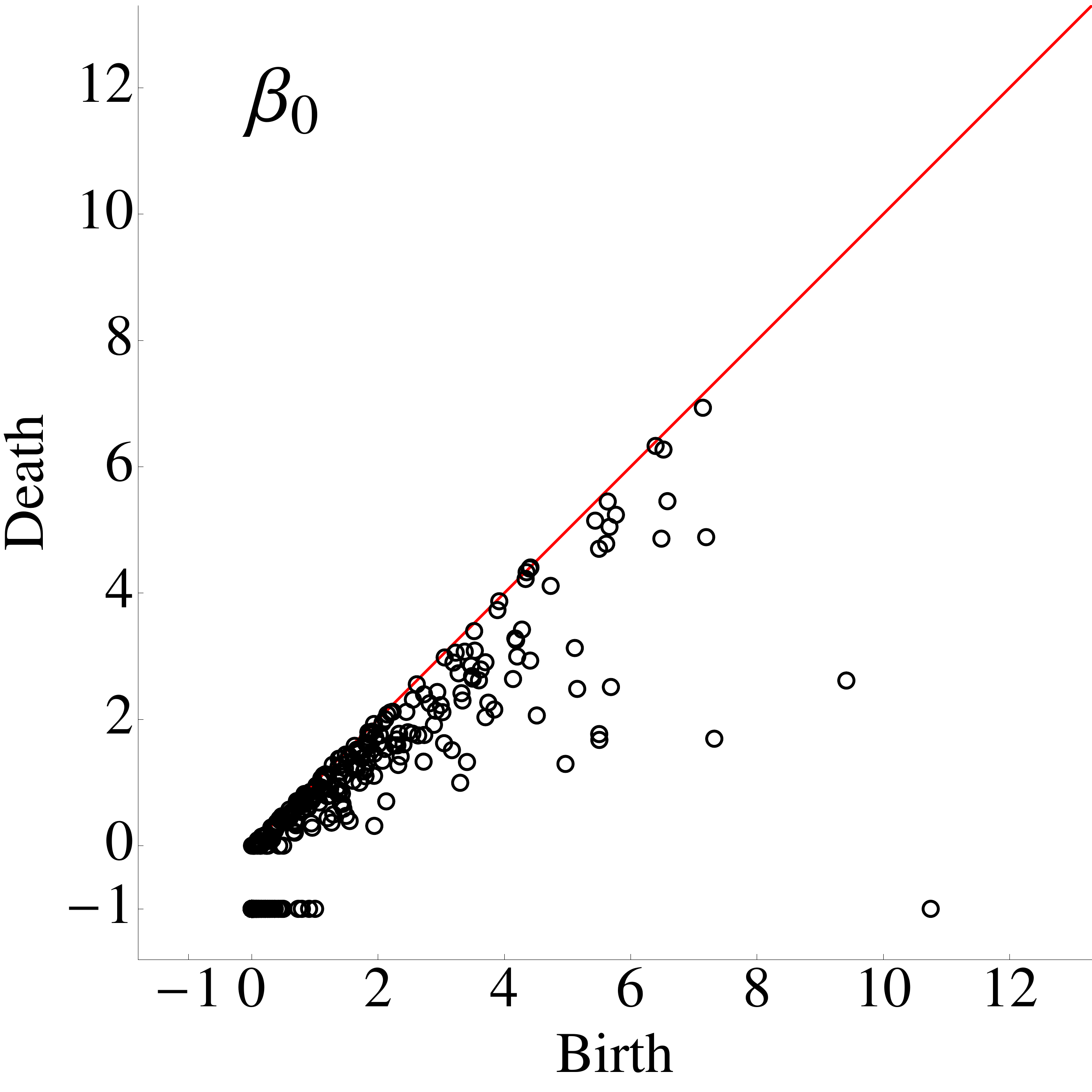}}
	\put(115,120){\includegraphics[width=1.5in]{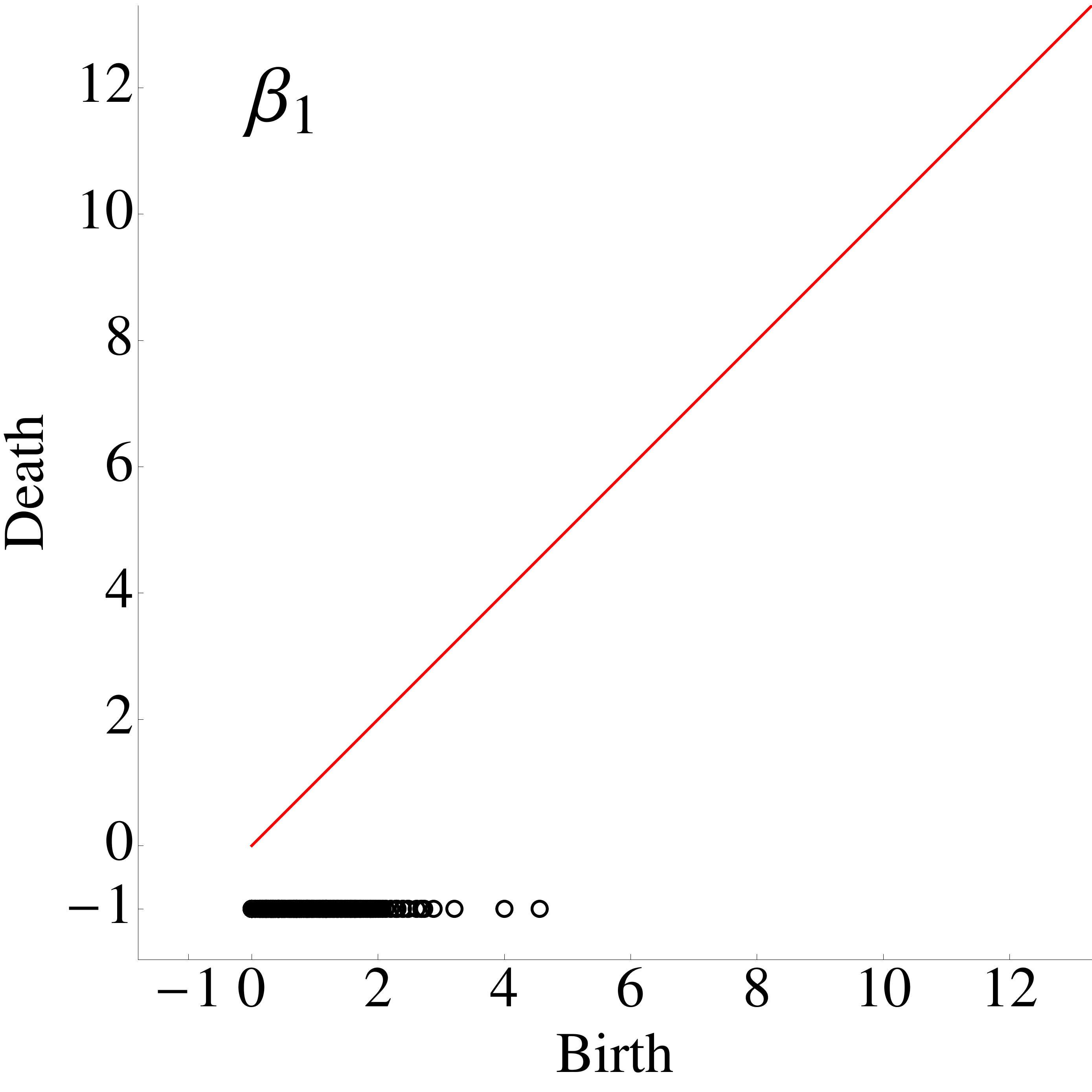}}

	 \put(-15,240){\includegraphics[width=1.5in]{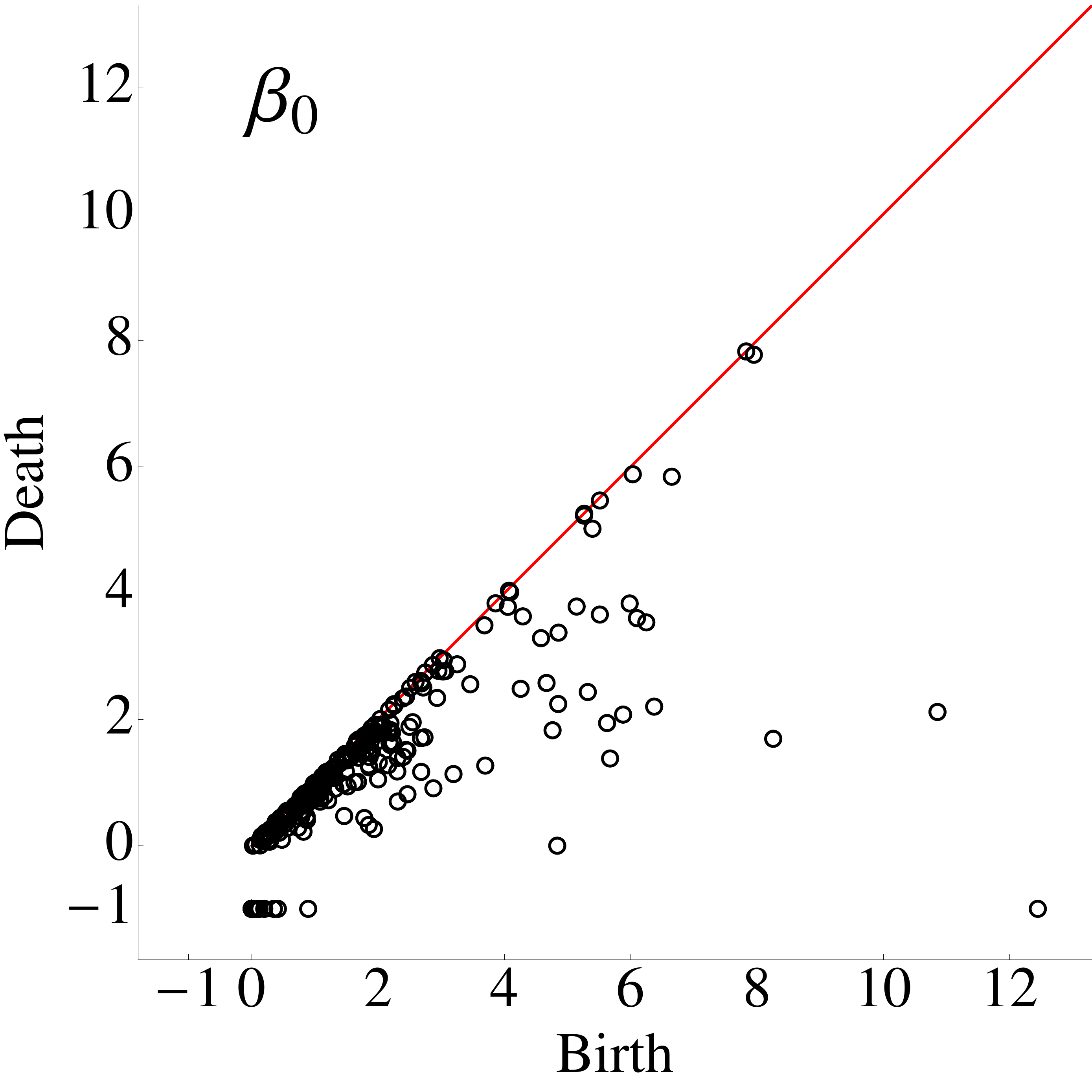}}
	\put(115,240){\includegraphics[width=1.5in]{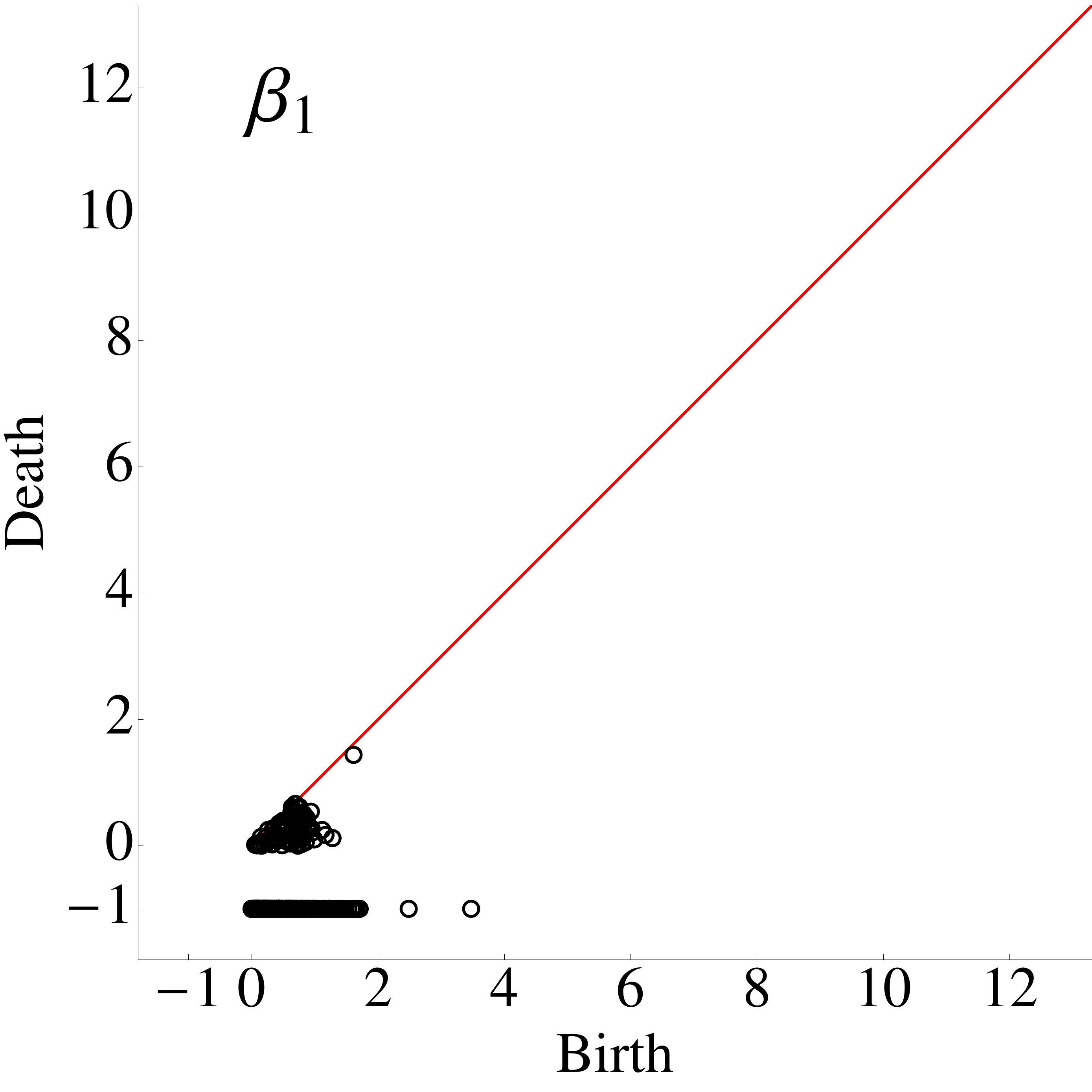}}

	\put(100,60){(c)}
	\put(100,180){(b)}
	\put(100,300){(a)}

	\end{picture}
\caption{
Persistence diagrams for $r_p = 0,~\mu = 0$ system at $\rho = 0.86$ shown in Figure~\ref{fig:r00mu00_networks}.  
Persistence diagrams for (a) digital force network  based on $1000\times 1000$ pixels, (b) position force network, and (c) non zero 
simplicies of the interaction force network.}
\label{fig:r00mu00_PD}
\end{figure}

\begin{figure}[t]
\centering
	\begin{picture}(400,275)
	\put(8,200){\includegraphics[width=1.5in]{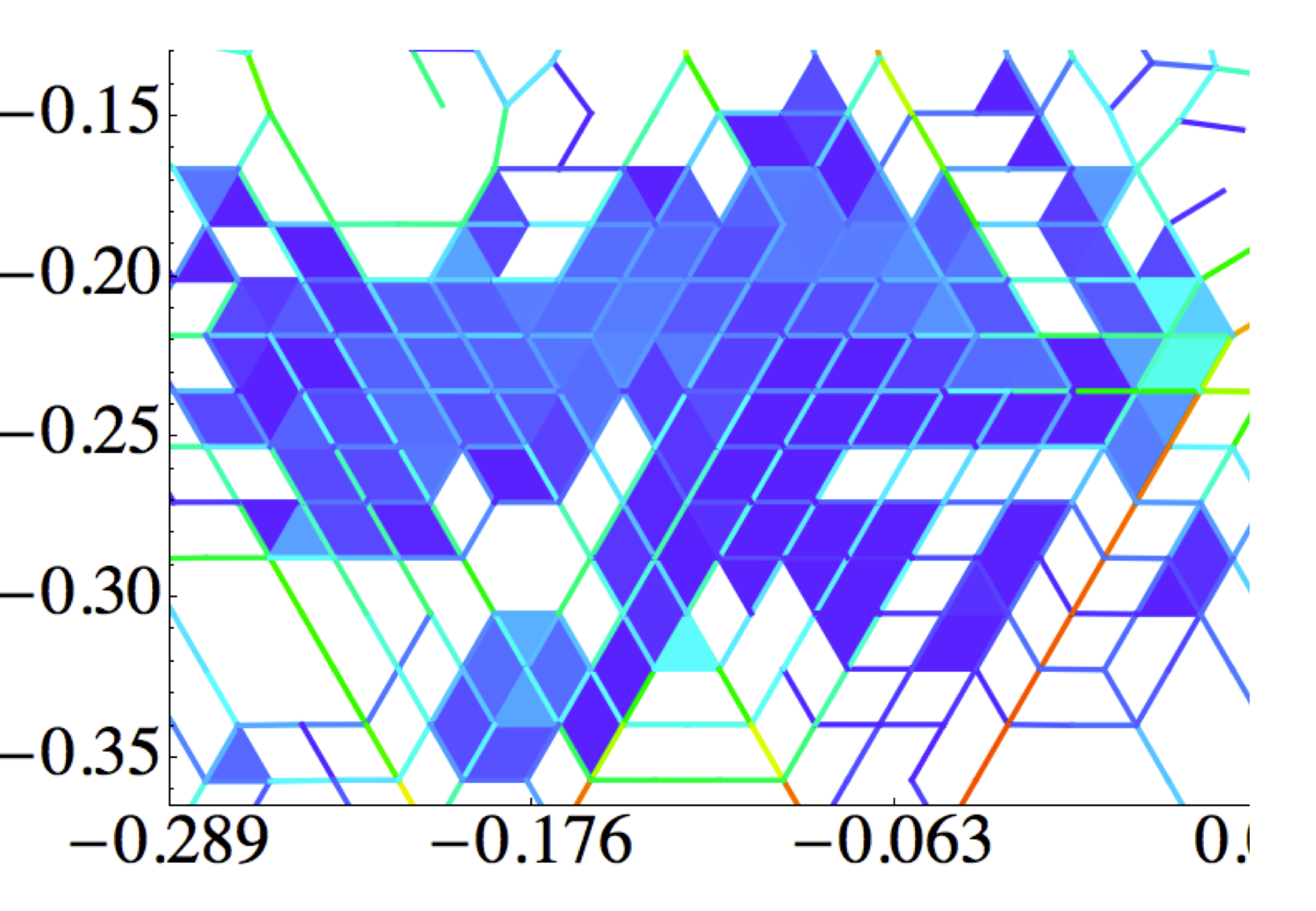}}
      	\put(133,200){\includegraphics[width=1.5in]{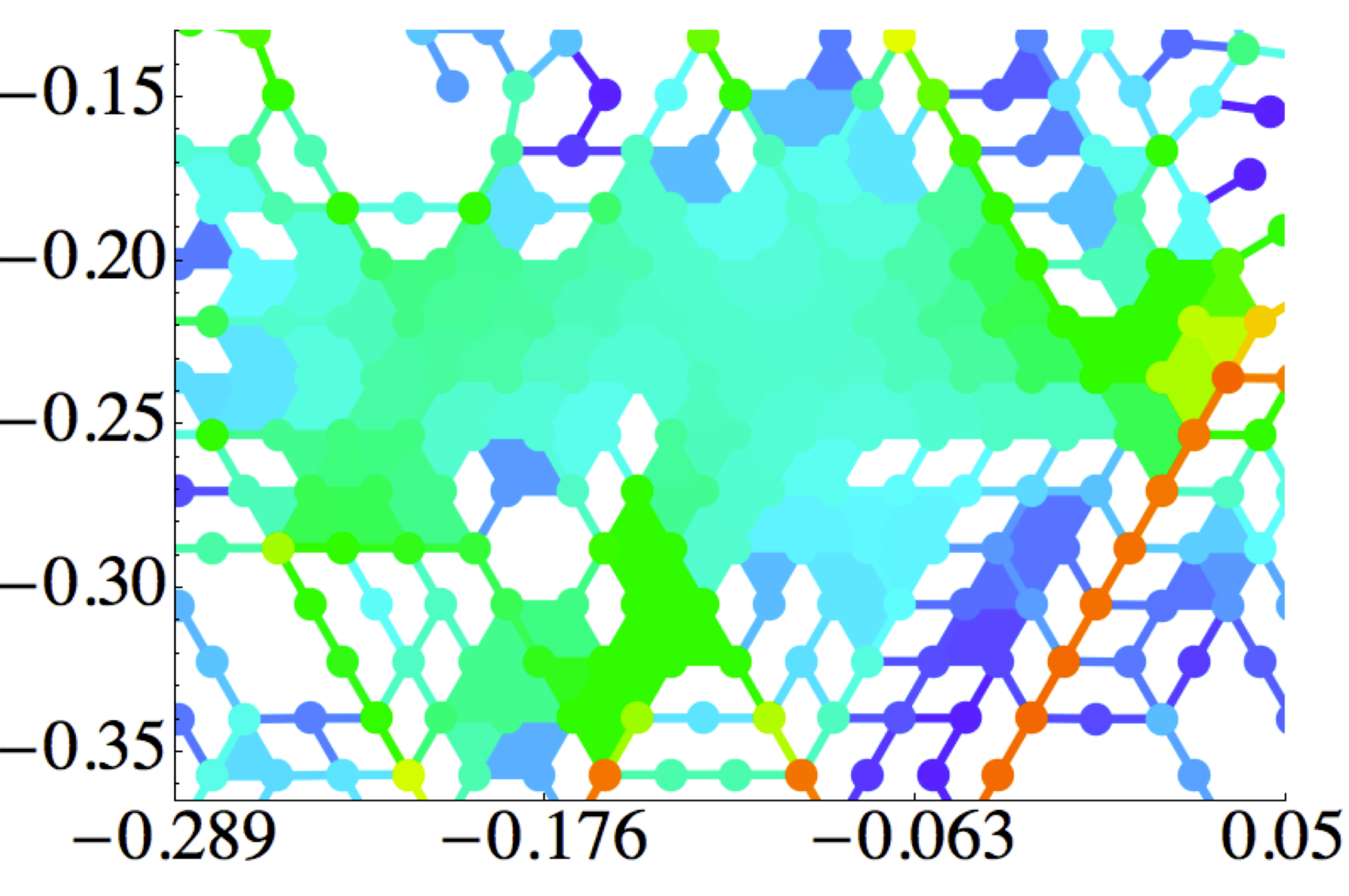}}
	
	\put(8,70){\includegraphics[width=1.5in]{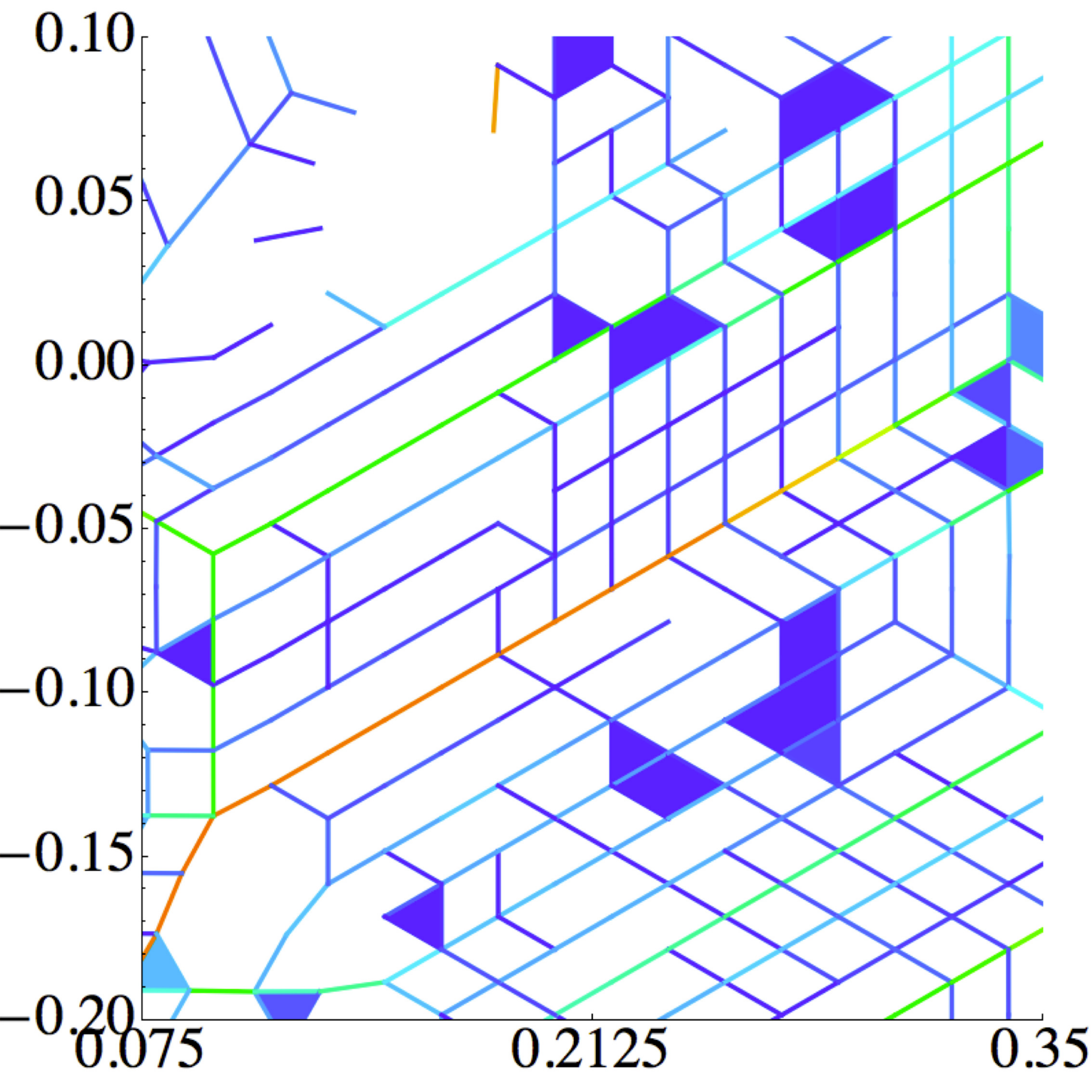}}
      	\put(133,70){\includegraphics[width=1.5in]{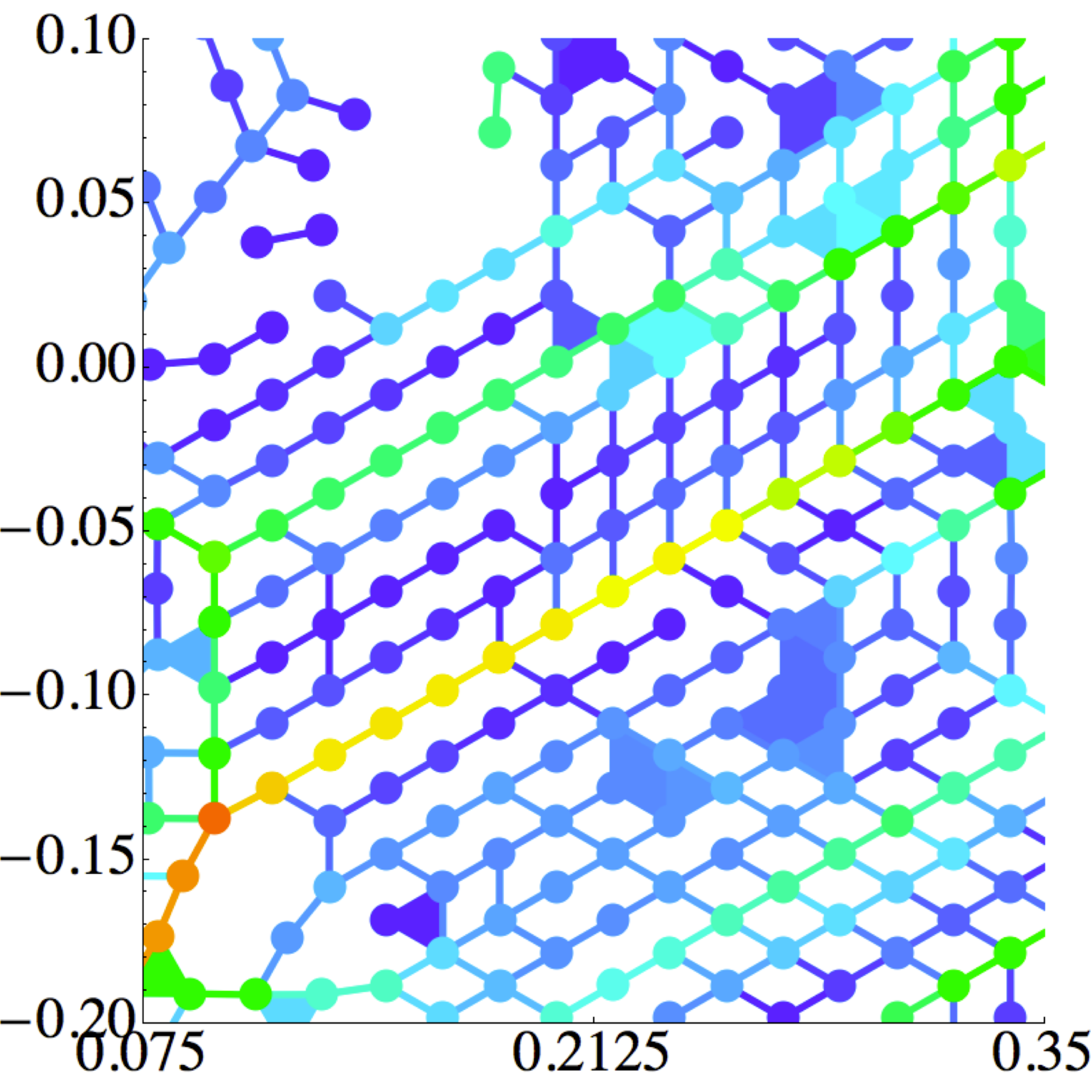}}
	\put(10,-5){\includegraphics[width=1.5in]{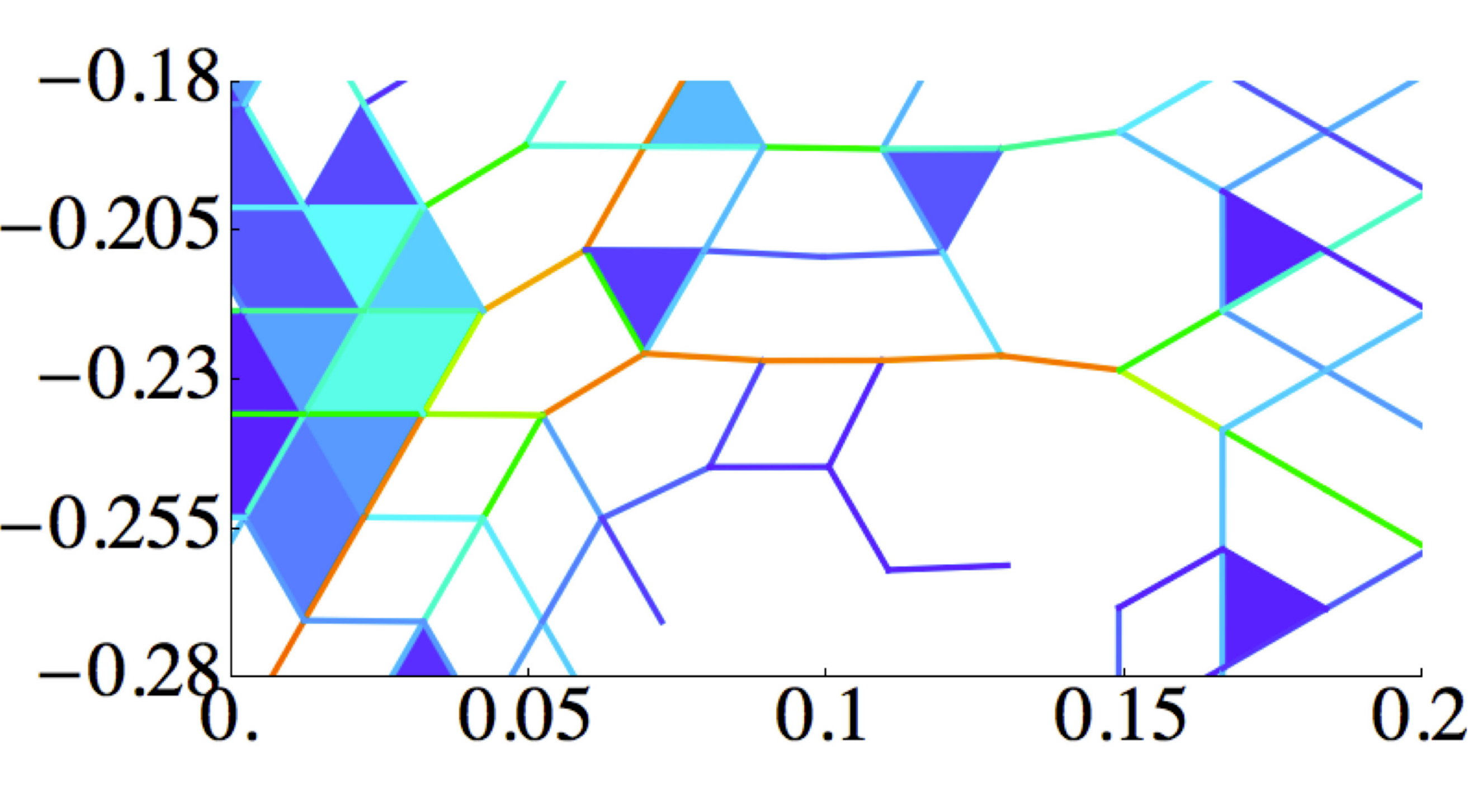}}
     	\put(135, -5){\includegraphics[width=1.5in]{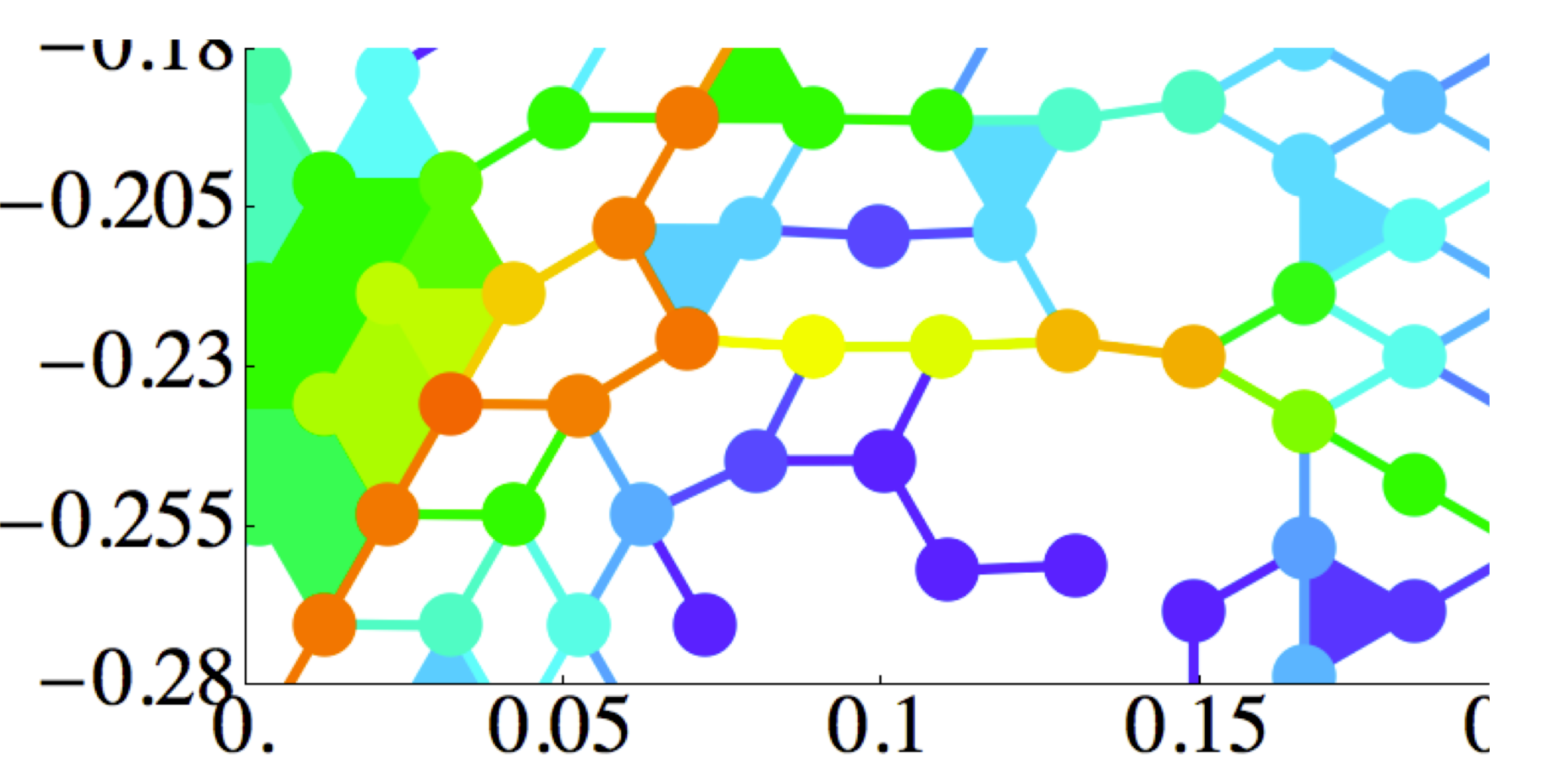}}
	
	\put(60,185){(a)}
	\put(185,185){(b)}
	\put(60,57){(c)}
	\put(185,57){(d)}
	\put(60,-15){(e)}
	\put(185,-15){(f)}

	\end{picture}
\caption{
Enlarged views, using the same color scheme, of three different subregions of the interaction (a), (c), (e)  and position (b), (d), (f)
force networks of Figures~\ref{fig:r00mu00_networks}(b) 
 and \ref{fig:r00mu00_networks}(c). }
\label{fig:NetworkDifferences}
\end{figure}
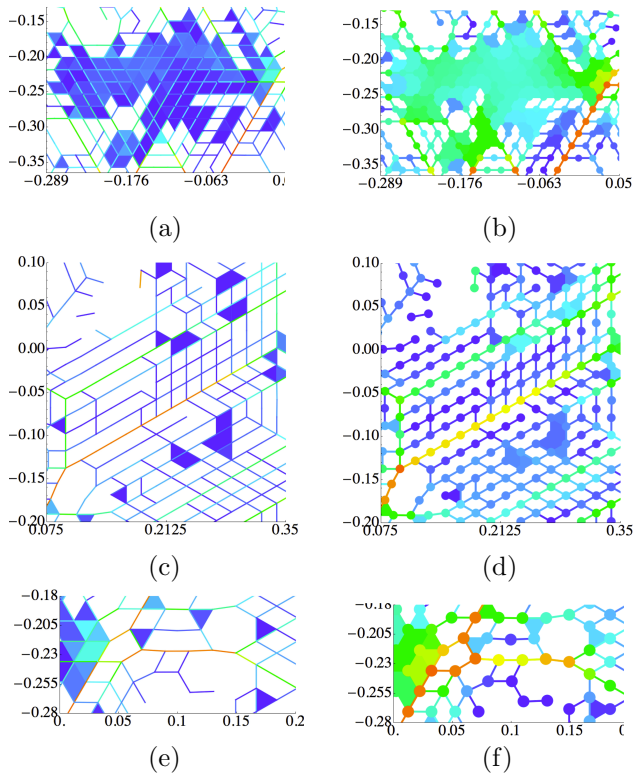

Figures~\ref{fig:r00mu00_networks}  and \ref{fig:r00mu00_PD} demonstrate that the digital, position and interaction force networks of a single system of particles can be quite different.  The idea behind the construction of the digital and  position  force networks is the same, the difference arises from the fact they are based on different complexes that provide different approximations of the geometry of the system of particles.  Thus, to focus on the essential differences we mostly restrict our discussion to a comparison of the position and interaction force networks. 
Figure~\ref{fig:NetworkDifferences}  provides an enlarged view  for three different subregions of the position and interaction force networks of Figures~\ref{fig:r00mu00_networks}(b)   and \ref{fig:r00mu00_networks}(c).   The position force network is defined in terms of the vertices and thus the corresponding figures include the magnitude of the force there.  The vertices are not highlighted in the interaction force network since the value of the force on the edges  is used to define the values on the vertices.
 
Figures~\ref{fig:NetworkDifferences}(a) and (b)  are typical of a region in which we see crystalline structure or equivalently a region over  which there are no defects.   Observe that in this crystalline region  the normal forces for the position force network are significantly  larger than those of the interaction force network. 
This
has to do with large number of contacts ($6$),  so the sum of the forces on  each particle  is high. 
Note that the forces are rather uniform in the crystalline zone and $\psi_i \approx  6\max_j\setof{\psi_{i,j}}$.  Let $f_I$ and $f_P$ denote the forces in the interaction and particle force networks, respectively.  Then 
 $$
 f_I(i) = \frac{\max_j\setof{\psi_{i,j}}}{\hat{f}_I}  = \frac{Z\max_j\setof{\psi_{i,j}}}{\hat{f}_P} \approx \frac{Z\psi_i}{6\hat{f}_P} = \frac{Z}{6}f_P(i).
 $$
 Except for the perfect crystal the value of  $Z$ is less than $6$. For the network shown in Figure~\ref{fig:r00mu00_networks} we computed that  $Z \approx 3$. Therefore  $f_I(i) \approx \frac{1}{2}f_P(i)$.

We now consider a part of the domain where we find sets of particles interacting by large forces, resembling a 'force chain.'
 In this case, as can be seen along the orange chain in Figure~\ref{fig:NetworkDifferences}(d), the position force network tends to report a lower magnitude of force than the interaction force network, compare with the red chain in Figure~\ref{fig:NetworkDifferences}(c).    Observe that along the red chain of particles in Figure~\ref{fig:NetworkDifferences}(c)  each particle typically has contact with 2 or 3 other particles. Therefore $\psi_{i} \leq Z \max_j\setof{\psi_{i,j}}$.
By  equation~\eqref{eq:Z}  and  the inequalities stated in this paragraph we obtain 
$$
f_I(i) = \frac{\max_j\setof{\psi_{i,j}}}{\hat{f}_I}  = \frac{Z\max_j\setof{\psi_{i,j}}}{\hat{f}_P} \geq   \frac{\psi_i}{\hat{f}_P} = f_P(i).\\
$$
An added effect is that a single  continuous  chain of strong force interactions in the interaction force network is reported to be a collection of shorter chains in the position force network (see
Figures~\ref{fig:NetworkDifferences}(e) and (f)).  
An immediate consequence  is that we expect to see more points
with relatively large birth values in the $\beta_0$ persistence diagram of  the position force network than in the 
$\beta_0$ persistence diagram of  the interaction force network. 
This is confirmed by counting the number of points in
the $\beta_0$ persistence diagrams of Figures~\ref{fig:r00mu00_PD}(b) and (c) 
with birth value greater than a given value.

Figures~\ref{fig:NetworkDifferences}(e) and (f) demonstrate another important difference between the position and interaction
force networks.  In Figure~\ref{fig:NetworkDifferences}(f)  there is a strong branching chain that forms a (red) loop.
The values at the edges forming the loop are stronger in the position force network. For the interaction force network, 
value at the edge next to the crystalline region is small.  A larger value in the position force network is caused by the presence
of particles in the crystalline region.  This  difference implies that loops are formed at lower force levels in the interaction force networks 
as compared to  the position force networks which, in turn, implies  that there should be fewer points with relatively
large  birth values in the $\beta_1$ persistence diagram of  the interaction force network as compared to  the position force network.  
This is corroborated by Figures~\ref{fig:r00mu00_PD}(b) and (c). 

In the rest of this paper we use the following convention. If the feature persists until the zero threshold  then  we set the death coordinate to minus  one. This only impacts the persistence diagram for the interaction force network, allowing for simple visual identification of the defects. This convention is solely  for visualization proposes and is not used for distance computations. 

Another striking difference between the position and interaction force
networks is that in the $\beta_1$ persistence diagram of  the position force network the death value
for all points is $-1$, i.e., once a loop is formed it never dies. This is not the case for  the $\beta_1$ persistence diagram of the interaction network.  There is no reason that loops in the position force network cannot die, but a possible
explanation  is as follows. The death of loops is associated with the appearance of $2$-simplices (triangles) that is indicative of crystalline structure.   
Consider a single $2$-simplex $\ang{v_i,v_j,v_k}$ and assume that 
\[
f(\ang{v_i}) > f(\ang{v_j}) > f(\ang{v_k}).
\]
Given the definition of the position force network, the vertex $\ang{v_i}$ appears first, followed by the
vertex $\ang{v_j}$ and the edge $\ang{v_i,v_j}$. Finally, the vertex $\ang{v_k}$, the edges $\ang{v_i,v_k}$ and
$\ang{v_j,v_k}$ and the $2$-simplex $\ang{v_i,v_j,v_k}$ are all included at the same step. Thus there is no opportunity
for a loop consisting of three edges to be generated. A similar argument can be made for the interaction network
and hence loops that appear in the persistence diagrams must involve multiple edges.  If we think of this sequence of edges as a `force chain', then
the previous argument suggests that for the position force network this chain is more likely to contain edges of lower
magnitude than in the interaction force network. At the same time, we have observed that in crystalline regions the force
magnitudes
at the particles are larger in the position than in the interaction force network. 
These two observations suggest that in the position force network  it is
difficult to construct a loop in a crystalline region that surrounds vertices with lower forces.

Let us now consider briefly the digital force networks.   
Though not obvious from Figure~\ref{fig:r00mu00_PD}
 the $\beta_0$ persistence diagrams indicate that  the digital force network have less points 
than those for the position or interaction network. 
This is due to the fact that, as is discussed in Section~\ref{sec:particle}, our construction of the digital network artificially inflates
area associated with each particle and  hence it is possible for distinct components in the position 
or interaction force network  to form a single component in the digital force network. 
This effect is particularly relevant in the context of rattlers, 
the particles that do not experience any force.   In contrast,  the number of persistence generators in $\beta_1$ persistence 
diagrams is larger for digital force networks due to the formation of the artificial loops, again as described in Section~\ref{sec:particle}.

In this section we have shown that the information that is available about a granular system influences
the detected properties of the force networks.     The interaction force networks, that are based on the information 
about the forces between interacting particles, provide most precise and reliable information and we concentrate
on these networks from now on.

\subsection{Comparison of different systems via persistence diagrams}
\label{sec:comparison_diagrams}

The most direct means of applying persistence diagrams is to use them to  distinguish and/or interpret  the global
force structures of DGM composed of particles characterized by different physical properties.    This approach 
was followed in our recent work~\cite{pre13}, where we considered the number of generators in different
parts of the persistence diagrams as the systems that differed  by their physical properties were compressed.
Here we describe how persistence diagrams can be used to extract a significant amount of information about
interaction force networks by considering only two snapshots, at $\rho = 0.86$, of the following two systems: 
a monodisperse  frictionless ($r_p = 0,~\mu = 0$) and a polydisperse frictional ($r_p = 0.4,~\mu = 0.5$) system. 
We already know that these two systems behave differently under compression~\cite{pre13}, 
and we use persistence diagrams to illustrate these differences.   The reader should note that the focus
here is on illustration of the technique and on interpretation of a limited set of results: more general (although 
less detailed) discussion that concentrates on physical interpretation can be found in~\cite{pre13}.     In particular,
note that the jamming points for these two systems differ, so that by considering the same $\rho$, we consider
two systems which are at  different distances from  the  jamming transition.

\begin{figure}[t]
\centering
	\begin{picture}(220,190)
	\put(0,0){\includegraphics[width=3.5in]{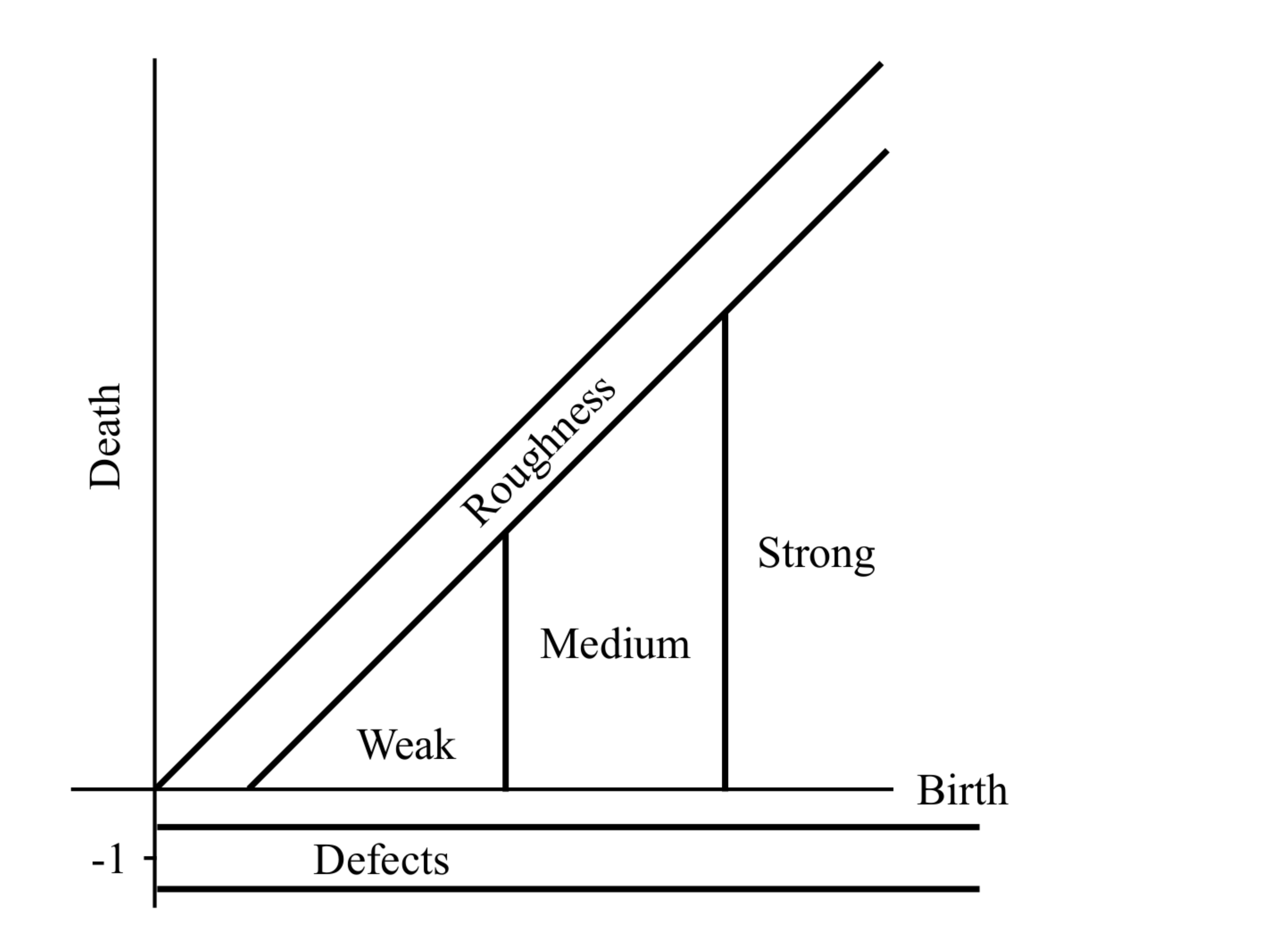}}
	\end{picture}
\caption{Persistence diagram divided into regions. Explanation of the regions is provided in the text.}
\label{fig:PD_division}
\end{figure}

We begin by assigning  physical meaning to the location of persistence points in the persistence diagrams.
Figure~\ref{fig:PD_division} shows a persistence diagram divided into five regions. With the exception of the region labelled
defects, the location of the division lines is intended to be either  system specific or  conceptual.   

\begin{figure}[t]
\centering
	\begin{picture}(220,350)
      	\put(0,130){\includegraphics[width=3in]{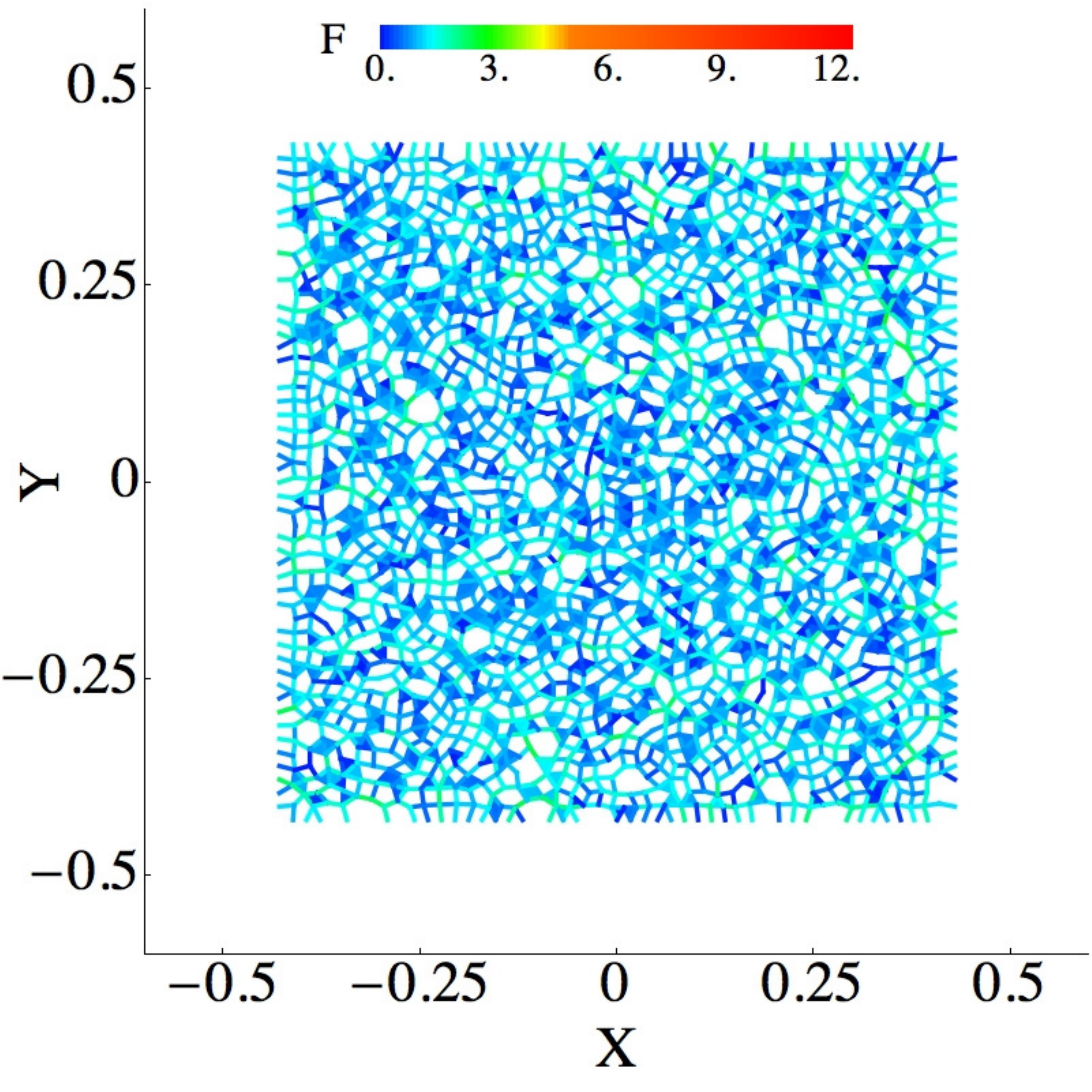}}
	\put(-15,5){\includegraphics[width=1.6in]{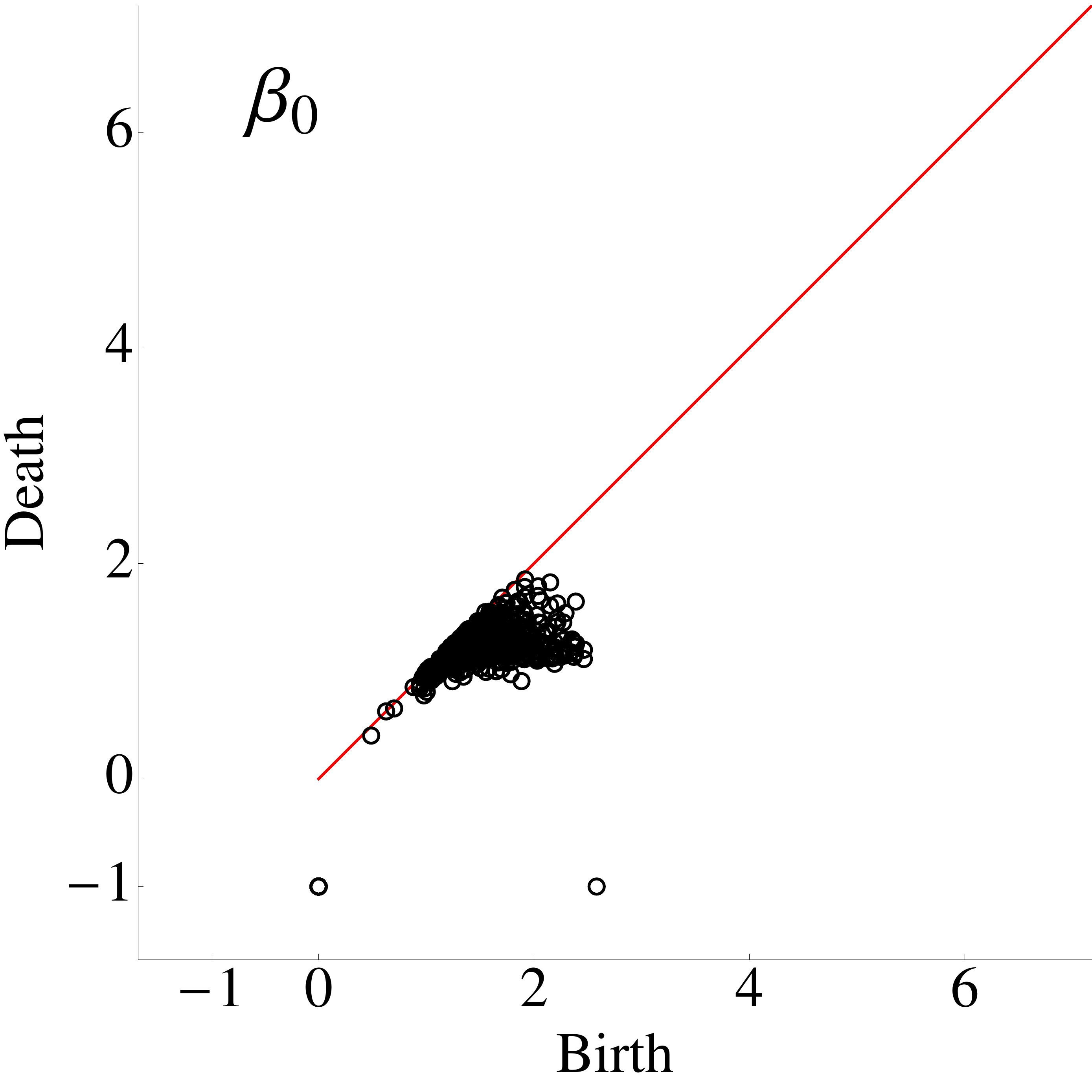}}
	\put(115,5){\includegraphics[width=1.6in]{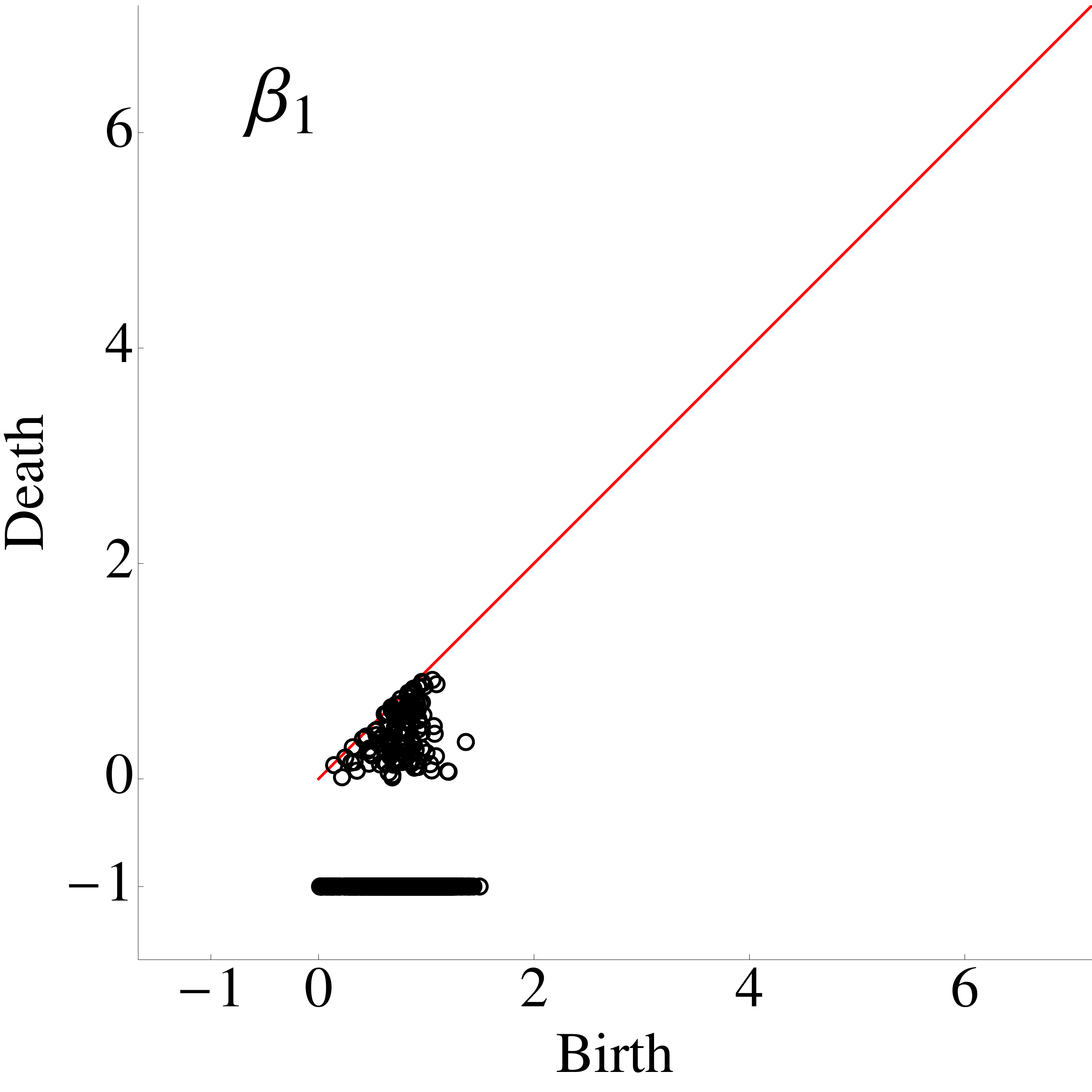}}

	\put(25,135){(a)}
	\put(0,0){(b)}
	\put(135,0){(c)}
	      
	\end{picture}
\caption{(a) Non-zero simplices of the interaction force network based on normal force. (b) $\beta_0$ (c) $\beta_1$ persistence diagrams for the 
$r_p = 0.4$, $\mu = 0.5$ system  at  $\rho=0.86$. 
}
\label{fig:r04mu05}
\end{figure}
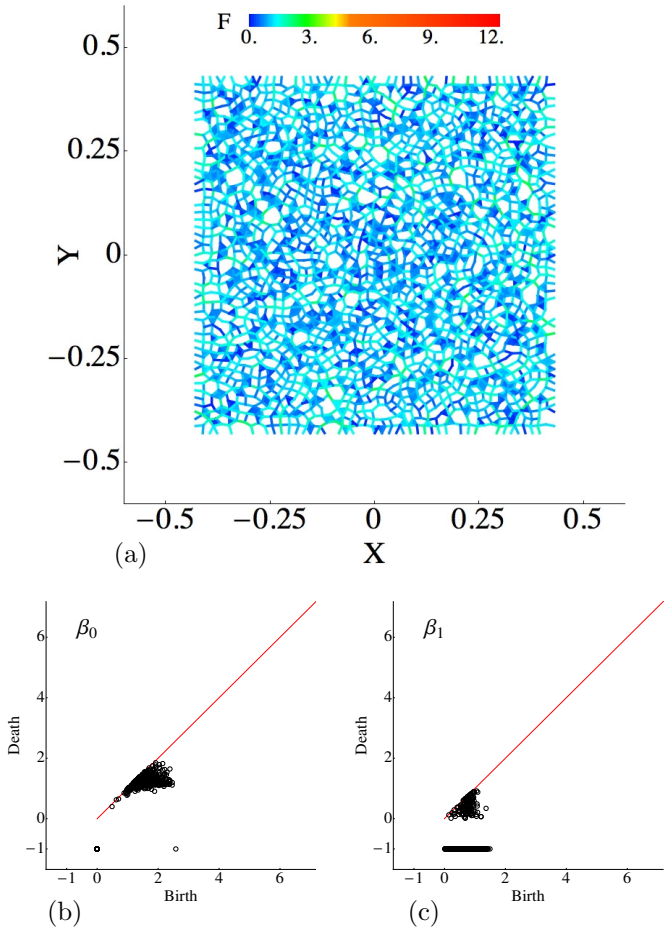

{\it Roughness.}
 The geometric features corresponding to the points in the region labeled roughness persist only over a small range of force values. For instance the point $(\theta_3,\theta_2)$ in the $\beta_0$ persistence diagram shown in Figure~\ref{fig:ToyDiagram1D} corresponds to a feature that persists over a relatively  short range compared with the other features in the same figure. There are at least two different interpretations of the points in the $\beta_0$ persistence diagram that lie
in this region. The first is to treat these points as noise, i.e.\ a byproduct of the imperfect 
measurements of the normal forces between particles. While this may be appropriate for many experimental settings, the 
data  represented in Figures~\ref{fig:r00mu00_networks}, \ref{fig:r00mu00_PD} and \ref{fig:r04mu05} come from  simulations, and the errors are small.
 This fact leads to the  second interpretation, which we adopt, that this region of the persistence diagram provides information about small fluctuations of the forces.
These fluctuations can be interpreted as a  measurement of how rough
or bumpy the normal force landscape is, e.g.\ should we view the surface of the landscape as being made of glass or 
sandpaper? Therefore by comparing Figures~\ref{fig:r00mu00_PD}(c) and \ref{fig:r04mu05} we  conclude that for the considered $\rho$,  
$r_p = 0.4,~\mu = 0.5$ system is rougher than  the $r_p = 0,~\mu = 0$.

{\it Strong.} To understand the region labelled as strong, observe that the image in  Figure~\ref{fig:r04mu05}(a) of the forces for  the $r_p = 0.4, \mu = 0.5$ system
does not contain any red simplices, implying that there are no  strong force interactions.  In contrast,
such red simplices are present in the $r_p = 0,~\mu = 0$ system displayed in Figure~\ref{fig:r00mu00_networks}(c).  
This difference can be inferred from the $\beta_0$ persistence diagrams 
 shown in Figures~\ref{fig:r00mu00_PD} and \ref{fig:r04mu05}. 
For the $r_p = 0.4,~\mu = 0.5$ system there are no persistence points with the birth value larger than $3$ (in terms of average interaction 
force) and only a few points with birth value larger than $2.5$.  Thus, depending on the exact cut-off there are no or at most few points in the 
region marked strong for the $r_p = 0.4,~\mu = 0.5$ system, in clear contrast to the $r_p = 0,~\mu = 0$ system, for the considered value of $\rho$.

{\it Medium.}
If we take the left division marker for the medium regime in Fig.~\ref{fig:PD_division}  to be $1$, then the persistence points in the medium  and strong regions provide information about the geometry  of strong contacts.
For the $r_p = 0.4,~\mu = 0.5$ system, we see a large number of $\beta_0$ persistence points  that are born between $1$ and $2.5$
and die before $0.8$.   
This suggests a landscape consisting of moderately high peaks separated by moderately high
valleys.  To continue the geographic metaphor, the $r_p = 0.4,~\mu = 0.5$  force chain network takes place on a high plateau. 
 In contrast,  the $r_p = 0,~\mu = 0$ system has fewer moderately high peaks, but they are separated by much deeper valleys since there are
points with death values below $0.6$.   Therefore, we conjecture that, for the considered $\rho$, 
landscape for the $r_p = 0,~\mu = 0$ system has fewer peaks (but some of them are strong) than that of the
$r_p = 0.4,~\mu = 0.5$  landscape, and these peaks are  in general
much more isolated and more likely to be separated by valleys of much weaker forces.

Let us summarize the previous paragraphs in the terms of 'force chains'. There are no force chains with force value 
exceeding three times the average value for the considered $r_p = 0.4,~\mu = 0.5$ system. These kind of force chains are present  
only  in the $r_p = 0,~\mu = 0$ system. On the other hand the number of `force chains' with strongest link exceeding the 
average force is much larger for $r_p = 0.4,~\mu = 0.5$ system. Fact that $\theta_d > 0.8$ for most of the points in the 
$\beta_0$ diagram   implies that in the $r_p = 0.4,~\mu = 0.5$ system there are links connecting the `force chains' with 
forces larger than $0.8$ times the average force. The connections between the `force chains' in the $r_p = 0,~\mu = 0$ 
system tend to be much weaker.

{\it Defects.}
In a $\beta_0$ persistence diagram, each point in this region  corresponds
to a distinct connected component of contact network.  
In the context of the $r_p = 0.4,~\mu = 0.5$ system, these points mostly correspond to rattlers. 
This conclusion is obtained by observing that aside from the single persistence point corresponding to 
a large birth force, that corresponds to the component containing most of the particles, the persistence points in the 
defects region have a birth value of $0$, indicating that they are not experiencing any normal force.  This is quite different from
the $r_p = 0,~\mu = 0$ system. In this case we have persistence points in the defects region with non-zero birth forces. 
This implies the existence of small clusters of particles (a single separated particle does not experience a force) that are not interacting with the 
dominant particle cluster.  Close inspection of the interaction force network in Figure~\ref{fig:r00mu00_networks}(c)
reveals these small components. 

{\it Weak.}
Finally, the points in the region labeled weak represent small clusters of particles, interacting weakly with the 
dominant particle cluster. Inspection of the $\beta_0$ persistence diagram reveals existence of these clusters  for  
the $r_p = 0,~\mu = 0$ system while they are virtually absent for the $r_p = 0.4,~\mu = 0.5$ system, for the considered $\rho$.

The defects region of the $\beta_1$ persistence diagrams provides additional information.
As indicated in Section~\ref{sec:function_complex},  $\beta_1$ persistence points lie  in the defects region if and only if
they correspond to loops that enclose non-crystalline regions.  
There are about  twice as many  persistence points in the defects region
in the $r_p = 0.4,~\mu = 0.5$ system as compared to the $r_p = 0,~\mu = 0$ one. 
This suggests that the $r_p = 0.4,~\mu = 0.5$ system is more likely to support defects, for the 
considered value of $\rho$.    At the same time there are $50\%$ more points in  $\beta_1$ persistence diagram that are not
in the defects region for the $r_p = 0.4,~\mu = 0.5$ system. These persistence points
correspond to loops that are filled in by $2$-dimensional simplices and 
thus must be contained within crystallized regions. Thus this
difference in the number of persistence points suggests
that the $r_p = 0.4,~\mu = 0.5$ system contains a multitude of small
crystalline regions as opposed to the $r_p = 0,~\mu = 0$  system.  
This is corroborated by a  careful examination of the force networks in  Figures~\ref{fig:r00mu00_networks}(c) and \ref{fig:r04mu05}.

The finding above may appear surprising since it is known that frictionless monodisperse systems may 
crystallize, as it was also shown in our earlier work~\cite{epl12}.  The resolution of this apparent contradiction is that polydisperse
systems considered appear to form large number of small (involving only a few particles) crystalline regions, in contrast to monodisperse
frictionless ones that are expected to form large crystalline zones involving many particles~\cite{epl12}.  

\begin{figure}[t]
\centering
	\begin{picture}(220,350)
      	\put(0,130){\includegraphics[width=3in]{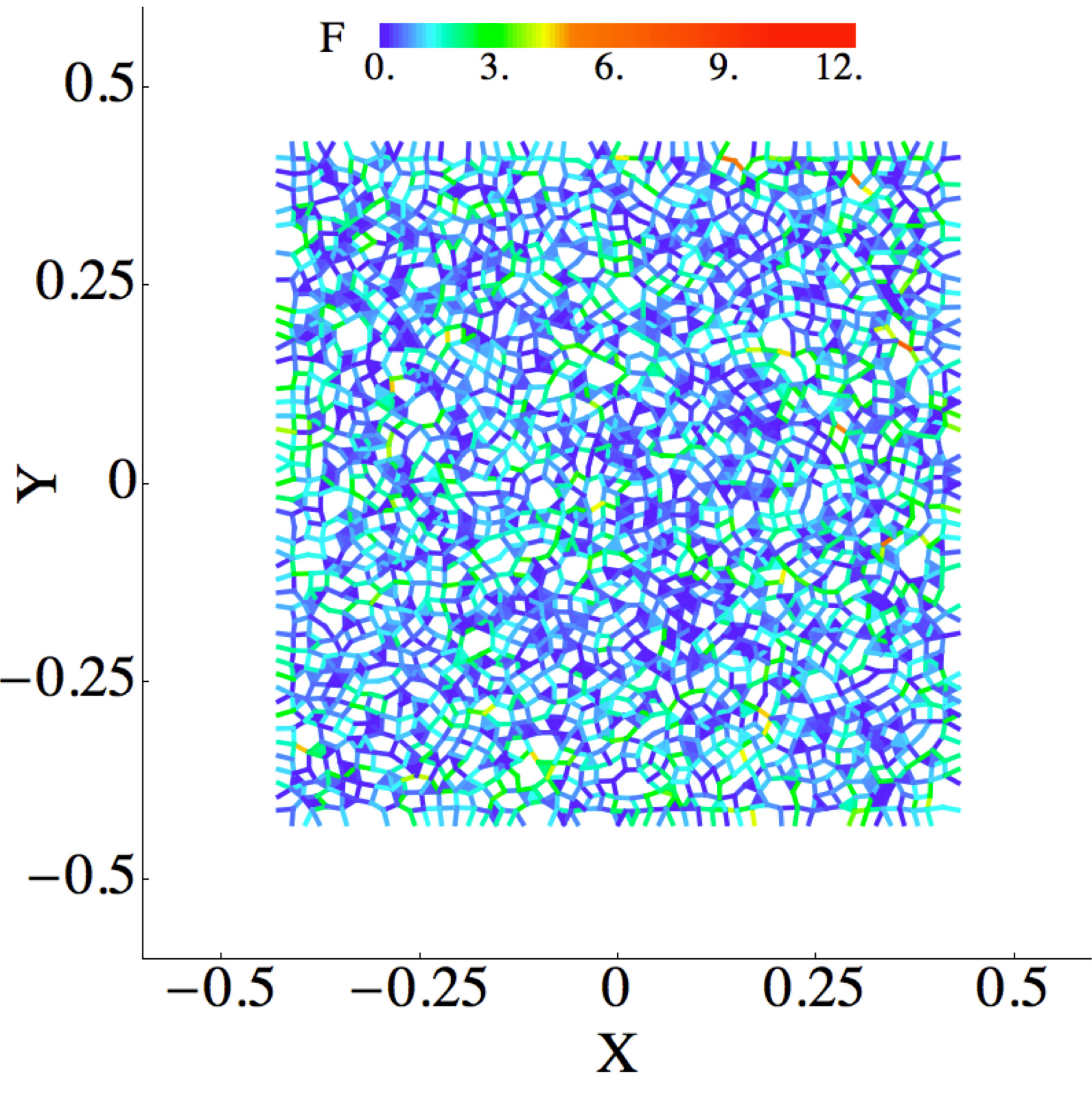}}
	\put(-15,5){\includegraphics[width=1.6in]{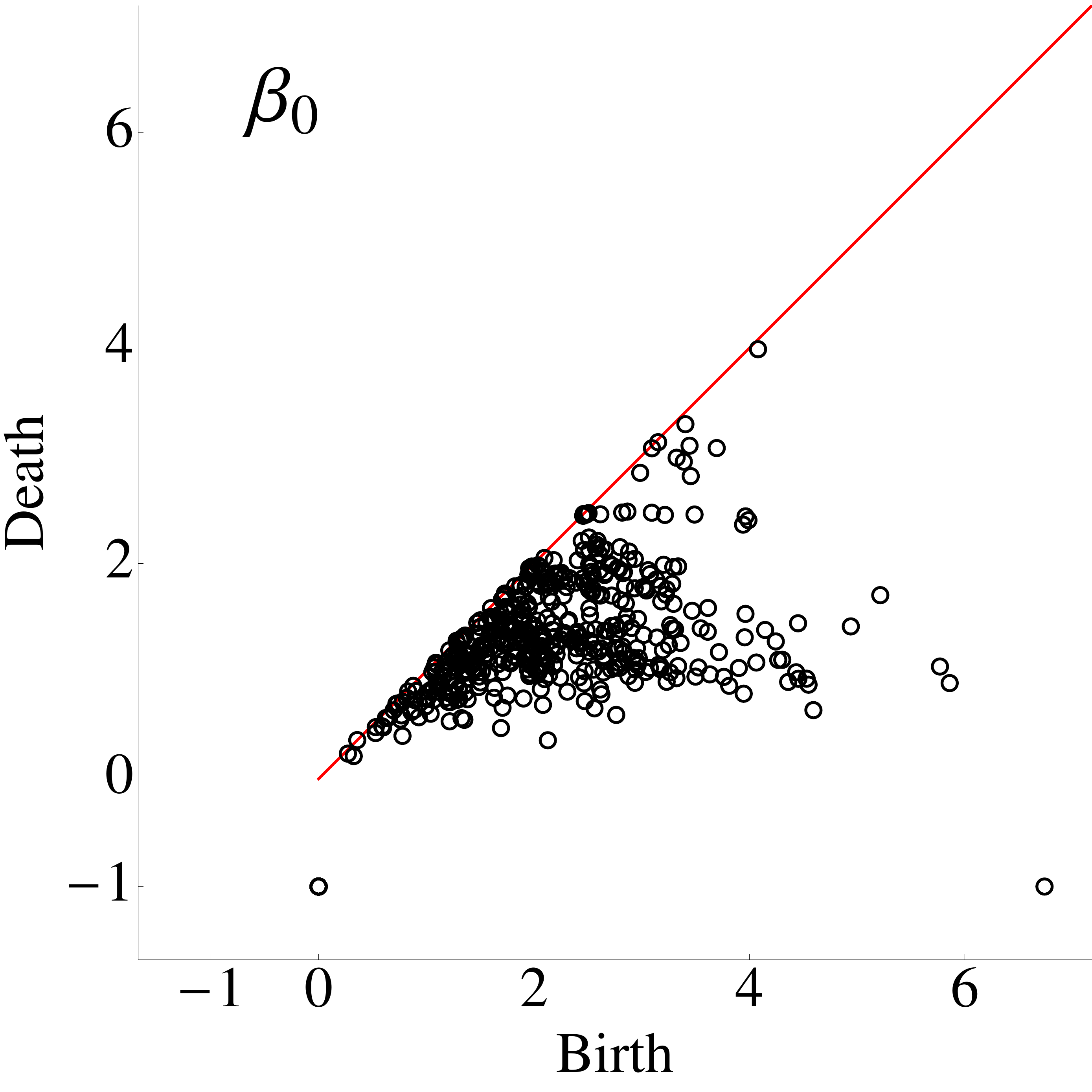}}
	\put(115,5){\includegraphics[width=1.6in]{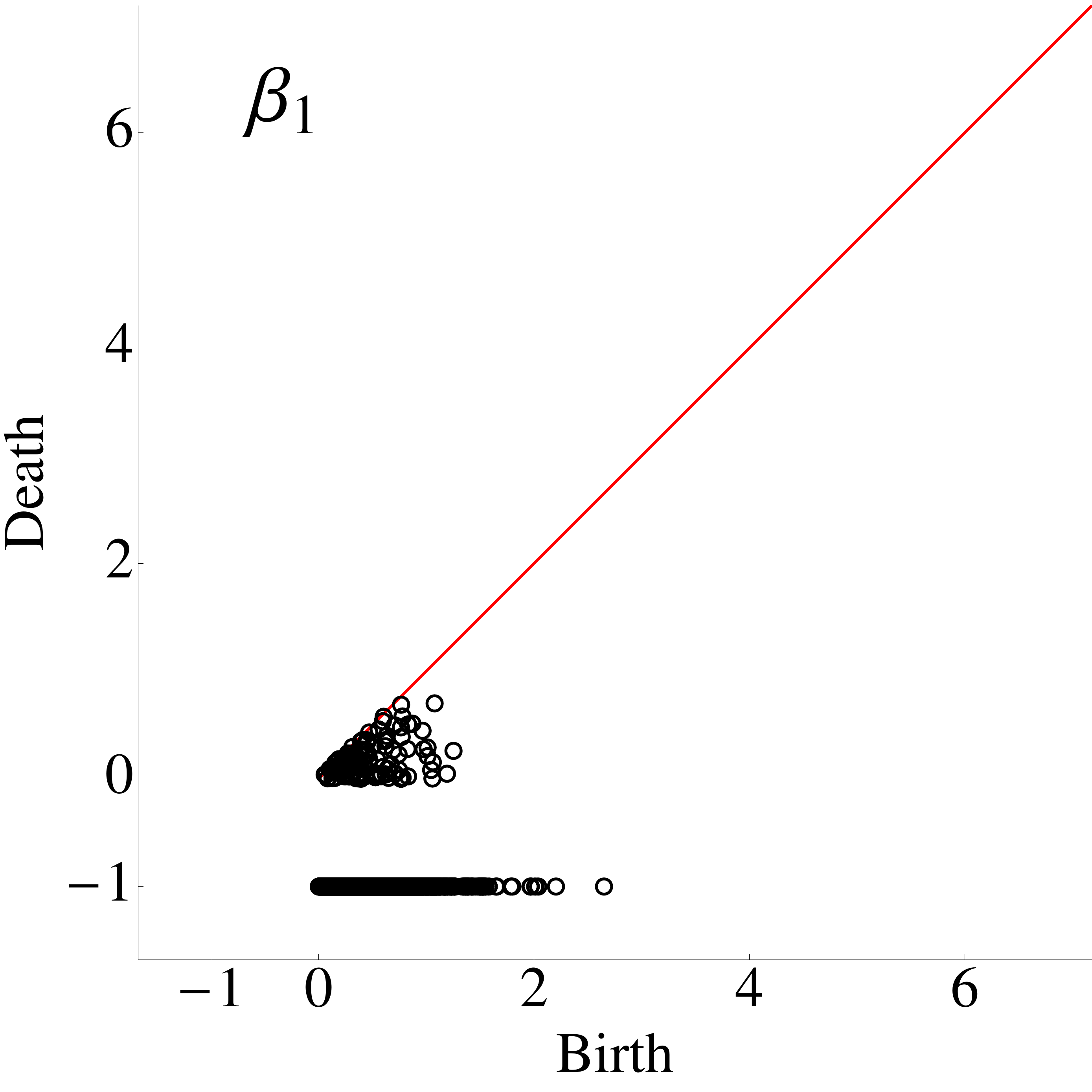}}
      	
	\put(25,135){(a)}
	\put(0,0){(b)}
	\put(135,0){(c)}

	\end{picture}
\caption{(a) 
Non-zero simplices of the interaction force network based on tangential force. (b) $\beta_0$ (c) $\beta_1$ persistence diagrams for the 
$r_p = 0.4$, $\mu = 0.5$ system  at  $\rho=0.86$ (the same particle configuration as in Fig.~\ref{fig:r04mu05}).   Note different range in (b) compared
to Fig.~\ref{fig:r04mu05}(b).  
 }
\label{fig:r04mu05T}
\end{figure}
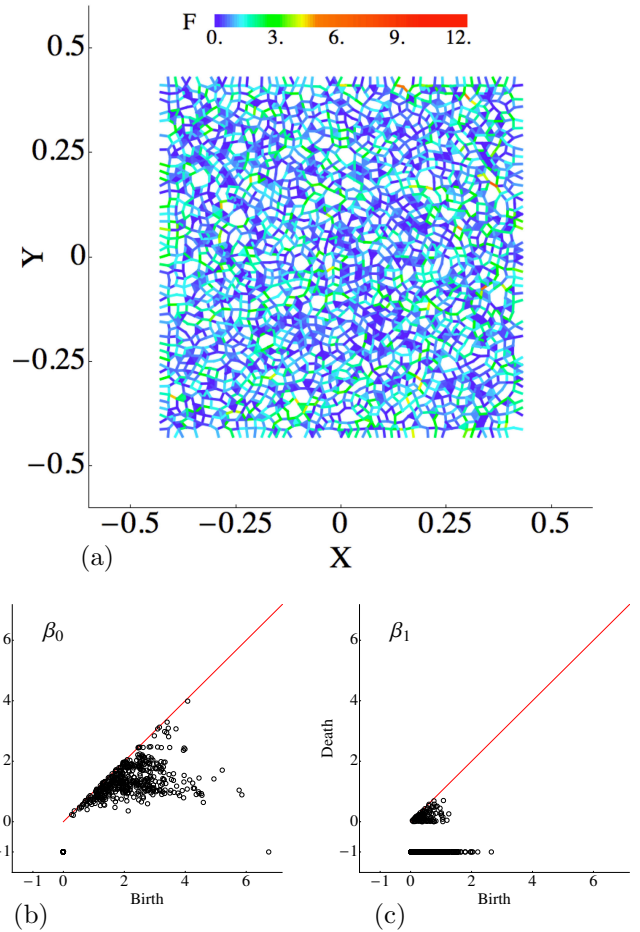

{\it Tangential forces.}
Analogous analysis can be done for the force network based on tangential forces. 
Figure~\ref{fig:r04mu05T} shows the tangential force network together with the persistence diagrams for the same
particle configuration as in  Figure~\ref{fig:r04mu05}; note that the tangential forces are normalized by the average
tangential force. While one could argue that visual comparison of Figure~\ref{fig:r04mu05}(a) and Figure~\ref{fig:r04mu05T}(a) is 
not particularly insightful, examining the persistence diagrams gives much more information.  The significant 
and crucial difference is that the tangential forces go to much larger values than the normal ones and consequently the
generators for connected components as well as loops cover much larger range.    This is particularly visible when considering
$\beta_0$ generators shown in Figure~\ref{fig:r04mu05T}(b).      Therefore, the analysis based on persistence diagrams 
suggest that the landscape defined by tangential forces is significantly more `mountainous' with much higher mountain tops.  
Here, we give only an example of tangential force results; future work should analyze how tangential forces evolve as system
is exposed to shear or compression, as done for normal forces in~\cite{pre13}.

\subsection{Comparison of different systems via distances.}
\label{sec:distances}

\begin{figure*}[t]
\begin{picture}(0,120)
{\includegraphics[width=7.4in]{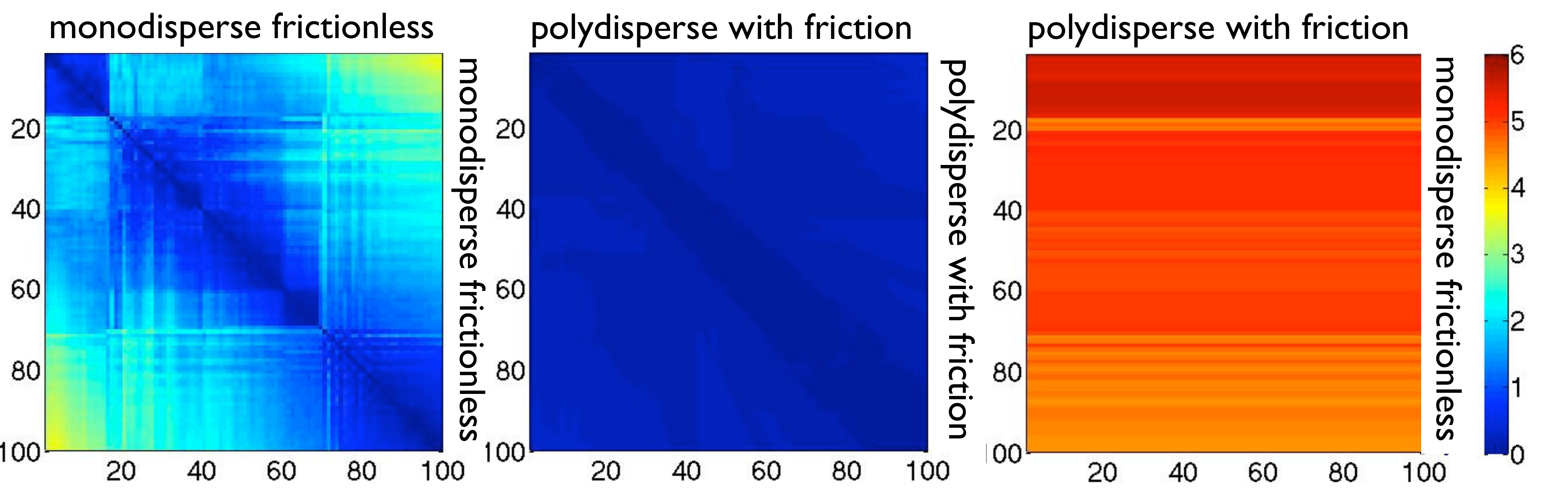}}

\put(-460,-8){(a)}
\put(-295,-8){(b)}
\put(-125,-8){(c)}
\end{picture}
\caption{Distance matrices for the bottleneck distance. The value $D(i,j)$ is the $d_B$ distance between the state 
$i$ and $j$ for the  (a) $r_p = 0,~\mu = 0$ system (b) (a) $r_p = 0.4,~\mu = 0.5$ system. (Note that by construction
$D$ is a symmetric matrix.)  (c) Comparison of the  
$r_p = 0,~\mu = 0$ system with the $r_p = 0.4,~\mu = 0.5$ system.  The states of the $r_p = 0.4,~\mu = 0.5$  
system change along the horizontal axis and the states of the $r_p = 0,~\mu = 0.5$  along the vertical axis.  }
\label{fig:B}
\end{figure*}

\begin{figure*}[t]
\begin{picture}(0,180)
{\includegraphics[width=7.4in]{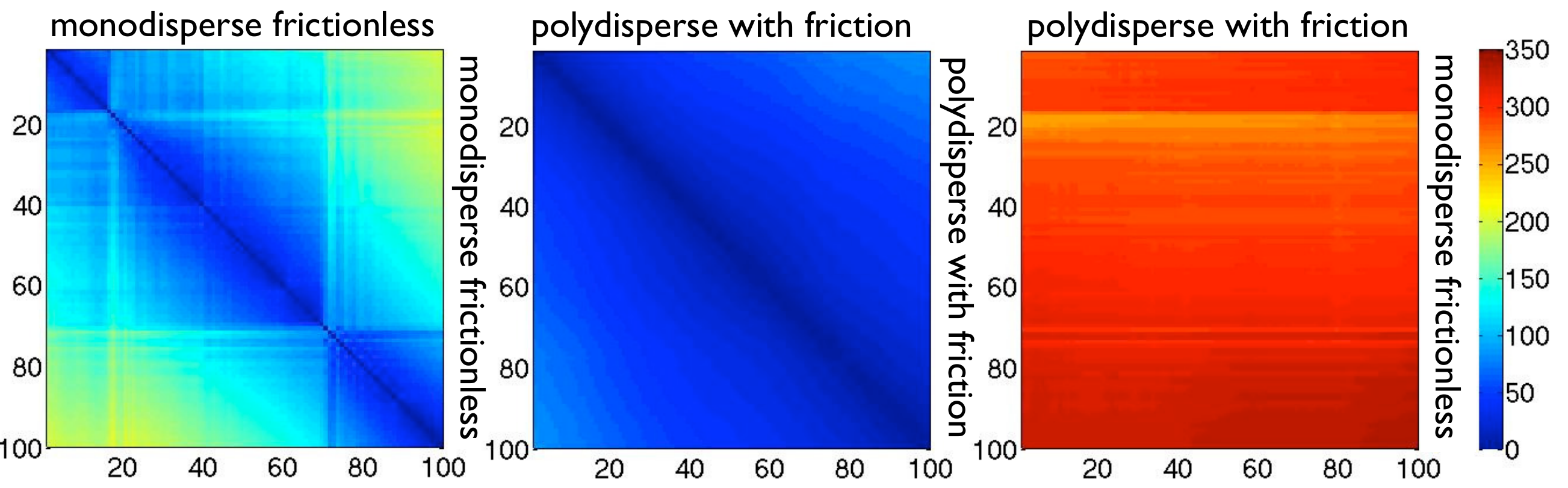}}

\put(-460,-8){(a)}
\put(-295,-8){(b)}
\put(-125,-8){(c)}
\end{picture}
\caption{
Distance matrices for the degree-$1$ Wasserstain distance, considering the same systems as in Fig.~\ref{fig:B}.}
\label{fig:W1}
\end{figure*}

Up to this point we have focussed on interpretation of individual persistence diagrams.  However,
we believe that the most significant value of this technique will come through the analysis of large sets of persistence diagrams. 
We present a simple example of this idea here, leaving more detailed investigations for future work.

A metric in the space of persistence diagrams measures the level of similarity between different states of DGM. If the states 
are similar the distance is small. By using different metrics we can access different notions of similarity. 
Bottleneck distance informs us about a single dominant change. Notion of similarity induced by this distance 
has a local character since it measures only the largest difference and ignores all the other changes. Overall 
similarity of two states is measured by Wasserstein $d_{W^1}$ distance which sums up all the differences 
between the states. Since large number of small differences can result in large $d_{W^1}$, this metric is sensitive to noise.  
This sensitivity can be mitigated by using $d_{W^q}$ for $q > 1$.

We use these metrics to analyze and compare the time evolution of the two considered DGM systems; to
facilitate the comparison, we chose two different $\rho$'s, that are at the similar  distance (in packing fraction) from the jamming
points.  In the present work,  the jamming point is loosely defined as a packing fraction, $\rho_J$, at which $Z \approx 3$. We use $\rho = 0.87$ for 
$r_p = 0$, $\mu = 0$ with  $\rho_J = 0.853$ and  $\rho = 0.786$ for $r_p = 0.4$, $\mu = 0.5$ with  $\rho_J = 0.786$.   The difference between
$\rho$'s of the  consecutive states is fixed to $\Delta \rho = 8\times 10^{-5}$ in both cases.

Figures~\ref{fig:B} and~\ref{fig:W1} show the distances between  all these states ($100$ of them).   Let us
focus first on the parts (a) of these two figures, showing the results for  $r_p = 0$, $\mu = 0$ system.
Figures~\ref{fig:B}(a) and \ref{fig:W1}(a) are the color coded distance matrices $D$ for the $d_B$ and $d_{W^1}$ distances.
The entry $D(i,j)$ is the distance between the state $i$ and $j$. Thus this is a symmetric matrix, $D(i,j)=D(j,i)$ and 
$D(i,i) = 0$ (black) for all $i,j$. 

Figures~\ref{fig:B}(a) and \ref{fig:W1}(a), upper left corners, show that for the first ($17$) states 
considered, the states are similar suggesting that the system is evolving slowly. 
Though less pronounced this is also the case for the last $20$ or so states (bottom right corner).  
The fact that the upper right corner is yellow indicates that the structure of the forces has evolved over the 
range of $\rho$'s ($10^{-3}$ in the considered case).    Therefore, the persistent homology can be used
to capture the evolution of the force networks.   

These comments are applicable to both the $d_B$ and $d_{W^1}$ distance matrices.
However, the metrics carry different information. The $d_B$ matrix implies that for the first 17 or last 20 steps
there is no single location at which a large change in the force structure occurs.
The corresponding $d_{W^1}$ matrix entries, which are approximately 50 times larger, suggests that (i) 
many small changes in the force structure are taking place from step to step, and (ii) that from step to step
roughly the same number of small changes are occurring.  To provide an analogy; there are ripples on the pond, but
the number of ripples are not changing with time.

The abrupt color change (increase in distance) at $D(17,18)$ indicates that there has been a significant structural
change of the forces of the DGM from state $17$ to $18$. Another significant transition can be seen between 
$70$ and $71$. The evolution in the region between the states  $20$ and $70$  has a similar character. 
Parts with slow evolution are separated by sudden transitions. The fact that we see these transitions more clearly
in the $d_B$ matrix as opposed to the $d_{W^1}$ matrix indicates that when significant changes occur, they are localized
in space. Conversely, since the two dominant transitions at $17$ and $70$ are clearly seen in the $d_{W^1}$ matrix
suggests that they are more global in  nature.

The distance matrices for $r_p = 0.4$, $\mu = 0.5$ system are shown in Figures~\ref{fig:B}(b) and \ref{fig:W1}(b).
The fact that distances between all the states are small and the relatively broad band of dark blue along the diagonal
implies that the evolution of this system is much slower and without any abrupt changes. This is profoundly different
from the behavior of the $r_p = 0$, $\mu = 0$ system, for the considered range of $\rho$'s.

We give a brief physical interpretation of the results shown in Figs.~\ref{fig:B} and~\ref{fig:W1}; 
more complete and detailed discussing will be given elsewhere.
The results shown in these figures suggest that at least for the considered narrow range of $\rho$'s, the evolution of 
$r_p =0,~\mu = 0$ system is abrupt, evolving through large scale rearrangements, while the evolution of 
$r_p =0.4,~\mu = 0.5$ is comparably smooth.   This is consistent with our earlier results~\cite{epl12, pre13}, where
we show that the $r_p =0,~\mu = 0$ system tends to form polycrystalline zones, while the $r_p =0.4,~\mu = 0.5$
system is much more disordered.   The large distances between the states $(17,18)$ and $(70,71)$ shown in
Figs.~\ref{fig:B} and~\ref{fig:W1} are related to breaking up of these polycrystalline zones.    

We end this section by directly comparing the two systems. The entry $D(i,j)$ of the distance matrices shown in  
Figure~\ref{fig:B}(c) and  \ref{fig:W1}(c) is the $d_B$ and $d_{W^1}$ distance, respectively, between the 
state $i$ of the system $r_p = 0$, $\mu = 0$ and state $j$ of the system $r_p = 0.4$, $\mu = 0.5$.
Since the differences between the systems are much larger than differences within a single system, the two systems 
can be clearly distinguished.

\section{Conclusion}

Based on different methods for collecting data we have defined three different chain complexes and used them to 
construct force networks for particulate systems. Using the force networks we compute persistence diagrams and discuss how one can 
use persistent homology to extract information about the geometric structure of the force distributions between the particles.
We provide both theoretical and numerical arguments to show that the persistence diagrams obtained 
from interaction force networks are the most robust with respect to numerical errors in input data.
Using numerical data obtained from discrete element simulations of a system of slowly compressed particles, we show that the
persistence diagrams associated to the different force networks can differ significantly. This in turn implies that
the geometry of the force distributions observed depends upon the  methods by which the system is sampled. 
We provide some intuition concerning how in general the sampling method affects the geometry.  We also demonstrate that
using persistent homology of any of the three force networks allows one to draw meaningful distinctions between the
properties of the force distributions of different systems.    These results now provide a solid mathematical background
for analysis of the force network in more complex configurations, including systems going through jamming transitions,
systems consisting of particles of different shapes, and particulate systems in 3D.   Our research in these directions is 
currently in progress.

\section*{Acknowledgments}
This work was partially supported by 
NSF-DMS-0835621, 0915019, 1125174, AFOSR and DARPA
(A.G., M.K., and K.M) and NSF Grant No. DMS-0835611, and DTRA Grant No. 1-10-1-0021 (A.G.\ and L.K.).

\bibliographystyle{ieeetr}

\begin{thebibliography}{10}

\bibitem{brujic03}
J.~Brujic, S.~Edwards, I.~Hopkinson, and H.~Makse, ``Measuring the distribution
  of interdroplet forces in a compressed emulsion system,'' {\em Physica A:
  Statistical Mechanics and its Applications}, vol.~327, p.~201, 2003.

\bibitem{cates98}
M.~E. Cates, J.~P. Wittmer, J.-P. Bouchaud, and P.~Claudin, ``Jamming, force
  chains and fragile matter,'' {\em Phys. Rev. Lett.}, vol.~81, p.~1841, 1998.

\bibitem{liu95}
C.~Liu, S.~R. Nagel, D.~A. Schecter, S.~N. Coppersmith, S.~Majumdar,
  O.~Narayan, and T.~A. Witten, ``Force fluctuations in bead packs,'' {\em
  Science}, vol.~269, p.~513, 1995.

\bibitem{majmudar05a}
T.~S. Majmudar and R.~P. Behringer, ``Contact force measurements and
  stress-induced anisotropy in granular materials,'' {\em Nature}, vol.~435,
  p.~1079, 2005.

\bibitem{alexander_physrep05}
S.~Alexander, ``Amorphous solids: {T}heir structure, lattice dynamics and
  elasticity,'' {\em Phys. Rep.}, vol.~296, p.~65, 1998.

\bibitem{radjai_96b}
F.~Radjai, M.~Jean, J.~J. Moreau, and S.~Roux, ``Force distribution in dense
  two-dimensional granular systems,'' {\em Phys. Rev. Lett.}, vol.~77, p.~274,
  1996.

\bibitem{radjai98b}
F.~Radjai, D.~E. Wolf, M.~Jean, and J.-J. Moreau, ``Bimodal character of stress
  transmission in granular packings,'' {\em Phys. Rev. Lett.}, vol.~80, p.~61,
  1998.

\bibitem{ostojic06}
S.~Ostojic, E.~Somfai, and B.~Nienhuis, ``Scale invariance and universality of
  force networks in static granular matter,'' {\em Nature}, vol.~439, p.~828,
  2006.

\bibitem{tighe_sm10}
B.~P. Tighe, J.~H. Snoeijer, T.~J.~H. Vlugt, and M.~van Hecke, ``The force
  network ensemble for granular packings,'' {\em Soft Matter}, vol.~6,
  pp.~2908--2917, 2010.

\bibitem{peters05}
J.~Peters, M.~Muthuswamy, J.~Wibowo, and A.~Tordesillas, ``Characterization of
  force chains in granular material,'' {\em Phys. Rev. E}, vol.~72, p.~041307,
  2005.

\bibitem{tordesillas_bob_pre12}
A.~Tordesillas, D.~M. Walker, G.~Froyland, J.~Zhang, and R.~Behringer,
  ``Transition dynamics and magic-number-like behavior of frictional granular
  clusters,'' {\em Phys. Rev. E}, vol.~86, p.~011306, 2012.

\bibitem{tordesillas_pre10}
A.~Tordesillas, D.~M. Walker, and Q.~Lin, ``Force cycles and force chains,''
  {\em Phys. Rev. E}, vol.~81, p.~011302, 2010.

\bibitem{daniels_pre12}
D.~Bassett, E.~Owens, K.~Daniels, and M.~Porter, ``Influence of network
  topology on sound propagation in granular materials,'' {\em Phys. Rev. E},
  vol.~86, p.~041306, 2012.

\bibitem{herrera_pre11}
M.~Herrera, S.~McCarthy, S.~Sotterbeck, E.~Cephas, W.~Losert, and M.~Girvan,
  ``Path to fracture in granular flows: Dynamics of contact networks,'' {\em
  Phys. Rev. E}, vol.~83, p.~061303, 2011.

\bibitem{walker_pre12}
D.~Walker and A.~Tordesillas, ``Taxonomy of granular rheology from grain
  property networks,'' {\em Phys. Rev. E}, vol.~85, p.~011304, 2012.

\bibitem{arevalo_pre10}
R.~Ar\'evalo, I.~Zuriguel, and D.~Maza, ``Topology of the force network in
  jamming transition of an isotropically compressed granular packing,'' {\em
  Phys. Rev. E}, vol.~81, p.~041302, 2010.

\bibitem{arevalo_pre13}
R.~Ar\'evalo, L.~Pugnaloni, I.~Zuriguel, and D.~Maza, ``Contact network
  topology in tapped granular media,'' {\em Phys. Rev. E}, vol.~87, p.~022203,
  2013.

\bibitem{epl12}
L.~Kondic, A.~Goullet, C.~O'Hern, M.~Kramar, K.~Mischaikow, and R.~Behringer,
  ``Topology of force networks in compressed granular media,'' {\em Europhys.
  Lett.}, vol.~97, p.~54001, 2012.

\bibitem{pre13}
M.~Kramar, A.~Goullet, L.~Kondic, and K.~Mischaikow, ``Persistence of force
  networks in compressed granular media,'' {\em Phys. Rev. E}, vol.~87,
  p.~042207, 2013.

\bibitem{carlsson}
G.~Carlsson, ``Topology and data,'' {\em Bull. Amer. Math. Soc. (N.S.)},
  vol.~46, p.~255, 2009.

\bibitem{edelsbrunner:harer}
H.~Edelsbrunner and J.~L. Harer, {\em Computational topology}.
\newblock Providence, RI: AMS, 2010.

\bibitem{trevijano_14}
S.~Ardanza-Trevijano, I.~Zuriguel, R.~Arevalo, and D.~Maza, ``A topological
  method to characterize tapped granular media from the position of the
  particles.'' preprint, 2013.

\bibitem{hartley_03}
R.~R. Hartley and R.~P. Behringer, ``Logarithmic rate dependence of force
  networks in sheared granular materials,'' {\em Nature}, vol.~421, p.~928,
  2003.

\bibitem{miro}
M.~Kramar, ``Chomp,'' 2013.
\newblock
  \url{http://chomp.rutgers.edu/TopologicalCharacterizationfOfDense/GranularMedia.html}.

\bibitem{perseus}
V.~Nanda, ``Perseus,'' 2012.
\newblock \url{http://www.math.rutgers.edu/~vidit/perseus.html}.

\bibitem{kaczynski:mischaikow:mrozek}
T.~Kaczynski, K.~Mischaikow, and M.~Mrozek, {\em Computational homology},
  vol.~157 of {\em Applied Mathematical Sciences}.
\newblock New York: Springer-Verlag, 2004.

\bibitem{cundall79}
P.~A. Cundall and O.~D.~L. Strack, ``A discrete numerical model for granular
  assemblies,'' {\em G\'eotechnique}, vol.~29, p.~47, 1979.

\bibitem{mischaikow:nanda:11}
K.~Mischaikow and V.~Nanda, ``Morse theory for filtrations and efficient
  computation of persistent homology,'' {\em Discrete \& Computational
  Geometry}, {to appear}.

\end{thebibliography}

\end{document}